\documentclass[twocolumn,showpacs,aps,floatfix,prd,nofootinbib]{revtex4}
\usepackage{graphicx}
\usepackage{dcolumn}
\usepackage{array, longtable}
\usepackage{epsfig}
\usepackage{amsmath}
\usepackage{colordvi}
\usepackage{hhline}

\newcommand{\BaBarYear}{07}
\newcommand{\BaBarNumber}{045}
\newcommand{\SLACPubNumber}{12753}

 \newcommand{\BaBarType}      {PUB}  

\input pubboard/babarsym

\def\Ecm          {\ensuremath {E_{\rm c.m.}}\xspace}
\def\fourpi       {\ensuremath {2(\pipi)}\xspace}
\def\fourpipn     {\ensuremath {2(\pipi)\piz}\xspace}
\def\fourpieta    {\ensuremath {2(\pipi)\eta}\xspace}

\def\KKpppn       {\ensuremath {\Kp\Km\pipi\piz}\xspace}
\def\KKppeta      {\ensuremath {\Kp\Km\pipi\eta}\xspace}
\def\chifourpi    {\ensuremath {\chi^2_{4\pi}}\xspace}
\def\chifourpipn  {\ensuremath {\chi^2_{4\pi\piz}}\xspace}
\def\chifourpieta {\ensuremath {\chi^2_{4\pi\eta}}\xspace}

\def\chiKKpppn    {\ensuremath {\chi^2_{2K2\pi\piz}}\xspace}
\def\chiKKppeta   {\ensuremath {\chi^2_{2K2\pi\eta}}\xspace}
\def\chiKKpppe    {\ensuremath {\chi^2_{2K2\pi\piz/\eta}}\xspace}
\def\ammm         {\ensuremath {(g\!-\!2)_\mu}\xspace}

\long\def\inst#1{\par\nobreak\kern 4pt\nobreak
    {\it #1}\par\vskip 10pt plus 3pt minus 3pt}

\begin{document}

\begin{flushleft}
arXiv:0708.2461 [hep-ex]\\
\babar-\BaBarType-\BaBarYear/\BaBarNumber \\
SLAC-PUB-\SLACPubNumber \\
Published in Phys.Rev.D.~{\bf 76}, 092005 (2007).
\end{flushleft}


\title{\large \bf
\boldmath
The $\epem\to \fourpipn$, \fourpieta, \KKpppn and \KKppeta
Cross Sections 
Measured with Initial-State Radiation 
} 

%
\author{B.~Aubert}
\author{M.~Bona}
\author{D.~Boutigny}
\author{Y.~Karyotakis}
\author{J.~P.~Lees}
\author{V.~Poireau}
\author{X.~Prudent}
\author{V.~Tisserand}
\author{A.~Zghiche}
\affiliation{Laboratoire de Physique des Particules, IN2P3/CNRS et Universit\'e de Savoie, F-74941 Annecy-Le-Vieux, France }
\author{J.~Garra~Tico}
\author{E.~Grauges}
\affiliation{Universitat de Barcelona, Facultat de Fisica, Departament ECM, E-08028 Barcelona, Spain }
% ...
\collaboration{The \babar\ Collaboration}
\noaffiliation

\date{\today}

\begin{abstract}

We study the processes $\epem\!\to \fourpipn\gamma$, $\fourpieta\gamma$, 
$\KKpppn\gamma$ and $\KKppeta\gamma$ 
with the hard photon radiated from the initial state.  
About 20000, 4300, 5500 and 375 fully reconstructed events, respectively, 
are selected from 232~\invfb of \babar\ data. 
The invariant mass of the hadronic final state defines the effective \epem
center-of-mass energy, so that 
the obtained cross sections from the threshold to about 5~\gev
can be compared with corresponding direct \epem measurements, 
currently available only for the $\eta\pipi$ and $\omega\pipi$
submodes of the $\epem\!\to \fourpipn$ channel. 
Studying the structure of these events, 
we find contributions from a number of intermediate states, 
and we extract their cross sections where possible.
In particular, we isolate the contribution from 
$\epem\!\to\omega(782)\pipi$ and study the $\omega(1420)$ and
$\omega(1650)$ resonances.
In the charmonium region,
we observe the $J/\psi$ in all these final states and several
intermediate states, 
as well as the $\psi(2S)$ in some modes,
and we measure the corresponding branching fractions.  

\end{abstract}


\vfill
\maketitle

\setcounter{footnote}{0}

\section{Introduction}
\label{sec:Introduction}

Electron-positron annihilation at fixed center-of-mass (c.m.) energies
has long been a mainstay of research in elementary particle physics.
The idea of utilizing initial-state radiation (ISR) to study \epem 
reactions below the nominal c.m.\ energies was outlined in Ref.~\cite{baier},
and discussed in the context of high-luminosity $\rm \phi$ and
$B$ factories in Refs.~\cite{arbus, kuehn, ivanch}.
At high energies, \epem annihilation is dominated by quark-level
processes producing two or more hadronic jets.
However, low-multiplicity processes dominate at energies
below about 2~\gev, and the region near charm threshold, 3.0--4.5~\gev,
features a number of resonances~\cite{PDG}.
These allow us to probe a wealth of physics parameters,
including cross sections, spectroscopy and form factors.

Of particular current interest are
several recently observed charmonium states and
a possible discrepancy between the measured value of the anomalous
magnetic moment of the muon \ammm
and that predicted by the Standard Model~\cite{g2new}.
Charmonium and other states with $J^{PC}=1^{--}$ can be observed as resonances
in the cross section, 
and intermediate states may be present in the hadronic system.
Measurement of the decay modes and their branching fractions is important
in understanding the nature of these states.
The prediction for \ammm is based on hadronic-loop corrections measured from
low-energy $\epem\!\to\,$hadrons data,
and these dominate the uncertainty on the prediction.
Improving this prediction requires not only more precise measurements,
but also measurements over the entire energy range and inclusion of
all the important subprocesses in order to understand possible
acceptance effects.
ISR events at $B$ factories provide independent measurements of
hadronic cross sections with complete coverage from the production
threshold to about 5~\gev.

The cross section for the radiation of a photon of energy $E_{\gamma}$ 
followed by the production of a particular hadronic final state $f$ 
is related to the corresponding direct $\epem\to f$ cross section 
$\sigma_f(s)$ by
\begin{equation}
\frac{d\sigma_{\gamma f}(s,x)}{dx} = W(s,x)\cdot \sigma_f(s(1-x))\ ,
\label{eq1}
\end{equation}
where $\sqrt{s}$ is the initial \epem c.m.\@ energy, 
$x\! =\! 2E_{\gamma}/\sqrt{s}$ is the fractional energy of the ISR photon 
and $\Ecm \!\equiv\! \sqrt{s(1-x)}$ is the effective c.m.\@ energy at
which the final state $f$ is produced. 
The probability density function $W(s,x)$ for ISR photon emission has
been calculated with better than 1\% precision (see e.g.\ Ref.~\cite{ivanch}).
It falls rapidly as $E_{\gamma}$ increases from zero, but has a long
tail, which combines with the increasing $\sigma_f(s(1-x))$ to produce
a sizable cross section at very low \Ecm.
The angular distribution of the ISR photon peaks along the beam directions, 
but 10--15\%~\cite{ivanch} of the photons are within a typical
detector acceptance.

Experimentally, the measured invariant mass of the hadronic final
state defines \Ecm.
An important feature of ISR data is that a wide range of energies is
scanned simultaneously in one experiment, 
so that no structure is missed and
the relative normalization uncertainties in
data from different experiments or 
accelerator parameters are avoided.
Furthermore, for large values of $x$ the hadronic system is collimated, 
reducing acceptance issues and allowing measurements at energies down to 
production threshold.
The mass resolution is not as good as a typical beam energy spread used in 
direct measurements,
but the resolution and absolute energy scale can be monitored
by the width and mass of well known resonances, such as the $J/\psi$
produced in the reaction $\epem \to J/\psi\gamma$. 
Backgrounds from $\epem \!\to\,$hadrons events at the nominal $\sqrt{s}$
and from other ISR processes can be 
suppressed by a combination of particle identification and 
kinematic fitting techniques.
Studies of $\epem\to\mumu\gamma$ and several multi-hadron ISR processes using
\babar\ data have been 
reported~\cite{Druzhinin1,isr3pi,isr4pi,isr6pi,isr2K2pi},
demonstrating the viability of such measurements.

The contributions to the \fourpipn final state from the $\eta\pipi$ and
$\omega\pipi$ channels have been measured directly by the
DM1~\cite{omegadm1}, DM2~\cite{etadm2,omegadm2}, CMD2~\cite{5picmd2}
and ND~\cite{etand}
collaborations for $\sqrt{s} <\! 2.2~\gev$. 
In this paper we present a comprehensive study of the \fourpipn final state 
along with new measurements of the \fourpieta, \KKpppn and \KKppeta 
final states.
In all cases we require detection of the ISR photon and perform a set
of kinematic fits.
We are able to suppress backgrounds sufficiently to study these
final states from their respective production thresholds up to 4.5~\gev.
In addition to measuring the overall cross sections, 
we study the internal structure of the events
and measure cross sections for a number of intermediate states.
We study the charmonium region, 
measuring several $J/\psi$ and $\psi(2S)$ branching fractions.

\section{\boldmath The \babar\ detector and dataset}
\label{sec:babar}

The data used in this analysis were collected with the \babar\ detector at
the \pep2\ asymmetric energy \epem\ storage rings. 
The total integrated luminosity used is 232~\invfb, 
which includes 211~\invfb collected at the $\Upsilon(4S)$ peak, 
$\sqrt{s}=10.58~\gev$, 
and 21~\invfb collected below the resonance, at $\sqrt{s}=10.54~\gev$.

The \babar\ detector is described elsewhere~\cite{babar}. 
Here we use charged particles reconstructed in the tracking system,
which compresed the five-layer silicon vertex tracker (SVT)
and the 40-layer drift chamber (DCH) in a 1.5 T axial magnetic field.
Separation of charged pions, kaons and protons uses a
combination of Cherenkov angles measured in the detector of internally
reflected Cherenkov light (DIRC) and specific ionization measured in
the SVT and DCH.
Here we use a kaon identification algorithm that provides 90--95\%
efficiency, depending on momentum, and rejects pions and protons
by factors of 20--100.
Photon and electron energies are measured in the CsI(Tl)
electromagnetic calorimeter (EMC).

To study the detector acceptance and efficiency, 
we use a simulation package developed for radiative processes.
The simulation of signal and background hadronic final states
is based on the approach suggested by Czy\.z and K\"uhn~\cite{kuehn2}.  
Multiple soft-photon emission from the initial-state charged particles is
implemented with a structure-function technique~\cite{kuraev, strfun},
and photon radiation from the final-state particles is simulated by
the PHOTOS package~\cite{PHOTOS}.  
The accuracy of the radiative corrections is about 1\%.

We simulate the \fourpipn final state both according to phase 
space and with models that include the $\eta\rho$, $\omega\pipi$ and
$\omega f_0(980)$ intermediate states,
and the \KKpppn final state both according to phase space and
including the intermediate $\phi$ and/or $\eta$ resonances.
The generated events go
through a detailed detector simulation~\cite{GEANT4}, 
and we reconstruct them with the same software chain as the experimental data. 
Variations in detector and background conditions are taken into account.

We generate a large number of background processes, 
including the ISR channels 
$\epem \!\!\to\! \pipi\pipi\gamma$, $\Kp\Km\pipi\gamma$, 
$2(\pipi)\ppz\gamma$, and $\Kp\Km\pipi\ppz\gamma$.
These can contribute due to a combination of particle misidentification,
and missing or spurious tracks or photons.
In addition, we study the non-ISR backgrounds
$\epem \!\!\to\! q \qbar$ $(q = u, d, s, c)$ generated by
JETSET~\cite{jetset} and
$\epem \!\!\to\! \tau^+\tau^-$ by KORALB~\cite{koralb}. 
The contribution from the \Y4S decays is negligible.
The cross sections for these processes are known with 
about 10\% accuracy or better, which is sufficient for these measurements.

\section{\boldmath Event Selection and Kinematic Fit}
\label{sec:Fits}

In the initial selection of candidate events, we consider
photon candidates in the EMC with energy above 0.03~\gev
and 
charged tracks reconstructed in the DCH or SVT or both that 
extrapolate within 0.25~cm of the beam axis and
within 3~cm of the nominal collision point along the axis.
We require a high-energy photon in the event with an energy in the
initial \epem c.m.\ frame of $E_\gamma > 3~\gev$,
and exactly four charged tracks with zero net charge that combine with a
pair of other photons to roughly balance the momentum of the 
highest-energy photon.
We fit a vertex to the set of charged tracks and use it as the point
of origin to calculate the photon directions.
Most events contain additional soft photons due to 
machine background or interactions in the detector material.

We subject each of these candidate events to a set of constrained 
kinematic fits, and use the fit results,
along with charged-particle identification,
both to select the final states of interest and to measure backgrounds
from other processes.
We assume the photon with the highest $E_\gamma$ is the ISR
photon, and the kinematic fits use its direction and energy along with the
four-momenta and covariance matrices of the initial \epem and the
set of selected tracks and photons.
The fitted three-momenta for each charged track and photon are used 
in further kinematical calculations.

We pair all non-ISR photon candidates and consider combinations with
invariant mass within $\pm$30~\mevcc of the $\pi^0$($\eta$) mass
as $\pi^0$($\eta$) candidates. 
For each candidate event, we perform a kinematic fit using the
momenta of the ISR photon, of the two additional photons and of the
four tracks (under the relevant mass hypotheses for each considered
final state), imposing five constraints (5C): the two photons
invariant mass must match the nominal \piz or $\eta$ mass, and the energy and
momentum of the whole event must match the energy and momentum of the
initial \epem state.
We retain the combination with the lowest \chisq,
either \chifourpipn or \chifourpieta, as a \fourpipn or \fourpieta
candidate, respectively.
If the four tracks include exactly one identified \Kp and one \Km,
we perform a set of similar fits under the \KKpppn and \KKppeta hypotheses 
and retain the two-photon combination with the lowest \chiKKpppe.
To reduce the large background from ISR \fourpi events~\cite{isr4pi}
with no additional neutral hadrons, 
we also fit each candidate event under the \fourpi hypothesis 
and require $\chifourpi\! >\! 20$.

\section{\boldmath The \fourpipn final state}
\subsection{Final Selection and Backgrounds}
\label{sec:selection1}

\begin{figure}[t]
\begin{center}
\includegraphics[width=0.9\linewidth]{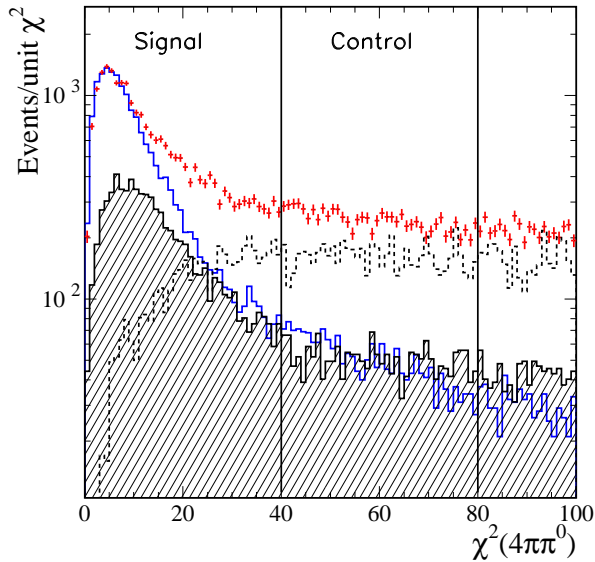}
\vspace{-0.4cm}
\caption{
       Distribution of \chisq from the five-constraint fit for
       \fourpipn candidates in the data (points) after subtracting 
the $\epem \!\!\to\! \qqbar$ background (hatched histogram).
       The open histogram is the distribution for simulated signal
       events, normalized as described in the text.
       The dashed histogram is the estimated backgrounds
       from other ISR channels, as described in the text.
}
\label{4pipi0_chi2_all}
\end{center}
\end{figure}
To suppress \KKpppn background, we require that no more than one
track in the event is identified as a kaon,
and we also fit under all four possible \KKpppn hypotheses and
require $\chiKKpppn\! >\! 30$.
To reject any $K^\pm\KS\pi^\mp\piz$ background,
we require all tracks to extrapolate within 2.5~mm of the beam axis.
The result of the 5C fit for the remaining events under
the $\fourpipn\gamma$ hypothesis 
with the $\fourpipn$ invariant mass up to 4.5~\gevcc
is used for the final
event selection and background estimate. We consider two types of
background: a non-ISR type background and ISR-type background.

The non-ISR type background comes from the $\epem \!\!\to\! \qqbar$ events
and we estimate it using the JETSET simulation.
It is dominated by events with a hard $\pi^0$ producing a fake ISR
photon, 
and the similar kinematics causes it to peak at low values of \chifourpipn.
We evaluate this background in a number of \Ecm ranges by
combining the ISR photon candidate with another photon candidate in 
both data and simulated events,
and comparing the \piz signals in the resulting $\gamma\gamma$
invariant mass distributions.
The simulation gives an \Ecm-dependence consistent with the data,
so we normalize it by an overall factor.
The hatched histogram in Fig.~\ref{4pipi0_chi2_all} represents
this background and we subtract it from the experimental distribution.

The \chifourpipn distribution for the remaining events is shown  
in Fig.~\ref{4pipi0_chi2_all} as points,
and the open histogram is the distribution for the simulated \fourpipn
events.
The simulated distribution is normalized to the data in the region 
$\chifourpipn\!\! <\! 10$ where the backgrounds and radiative corrections
are smallest.
The experimental distribution has contributions from 
ISR-type background processes,
but the simulated distribution is also 
broader than the expected 5C \chisq distribution. 
This is due to multiple soft-photon emission from the initial state
and radiation from the final-state charged particles, 
which are not taken into account by the fit.
The shape of the \chisq distribution at high values was studied in 
detail~\cite{isr4pi,isr6pi} using ISR processes for which very clean
samples can be obtained without any limit on the \chisq value and MC
signal events have been found to accurately simulate it. 

All ISR-type background sources are consistent with having a \chifourpipn
distribution that is nearly uniform over the range shown in
Fig.~\ref{4pipi0_chi2_all}. 
As an example, the \chifourpipn distribution predicted from our
simulations of other ISR channels (see Sec.~\ref{sec:babar}) is shown as the
dashed histogram, with the main contribution from the 
$2(\pipi)\ppz\gamma$ process~\cite{isr6pi}.
We therefore determine the \chifourpipn distribution of the ISR-type
background ($\chisq_{ISRbkg}$) from the 
data distribution, by subtracting the \chifourpipn distributions of simulated
signal and of the $q \qbar$ backgrounds, both of which are
normalized to data as mentioned above.
The obtained $\chisq_{ISRbkg}$ distribution
is in agreement with simulation in shape, but contains
events from the processes which are not included into simulation. 

In order to determine the mass spectrum of the genuine \fourpipn
events, we define signal ($\chifourpipn<$40) and control
(40$<\chifourpipn<$80) regions as shown in
Fig.~\ref{4pipi0_chi2_all}. 
The signal region of Fig.~\ref{4pipi0_chi2_all} contains 30776 data
and 17477 simulated signal events, and the control region contains
11829 data and 2012 simulated events.
For each mass bin, the number 
of signal and ISR-background events in the signal region are
extracted using the observed numbers of events in the two regions with
the $q\qbar$ background subtracted, and the two ratios of
contributions expected from the shapes of the simulated signal and the
$\chisq_{ISRbkg}$ distributions. The $q\qbar$
subtraction is actually performed using a smooth function
interpolating the simulated mass distribution. 
\begin{figure}[tbh]
\includegraphics[width=0.9\linewidth]{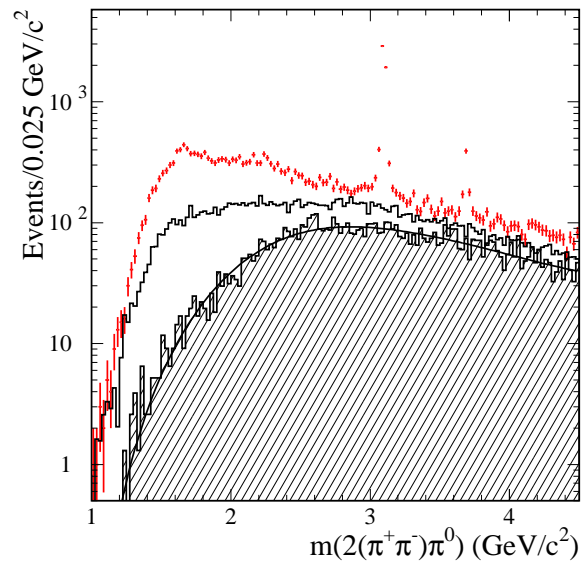}
\vspace{-0.4cm}
\caption{
   Invariant mass distribution for selected \fourpipn events in the
   data (points).
   The hatched and open histograms represent
   the non-ISR background and 
sum of all backgrounds respectively.
The smooth line approximates the non-ISR background, as described in the text.
   }
\label{4pipi0_babar}
\end{figure}

Figure~\ref{4pipi0_babar} shows the \fourpipn invariant mass distribution 
from threshold up to 4.5~\gevcc for the experimental events in the
signal region of Fig.~\ref{4pipi0_chi2_all}. 
Narrow peaks are apparent at the $J/\psi$ and $\psi(2S)$ masses.
The hatched histogram represents the \qqbar background,
which is negligible at low mass but becomes large at higher masses.
The open histogram represents the sum of all backgrounds,
including those estimated from the control region.
They total about 20\% at low mass but account for 60-80\% of the
observed data near 4 \gevcc.

Considering uncertainties in the cross sections for the background processes, 
the normalization of events in the control region and 
the simulation statistics,
we estimate a systematic uncertainty on the signal yield that is
about 5\% in the 1.5--1.8~\gevcc mass region, but 
increases to 20\% at 2.5~\gevcc and to more than 50\% in the region
above 3.5~\gevcc.

\subsection{Selection Efficiency}
\label{sec:eff1}

\begin{figure}[t]
\includegraphics[width=0.9\linewidth]{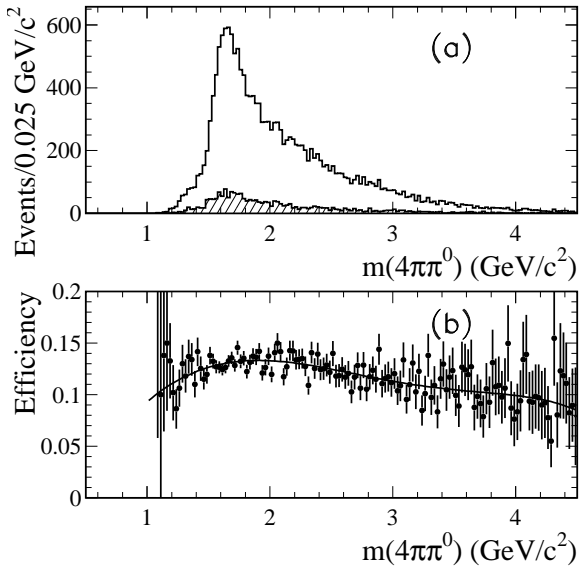}
\vspace{-0.4cm}
\caption{
  (a) Invariant mass distributions for simulated \fourpipn events in the phase
  space model, reconstructed in the signal (open) and control 
  (hatched) regions of Fig.~\ref{4pipi0_chi2_all}.
  (b) Net reconstruction and selection efficiency as a function of
  mass obtained from this simulation. 
  The curve represents a $3^{\rm rd}$ order polynomial fit.
}
\label{mc_acc1}
\end{figure}

The selection procedures applied to the data are also applied to the         
simulated signal samples.
The resulting \fourpipn invariant-mass distributions in the signal and
control regions of Fig.~\ref{4pipi0_chi2_all} are shown in
Fig.~\ref{mc_acc1}(a) for the phase space  simulation.
The fraction of simulated events in the \chifourpipn control region remains
constant over mass,
supporting the assumption of mass-independent \chisq shape.
The broad, smooth mass distribution is chosen to facilitate the estimation 
of the efficiency as a function of mass,
and this model reproduces the observed distributions of pion
momenta and polar angles.
We divide the number of reconstructed simulated events in each 25~\mevcc  
mass interval by the number generated in that interval to obtain the
efficiency shown as the points in Fig.~\ref{mc_acc1}(b); 
the curve represents a 3$^{\rm rd}$ order polynomial fit to the points.
We simulate events with the ISR photon confined to the
angular range of EMC acceptance.
The computed efficiency is for this fiducial region, 
but includes the acceptance for the final-state hadrons, 
the inefficiencies of the detector subsystems, 
and event loss due to additional soft-photon emission.

The simulations including the $\omega\pipi$ and/or $\eta\pipi$
channels have very different distributions of mass and angles in the  
\fourpipn rest frame.
However, the angular acceptance is quite uniform for ISR events,
and the efficiencies are consistent with those from the phase space
simulation within 3\%. 

\begin{table*}
\caption{Measurements of the $\ep\en\to 2(\pipi)\pi^0$ 
cross section (errors are statistical only).}
\label{4pipi0_tab}
\begin{ruledtabular}
\hspace{-1.8cm}
\begin{tabular}{ c c c c c c c c }
$E_{\rm c.m.}$ (GeV) & $\sigma$ (nb)  
& $E_{\rm c.m.}$ (GeV) & $\sigma$ (nb) 
& $E_{\rm c.m.}$ (GeV) & $\sigma$ (nb) 
& $E_{\rm c.m.}$ (GeV) & $\sigma$ (nb)  
\\
\hline

 1.0125 &  0.02 $\pm$  0.03 & 1.8875 &  1.64 $\pm$  0.21 & 2.7625 &  0.39 $\pm$  0.13 & 3.6375 &  0.16 $\pm$  0.08 \\
 1.0375 & -0.01 $\pm$ 0.04~ & 1.9125 &  1.96 $\pm$  0.21 & 2.7875 &  0.20 $\pm$  0.12 & 3.6625 &  0.38 $\pm$  0.09 \\
 1.0625 &  0.03 $\pm$  0.05 & 1.9375 &  1.88 $\pm$  0.21 & 2.8125 &  0.38 $\pm$  0.12 & 3.6875 &  1.55 $\pm$  0.12 \\
 1.0875 & -0.01 $\pm$ 0.05~ & 1.9625 &  1.76 $\pm$  0.20 & 2.8375 &  0.25 $\pm$  0.12 & 3.7125 &  0.40 $\pm$  0.09 \\
 1.1125 &  0.04 $\pm$  0.06 & 1.9875 &  1.47 $\pm$  0.20 & 2.8625 &  0.07 $\pm$  0.12 & 3.7375 &  0.17 $\pm$  0.08 \\
 1.1375 &  0.02 $\pm$  0.05 & 2.0125 &  1.66 $\pm$  0.20 & 2.8875 &  0.23 $\pm$  0.12 & 3.7625 &  0.14 $\pm$  0.07 \\
 1.1625 &  0.09 $\pm$  0.07 & 2.0375 &  1.57 $\pm$  0.20 & 2.9125 &  0.30 $\pm$  0.12 & 3.7875 &  0.14 $\pm$  0.07 \\
 1.1875 &  0.20 $\pm$  0.08 & 2.0625 &  1.77 $\pm$  0.20 & 2.9375 &  0.29 $\pm$  0.12 & 3.8125 &  0.08 $\pm$  0.07 \\
 1.2125 &  0.13 $\pm$  0.09 & 2.0875 &  1.33 $\pm$  0.19 & 2.9625 &  0.35 $\pm$  0.12 & 3.8375 &  0.27 $\pm$  0.07 \\
 1.2375 & -0.02 $\pm$ 0.10~ & 2.1125 &  1.44 $\pm$  0.19 & 2.9875 &  0.45 $\pm$  0.11 & 3.8625 &  0.08 $\pm$  0.06 \\
 1.2625 &  0.25 $\pm$  0.12 & 2.1375 &  1.42 $\pm$  0.19 & 3.0125 &  0.31 $\pm$  0.12 & 3.8875 &  0.09 $\pm$  0.07 \\
 1.2875 &  0.31 $\pm$  0.13 & 2.1625 &  1.76 $\pm$  0.19 & 3.0375 &  0.39 $\pm$  0.12 & 3.9125 &  0.18 $\pm$  0.07 \\
 1.3125 &  0.50 $\pm$  0.14 & 2.1875 &  1.38 $\pm$  0.18 & 3.0625 &  1.55 $\pm$  0.15 & 3.9375 &  0.15 $\pm$  0.07 \\
 1.3375 &  0.72 $\pm$  0.16 & 2.2125 &  1.12 $\pm$  0.18 & 3.0875 & 17.13 $\pm$  0.35 & 3.9625 &  0.14 $\pm$  0.06 \\
 1.3625 &  0.91 $\pm$  0.17 & 2.2375 &  1.75 $\pm$  0.19 & 3.1125 & 10.76 $\pm$  0.29 & 3.9875 &  0.03 $\pm$  0.06 \\
 1.3875 &  0.95 $\pm$  0.18 & 2.2625 &  1.56 $\pm$  0.18 & 3.1375 &  0.90 $\pm$  0.13 & 4.0125 &  0.10 $\pm$  0.06 \\
 1.4125 &  1.58 $\pm$  0.21 & 2.2875 &  1.20 $\pm$  0.17 & 3.1625 &  0.32 $\pm$  0.11 & 4.0375 &  0.11 $\pm$  0.06 \\
 1.4375 &  1.65 $\pm$  0.22 & 2.3125 &  1.04 $\pm$  0.16 & 3.1875 &  0.23 $\pm$  0.11 & 4.0625 &  0.09 $\pm$  0.06 \\
 1.4625 &  1.67 $\pm$  0.22 & 2.3375 &  1.25 $\pm$  0.17 & 3.2125 &  0.31 $\pm$  0.10 & 4.0875 &  0.05 $\pm$  0.06 \\
 1.4875 &  1.92 $\pm$  0.24 & 2.3625 &  0.92 $\pm$  0.16 & 3.2375 &  0.25 $\pm$  0.10 & 4.1125 &  0.04 $\pm$  0.06 \\
 1.5125 &  2.20 $\pm$  0.24 & 2.3875 &  0.82 $\pm$  0.15 & 3.2625 &  0.15 $\pm$  0.09 & 4.1375 &  0.13 $\pm$  0.06 \\
 1.5375 &  2.20 $\pm$  0.24 & 2.4125 &  1.13 $\pm$  0.15 & 3.2875 &  0.17 $\pm$  0.10 & 4.1625 &  0.16 $\pm$  0.06 \\
 1.5625 &  2.37 $\pm$  0.25 & 2.4375 &  0.58 $\pm$  0.14 & 3.3125 &  0.40 $\pm$  0.10 & 4.1875 &  0.18 $\pm$  0.06 \\
 1.5875 &  2.36 $\pm$  0.25 & 2.4625 &  0.81 $\pm$  0.15 & 3.3375 &  0.07 $\pm$  0.09 & 4.2125 &  0.14 $\pm$  0.06 \\
 1.6125 &  3.34 $\pm$  0.27 & 2.4875 &  0.64 $\pm$  0.15 & 3.3625 &  0.10 $\pm$  0.09 & 4.2375 &  0.11 $\pm$  0.06 \\
 1.6375 &  3.29 $\pm$  0.27 & 2.5125 &  0.77 $\pm$  0.14 & 3.3875 &  0.24 $\pm$  0.09 & 4.2625 &  0.10 $\pm$  0.06 \\
 1.6625 &  3.77 $\pm$  0.27 & 2.5375 &  0.52 $\pm$  0.14 & 3.4125 &  0.28 $\pm$  0.09 & 4.2875 &  0.05 $\pm$  0.06 \\
 1.6875 &  3.20 $\pm$  0.26 & 2.5625 &  0.50 $\pm$  0.14 & 3.4375 &  0.15 $\pm$  0.09 & 4.3125 &  0.11 $\pm$  0.05 \\
 1.7125 &  2.45 $\pm$  0.25 & 2.5875 &  0.42 $\pm$  0.13 & 3.4625 &  0.03 $\pm$  0.08 & 4.3375 &  0.08 $\pm$  0.05 \\
 1.7375 &  2.71 $\pm$  0.24 & 2.6125 &  0.39 $\pm$  0.13 & 3.4875 &  0.13 $\pm$  0.08 & 4.3625 &  0.11 $\pm$  0.05 \\
 1.7625 &  2.53 $\pm$  0.24 & 2.6375 &  0.65 $\pm$  0.14 & 3.5125 &  0.24 $\pm$  0.09 & 4.3875 &  0.08 $\pm$  0.05 \\
 1.7875 &  2.31 $\pm$  0.23 & 2.6625 &  0.49 $\pm$  0.13 & 3.5375 &  0.06 $\pm$  0.08 & 4.4125 &  0.03 $\pm$  0.05 \\
 1.8125 &  2.50 $\pm$  0.23 & 2.6875 &  0.51 $\pm$  0.13 & 3.5625 &  0.18 $\pm$  0.08 & 4.4375 &  0.11 $\pm$  0.05 \\
 1.8375 &  2.14 $\pm$  0.22 & 2.7125 &  0.74 $\pm$  0.13 & 3.5875 &  0.16 $\pm$  0.08 & 4.4625 &  0.07 $\pm$  0.05 \\
 1.8625 &  1.93 $\pm$  0.21 & 2.7375 &  0.29 $\pm$  0.12 & 3.6125 &  0.07 $\pm$  0.08 & 4.4875 &  0.15 $\pm$  0.06 \\

\end{tabular}
\end{ruledtabular}
\end{table*}

\begin{figure}[tbh]
\includegraphics[width=0.9\linewidth]{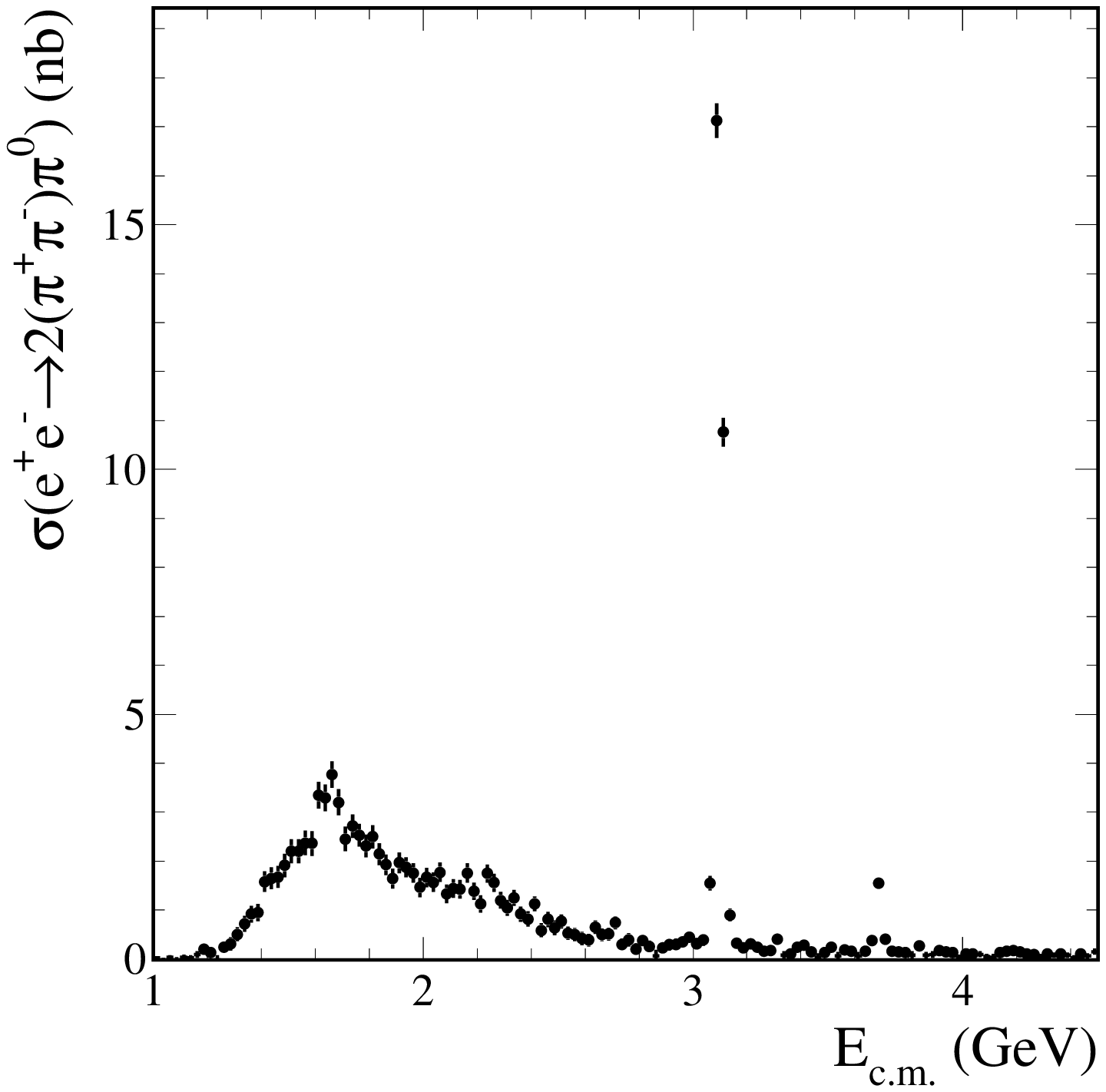}
\vspace{-0.4cm}
\caption{
   The $\epem \!\!\to\! \fourpipn$ cross section as a function of 
   \epem c.m.\@ energy measured with ISR data.
   Only statistical errors are shown.
}
\label{4pipi0_ee_babar}
\end{figure} 

We study the shape of the \chifourpipn distribution using events in
the large $J/\psi$ peak.
By comparing $J/\psi$ yields in data and simulation for 
$\chifourpipn\!\! <\! 40$ and $\chifourpipn\!\! <\! 200$,
we limit any mis-modelling of the efficiency to 3\%.
We correct the track finding efficiency following the procedures described in
Ref.~\cite{isr4pi}, with a much larger sample of $2(\pipi)$ events.
We consider data and simulated events that contain a high-energy photon plus 
exactly three charged tracks and satisfy a set of kinematic criteria, 
including a good $\chi^2$ from a kinematic fit under the hypothesis
that there is exactly one missing track in the event.
We find that the simulated track-finding efficiency is overestimated by
$(0.8\pm0.5)\%$ per track, so we apply a correction of $+(3\pm2)\%$ 
to the signal yield.
We correct the \piz-finding efficiency using the procedure described in
detail in Ref.~\cite{isr6pi}.
From ISR $\epem \!\!\to\! \omega\piz\gamma \!\!\to\! \pipi\ppz\gamma$ events
selected with and without the \piz from the $\omega$ decay,
we find an excess of simulated efficiency for one \piz of ($3 \pm 2$)\%.

\begin{table}[b]
\caption{
  Summary of systematic uncertainties on the $\epem \!\!\to\! \fourpipn$  cross
  section.
  The total uncertainty is the sum in quadrature of the components.
  }
\label{error_tab}
\begin{ruledtabular}
\begin{tabular}{l c rr} 
     Source         & Correction~~~~& \multicolumn{2}{c}{Uncertainty~~~~~~~~~~}  \\
\hline
                    &            &       &          \\[-0.2cm]
Rad. Corrections    &  --        &  1\%  &          \\
Backgrounds         &  --        & 5\% &          $\Ecm\! =\! 1.7~\gev$ \\
                    &            & 20\%  &  $ \!\Ecm\! =\! 2.5~\gev$ \\
                    &            & 50\%  &          $\Ecm\! =\! 3.5~\gev$ \\
Model Dependence    &  --        &  3\%  &          \\
\chifourpipn Distn. &  --        &  3\%  &          \\ 
Tracking Efficiency & $+3\%$     &  2\%  &          \\
$\pi^0$ Efficiency  & $+3\%$     &  2\%  &          \\
ISR Luminosity      &  --        &  3\%  &          \\[0.1cm]
\hline
                    &            &       &          \\[-0.2cm]
Total               & $+6\%$     &  7\%  &          $\Ecm\! =\! 1.7~\gev$ \\
                    &            & 20\%  & $\!\Ecm\! =\! 2.5~\gev$ \\
                    &            & 50\%  &          $\Ecm\! =\! 3.5~\gev$ \\
\end{tabular}
\end{ruledtabular}
\end{table}

\subsection{\boldmath Cross Section for $\epem \!\to \fourpipn$}
\label{sec:xs4pipi0}

We calculate the cross section as a function of effective c.m.\ energy
for the reaction $\epem \!\!\to\! \fourpipn$ from
\begin{equation}
    \sigma_{\fourpipn}(\Ecm)
  = \frac{dN_{\fourpipn\gamma}(\Ecm)}
         {d{\cal L}(\Ecm) \cdot \epsilon_{\fourpipn}(\Ecm)}\ ,
  \label{xseqn}
\end{equation}
where 
$\Ecm \equiv m_{\fourpipn}c^2$, 
$m_{\fourpipn}$ is the measured invariant mass of the \fourpipn system,
$dN_{\fourpipn\gamma}$ is the number of selected events after background 
subtraction in the interval $d\Ecm$, 
and
$\epsilon_{\fourpipn}(\Ecm)$ is the corrected detection efficiency.
We calculate the differential luminosity, $d{\cal L}(\Ecm)$, 
in each interval $d\Ecm$ using integrated \babar~ luminosity and 
the probability density function from Eq.~\ref{eq1}. We compare
the experimental di-muon mass spectrum from ISR $\mumu\gamma$ events,
selected with the help of the instrumented flux return (IFR),
with the calculated one~\cite{isr4pi}
and conservatively estimate 
a systematic uncertainty on $d{\cal L}$ of 3\%. 
 This $d{\cal L}$ has been corrected for vacuum polarization (VP),
so the obtained cross section includes contribution from VP which
should be excluded when using these data in calculations
of \ammm~\cite{g2new}. The initial- and part of the final-state soft-photon
emissions are canceled out in the ratio. 

We show the cross section as a function of energy in
Fig.~\ref{4pipi0_ee_babar}, with statistical errors only,
and provide a list of our results in Table~\ref{4pipi0_tab}. 
There is no direct \epem measurement of inclusive \fourpipn final
state for a comparison. 
The applied corrections and systematic uncertainties are summarized in
Table~\ref{error_tab}.

The cross section rises from threshold to a peak value of about 4.0~\nb near 
1.65~\gev, then generally decreases with increasing energy except for
prominent peaks at the $J/\psi$ and $\psi(2S)$ masses.
Gaussian fits to the simulated line shapes give a resolution on the 
measured \fourpipn mass that varies between 6.8~\mevcc in the
1.5--2.5~\gevcc region and 8.8~\mevcc in the 2.5--3.5~\gevcc region. 
The resolution function is not purely Gaussian due to soft-photon
radiation, 
but less than 20\% of the signal is outside the 25~\mevcc mass bin.
Since the cross section has no sharp structure other than the $J/\psi$
and $\psi(2S)$ peaks discussed in Sec.~\ref{sec:charmonium} below, 
we apply no correction for the resolution.

\begin{figure}[t]
\begin{center}
\includegraphics[width=0.48\linewidth]{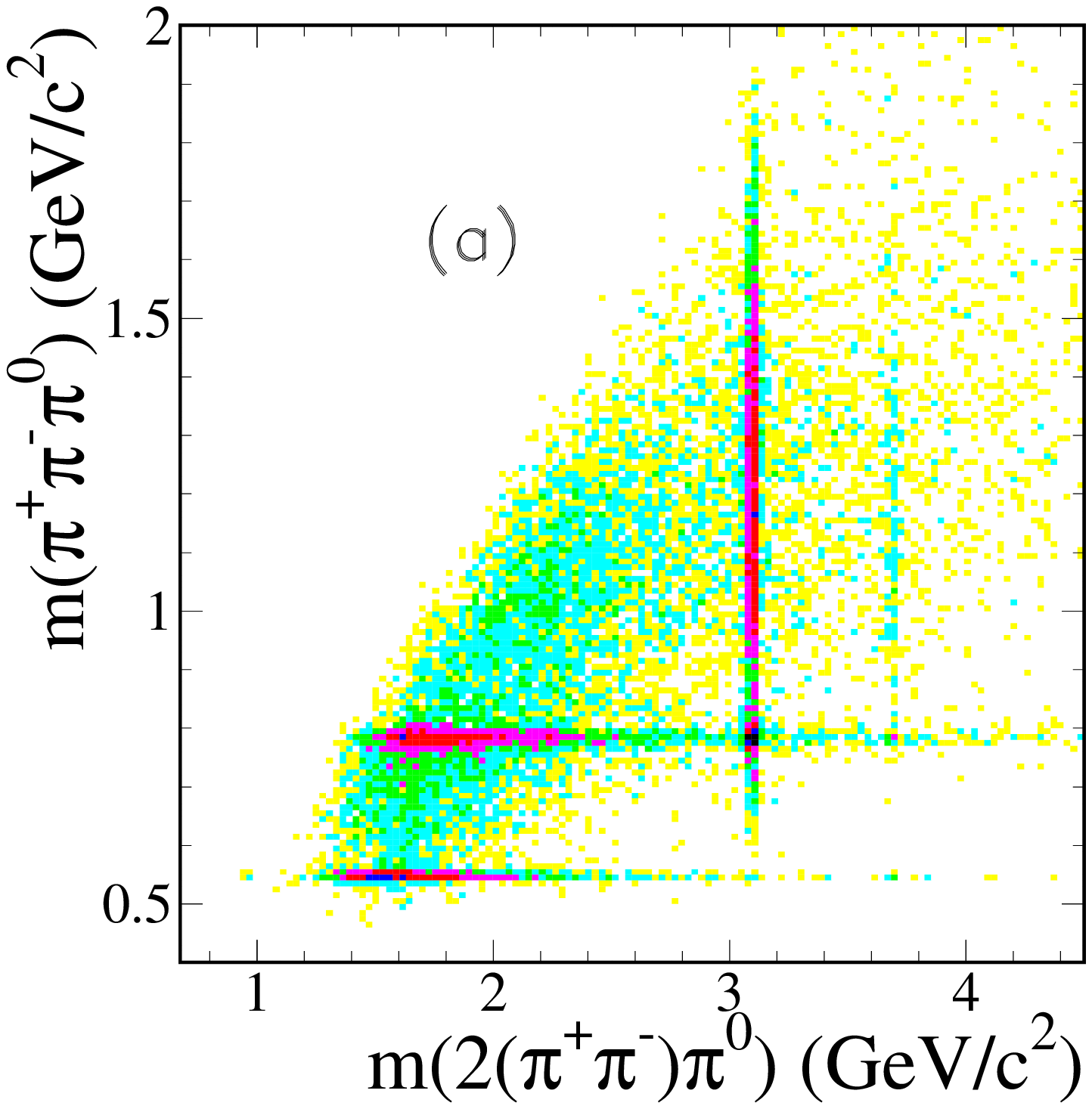}
\includegraphics[width=0.48\linewidth]{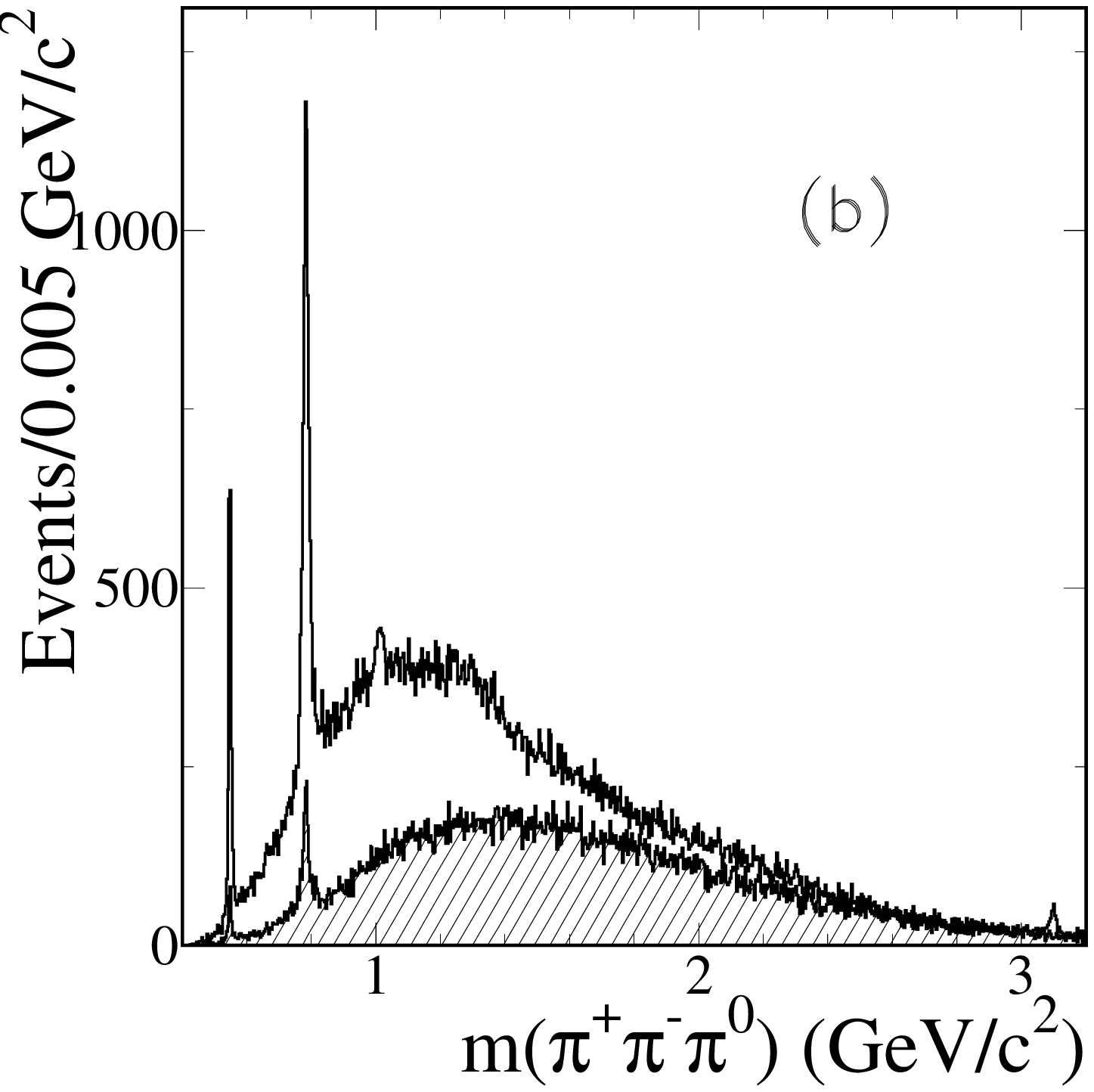}
\vspace{-0.2cm}
\caption{
  (a) The smallest $\pipi\piz$ mass in each selected \fourpipn event 
  versus the five-pion mass.
  (b) The full $\pipi\piz$ mass distribution (four entries per event)
  in these events.
  The cross-hatched histogram represents the estimated non-ISR background.
  }
\label{3pivs5pi}
\end{center}
\end{figure}

\begin{figure}[t]
\begin{center}
\includegraphics[width=0.48\linewidth]{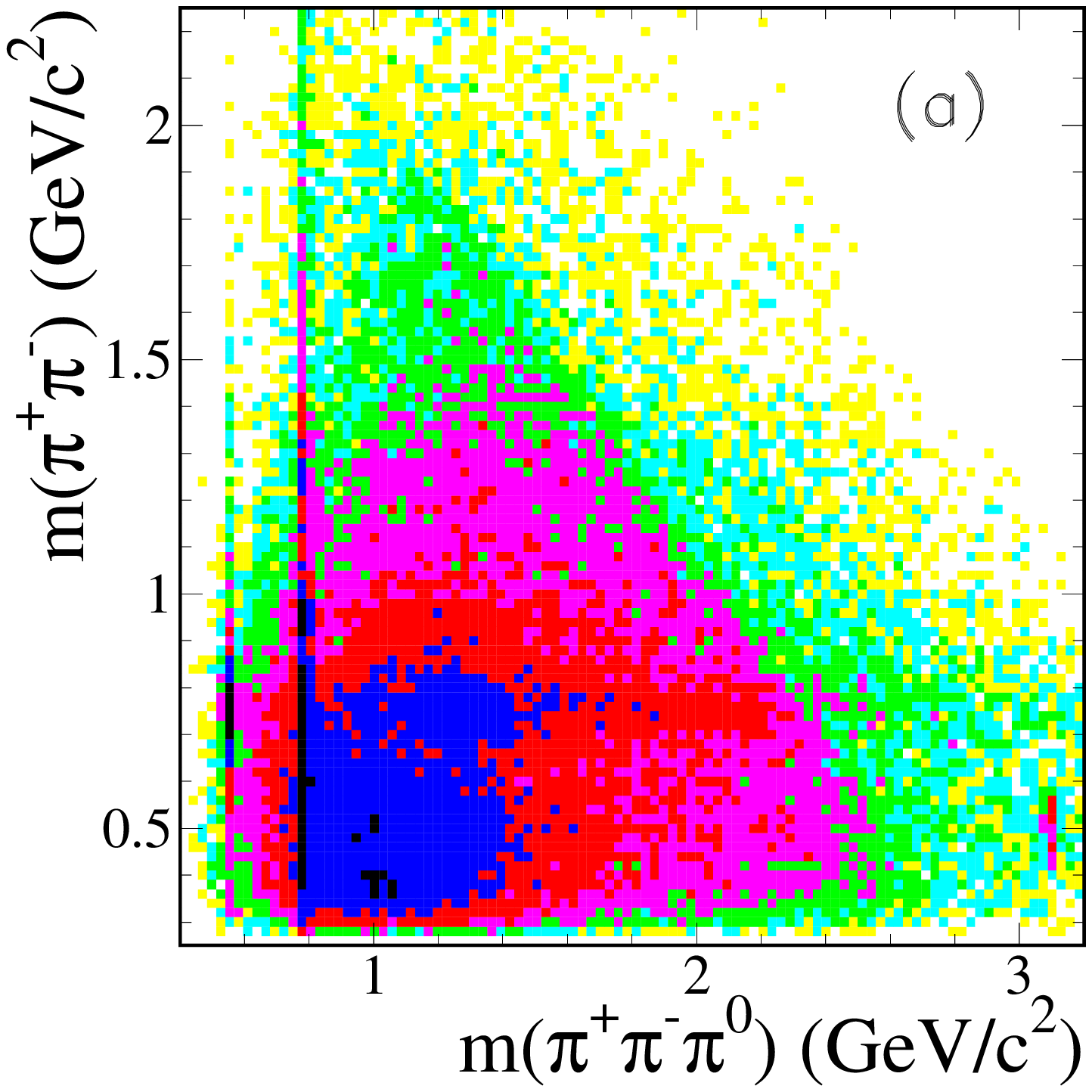}
\includegraphics[width=0.48\linewidth]{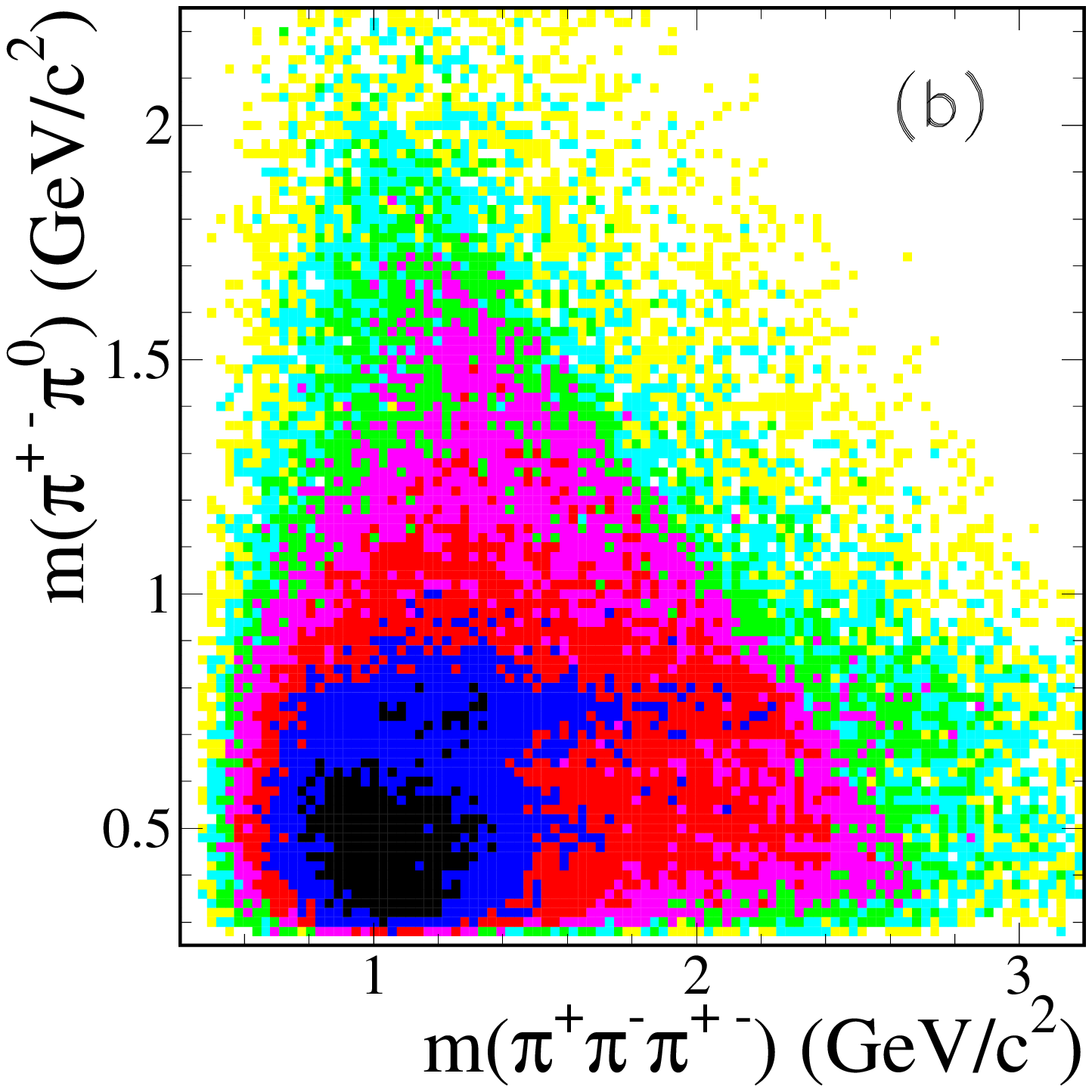}
\vspace{-0.2cm}
\caption{
   Scatter plots of (a) each $\pipi$ mass versus the mass of the
   remaining $\pipi\piz$ and
   (b) each $\pi^\pm\pi^0$ mass versus the mass of the remaining 
   $\pipi\pi^\mp$ for the selected $2(\pipi)\pi^0$ candidates
   (four entries per plot per event).
   }
\label{2pivs3pi}  
\end{center}
\end{figure}

\subsection{\boldmath Substructure in the \fourpipn Final State}

The $2(\pipi)\pi^0$ final state has a rich internal structure. 
Figure~\ref{3pivs5pi}(a) shows a scatter plot of the smallest
$\pipi\pi^0$ mass in each candidate event versus the five-pion mass. 
There are horizontal bands corresponding to the
$\eta\pipi$ and $\omega\pipi$ channels
as well as vertical bands from the $J/\psi$ and $\psi(2S)$.
Figure~\ref{3pivs5pi}(b) shows the full $\pipi\piz$ mass distribution
(four entries per event) for selected events as the open histogram and
for the estimated non-ISR background as the cross-hatched histogram.
There is also a small signal for the $\phi\pipi$ channel,
and a peak at the $J/\psi$ mass, which is due to the
$\psi(2S)\! \to\! J/\psi\pipi$, $J/\psi\!\to\! \pipi\piz$ decay chain.

Figure~\ref{2pivs3pi}(a) shows a scatter plot of all four 
$m_{\pipi}$ vs.\ $m_{\pipi\piz}$ combinations in each event.
There is a horizontal band corresponding to the $\rho^0(770)$ and an
enhancement where it crosses the vertical $\eta$ band.
A $\rho^\pm(770)$ band is similarly visible in Fig.~\ref{2pivs3pi}(b),
a scatter plot of all four $m_{\pi^\pm\piz}$ vs.\ $m_{\pipi\pi^\mp}$ 
combinations in each event.
There is a suggestion of structure along these bands and in 
Fig.~\ref{3pivs5pi}(b) around 1.2--1.3~\gevcc, 
which could correspond to the $a_1(1260)$, $\pi(1300)$ or $a_2(1320)$
resonances.
We now study events containing an $\eta$, $\omega$ or $\rho$ in detail.

\begin{figure}[tbh]
\begin{center} 
\includegraphics[width=0.9\linewidth]{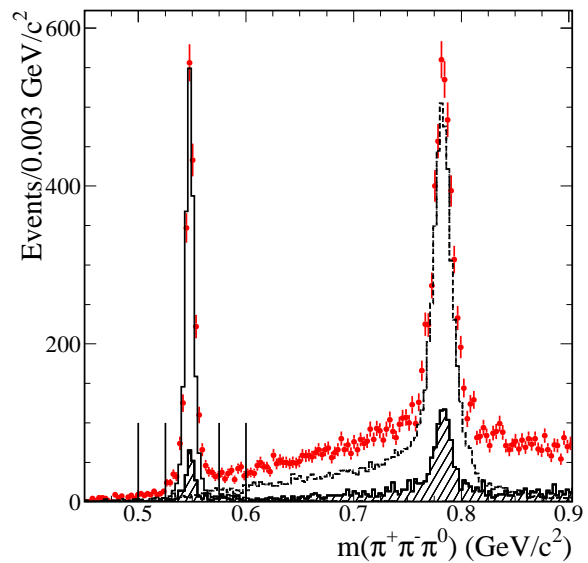}
\vspace{-0.4cm}
\caption{
  Distribution of the $\pipi\pi^0$ mass closest to the $\eta$ mass in
  the data (points).
  The histograms represent the distributions from simulated
  $\eta\rho$ (open), $\omega\pipi$ (dashed) and $uds$ events (hatched),
  normalized as described in the text.
  The inner (inner and outer) vertical lines delimit the $\eta\pipi$  
  signal (sideband) region.
  }
\label{3pietamass}  
\end{center}
\end{figure}
\begin{figure}[tbh]
\begin{center} 
\includegraphics[width=0.9\linewidth]{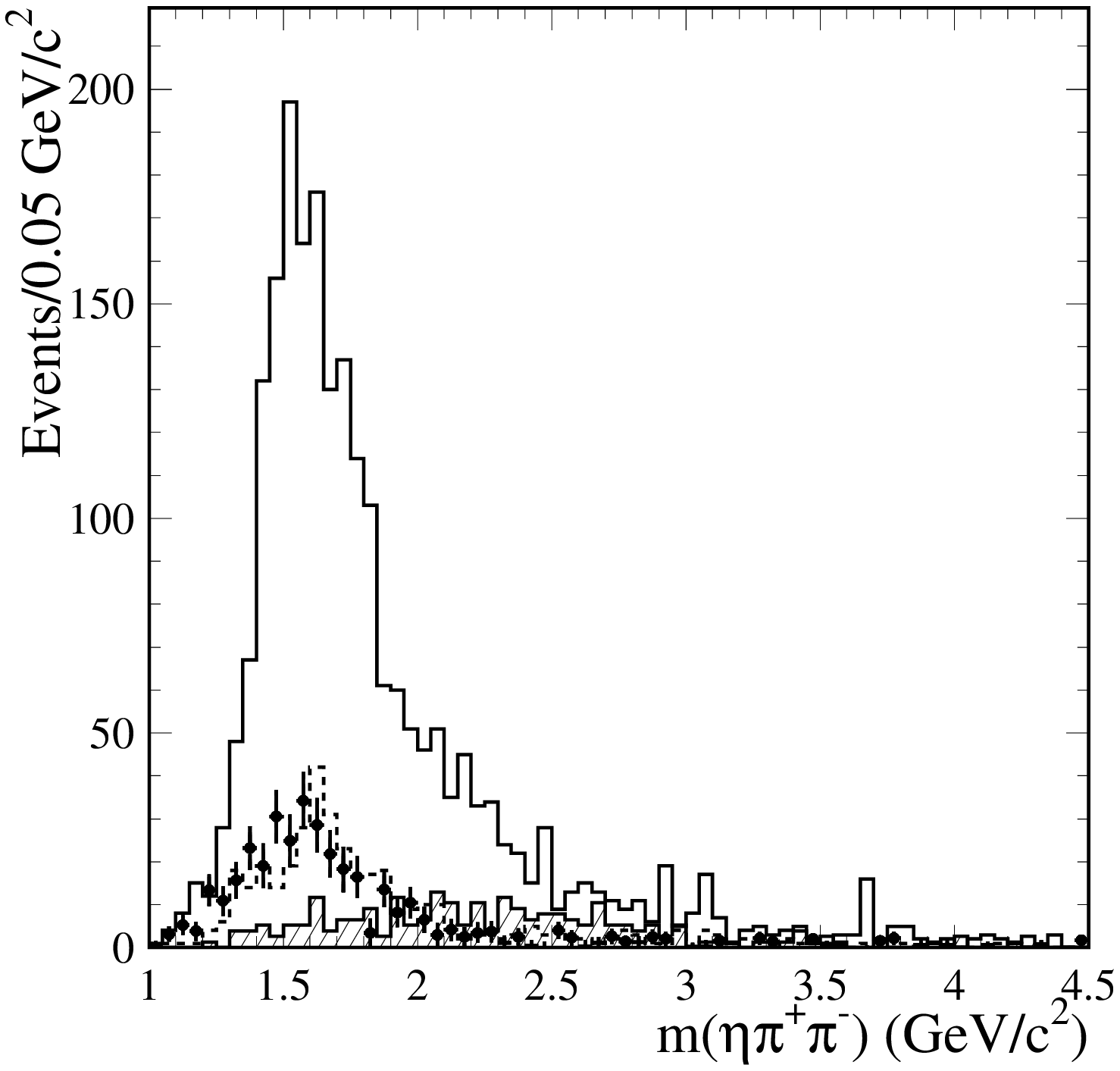}
\hspace{-4.5cm}
\includegraphics[width=0.45\linewidth]{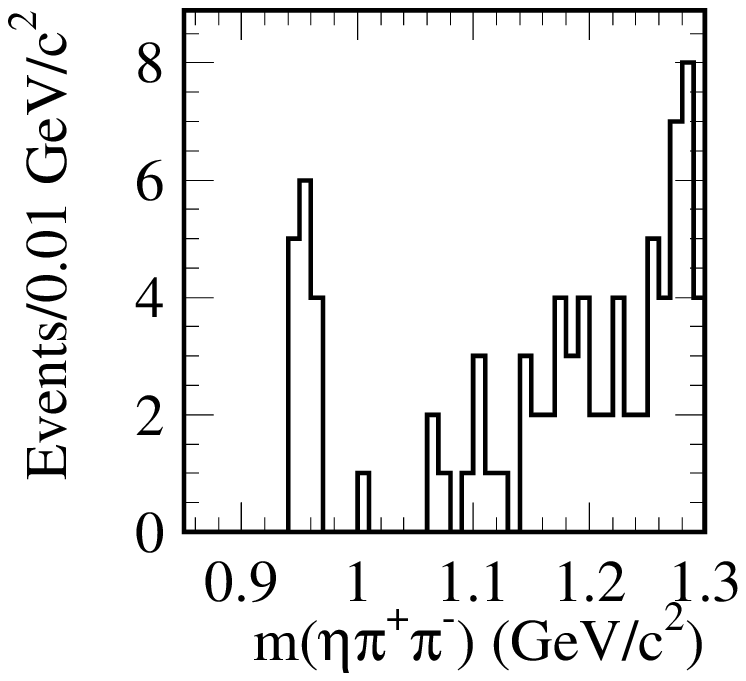}
\vspace{-0.4cm}
\caption{
  Invariant mass distribution for events in the $\eta\pipi$ 
  signal (open histogram) and sideband (dashed) regions 
  (see Fig.~\ref{3pietamass}).
  The points represent the background obtained from the \chifourpipn
  control region, and the hatched histogram is that from $uds$ events.
  The inset shows the low end of the mass distribution, where direct
  $\epem\!\! \to\! \eta^{\prime}\gamma$ events are visible.
  }
\label{neveta}  
\end{center}
\end{figure}
\begin{figure}[tbh]
\begin{center}
\includegraphics[width=0.9\linewidth]{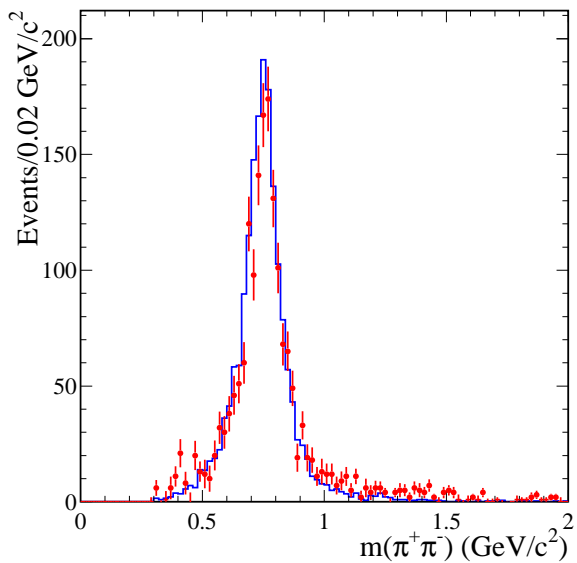}
\vspace{-0.4cm}
\caption{
  Invariant mass distribution of the $\pipi$ pair not from the $\eta$
  in selected $\eta\pipi$ events in the data (points) and in simulated
  $\eta\rho$ events (histogram).
  }
\label{neveta2}  
\end{center}
\end{figure}
\subsection{\bf\boldmath The $\eta\pipi$ and $\eta\rho$ Intermediate States}
\label{Sec:etapipi}
To extract the contribution from the $\eta\pipi$ intermediate state we
select the $\pipi\pi^0$ combination in each event 
(from four possible combinations)
with mass closest to the $\eta$ mass. 
Figure~\ref{3pietamass} shows the distribution of this mass in the
data as points, along with various simulated distributions.
The open histogram is for simulated $\eta\rho\! \to\! 2(\pipi)\pi^0$ events,
is normalized to data, and shows only a narrow $\eta$ peak.
The dashed histogram for simulated $\omega\pipi$ events shows a
strong $\omega$ peak with a tail toward lower masses that contributes a small
number of events in the $\eta$ region.
The hatched histogram for simulated $uds$ events is normalized as
described in Sec.~\ref{sec:selection1} and shows both
$\eta$ and $\omega$ signals over a small combinatoric contribution.

We define an $\eta$ signal region as mass in the range 525--575~\mevcc,
indicated by the inner vertical lines in Fig.~\ref{3pietamass},
and two sidebands, 500--525 and 575--600~\mevcc, indicated by the
outer vertical lines.
The $\eta$ signal region contains 1897 data events,
and we show their \fourpipn invariant mass distribution as the open 
histogram in Fig.~\ref{neveta}.
The hatched histogram is the $uds$ background, 
evaluated as described in section~\ref{sec:selection1},
which contributes mostly at higher masses.
We evaluate the remaining background in two ways.
Using the control region of the \chifourpipn distribution,
as discussed in Sec.~\ref{sec:selection1} but only in the $\eta$ 
signal region, 
we obtain the mass distribution shown as the points in Fig.~\ref{neveta}. 
Alternatively, the mass distribution for events in the $\eta$ sidebands 
is shown as the dashed histogram.
These two distributions are consistent,
indicating that very few non-$\eta$ \fourpipn events are present.
Here we use the sideband distribution since it is more precise and
contains all backgrounds.
The inset in Fig.~\ref{neveta} shows the distribution in the region
below 1.3~\gevcc with finer binning.
A signal from the direct $\epem\!\!\to\! \eta '\gamma$ process is
visible; these events  were studied  in our previous 
measurement~\cite{etaprimgamma}.

\begin{figure}[tbh]
\begin{center}
\includegraphics[width=0.9\linewidth]{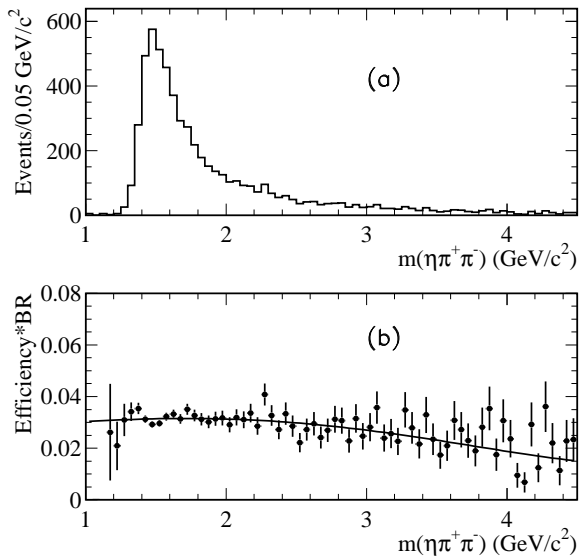}
\vspace{-0.4cm}
\caption{
  (a) Invariant mass distribution for selected simulated
  $\eta\rho$ events, and
  (b) the detection efficiency vs.\ mass, 
  including the $\eta\! \to\! \pipi\pi^0$ branching fraction.
  }
\label{xsetapipi}  
\end{center}
\end{figure}

The invariant mass distribution of the \pipi pair not from the $\eta$
is shown after subtraction of the $uds$ and $\eta$-sideband backgrounds
as the points in Fig.~\ref{neveta2},
and has a strong peak in the $\rho(770)$ region.
The histogram in Fig.~\ref{neveta2} is the distribution for simulated 
$\eta\rho\!\to\! 2(\pipi)\pi^0$ events, 
and its similarity to the data indicates that this channel
dominates the $\eta\pipi$ intermediate state.
We therefore use the simulated $\eta\rho$ events to estimate the
detection efficiency for $\eta\pipi$ events, 
although the other simulations give consistent results.
Figure~\ref{xsetapipi}(a) shows the simulated invariant mass 
distribution for selected events, 
and Fig.~\ref{xsetapipi}(b) shows the simulated efficiency,
which includes the 22.6\% branching fraction of 
$\eta\! \to\! \pipi\pi^0$.  

\begin{table*}
\caption{Measurements of the $\ep\en\to\eta\pipi$ 
cross section (errors are statistical only).}
\label{etapipi_tab}
\begin{ruledtabular}
\hspace{-1.8cm}
\begin{tabular}{ c c c c c c c c }
$E_{\rm c.m.}$ (GeV) & $\sigma$ (nb)  
& $E_{\rm c.m.}$ (GeV) & $\sigma$ (nb) 
& $E_{\rm c.m.}$ (GeV) & $\sigma$ (nb) 
& $E_{\rm c.m.}$ (GeV) & $\sigma$ (nb)  
\\
\hline

 1.0250 &  0.00 $\pm$  0.05 & 1.5250 &  4.29 $\pm$  0.37 & 2.0250 &  0.54 $\pm$  0.14 & 2.5250 &  0.00 $\pm$  0.07 \\
 1.0750 &  0.11 $\pm$  0.08 & 1.5750 &  3.13 $\pm$  0.34 & 2.0750 &  0.50 $\pm$  0.15 & 2.5750 &  0.06 $\pm$  0.07 \\
 1.1250 &  0.05 $\pm$  0.10 & 1.6250 &  2.83 $\pm$  0.35 & 2.1250 &  0.36 $\pm$  0.12 & 2.6250 &  0.11 $\pm$  0.07 \\
 1.1750 &  0.20 $\pm$  0.13 & 1.6750 &  2.13 $\pm$  0.29 & 2.1750 &  0.64 $\pm$  0.12 & 2.6750 &  0.02 $\pm$  0.07 \\
 1.2250 &  0.10 $\pm$  0.13 & 1.7250 &  2.33 $\pm$  0.28 & 2.2250 &  0.32 $\pm$  0.11 & 2.7250 &  0.05 $\pm$  0.06 \\
 1.2750 &  0.66 $\pm$  0.18 & 1.7750 &  1.90 $\pm$  0.25 & 2.2750 &  0.42 $\pm$  0.10 & 2.7750 &  0.00 $\pm$  0.06 \\
 1.3250 &  0.76 $\pm$  0.24 & 1.8250 &  1.57 $\pm$  0.23 & 2.3250 &  0.16 $\pm$  0.10 & 2.8250 &  0.05 $\pm$  0.06 \\
 1.3750 &  1.37 $\pm$  0.26 & 1.8750 &  0.80 $\pm$  0.18 & 2.3750 &  0.15 $\pm$  0.09 & 2.8750 &  0.00 $\pm$  0.05 \\
 1.4250 &  2.88 $\pm$  0.33 & 1.9250 &  0.77 $\pm$  0.17 & 2.4250 &  0.05 $\pm$  0.08 & 2.9250 &  0.20 $\pm$  0.06 \\
 1.4750 &  3.59 $\pm$  0.34 & 1.9750 &  0.69 $\pm$  0.15 & 2.4750 &  0.25 $\pm$  0.09 & 2.9750 &  0.00 $\pm$  0.05 \\
\end{tabular}
\end{ruledtabular}
\end{table*}

\begin{figure}[tbh]
\begin{center} 
\includegraphics[width=0.9\linewidth]{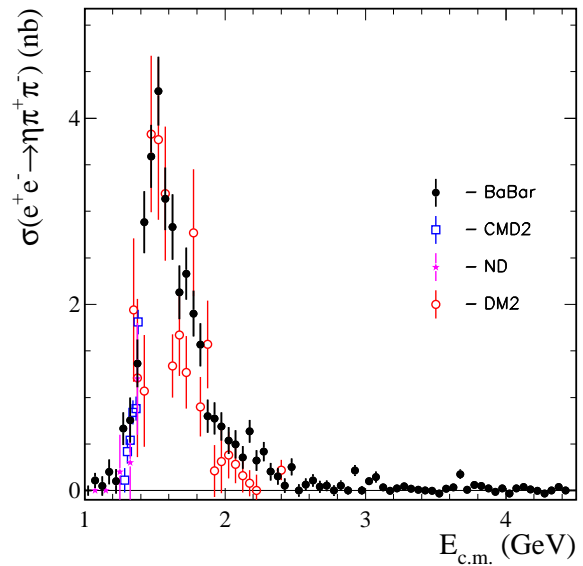}
\vspace{-0.4cm}
\caption{
  The $\epem\!\! \to\! \eta\pipi$ cross section as a function of c.m.\
  energy obtained via ISR at \babar. 
  The direct measurements from DM2, ND and CMD2 are also shown.
   Only statistical errors are shown.
  }
\label{xsetapipi2}  
\end{center}
\end{figure}

Subtracting the backgrounds and dividing by the ISR luminosity and
efficiency, 
parametrized by the third order polynomial fit shown in 
Fig.~\ref{xsetapipi}(b) and corrected as discussed above,
we obtain the $\epem\!\! \to\! \eta\pipi$ cross section 
shown in Fig.~\ref{xsetapipi2}.
Also shown are the previous direct \epem measurements from the
DM2~\cite{etadm2}, CMD2~\cite{5picmd2} and ND~\cite{etand} experiments. 
All measurements are consistent, 
and ours covers the widest energy range and is by far the most precise
above 1.4~\gev. 
 
The cross section shows a steep rise from $\eta\rho(770)$ threshold,
followed by a general decrease with increasing energy.
Possible structures near 1.6 and 1.8~\gev cannot be resolved with the
current statistics.
We list the cross section in Table~\ref{etapipi_tab} for c.m.\ energies 
up to 3~\gev with statistical errors only.
The systematic uncertainties are the same as those discussed in
section~\ref{sec:xs4pipi0}, 
totalling about 8\% below 3~\gev. 
Above 3~\gev the cross section is consistent with zero within the
current statistical errors, 
except for the $J/\psi$ peak, which is discussed below.

\begin{figure}[tbh]
\begin{center}
\includegraphics[width=0.9\linewidth]{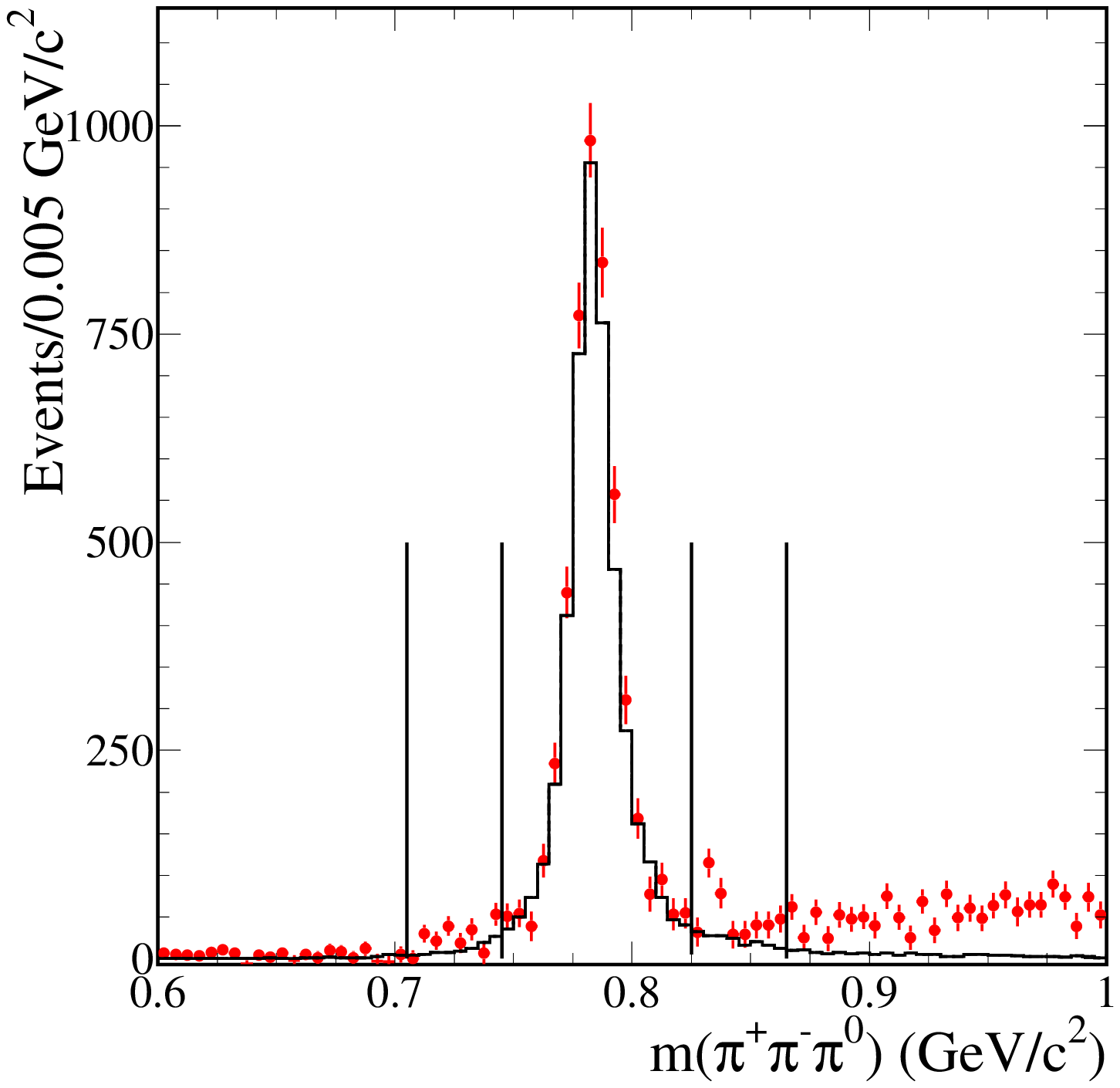}
\vspace{-0.4cm}
\caption{
  Distribution of the $\pipi\pi^0$ mass closest to the $\omega$ mass
  in the data (points) after subtraction of $uds$ and ISR-type backgrounds.
  The histogram is the distribution for simulated $\omega\pipi$ events.
  The inner (inner and outer) vertical lines delimit the $\omega$ signal 
  (sideband) region.
  }
\label{3piommass}  
\end{center}
\end{figure}
\begin{figure}[tbh]
\begin{center} 
\includegraphics[width=0.9\linewidth]{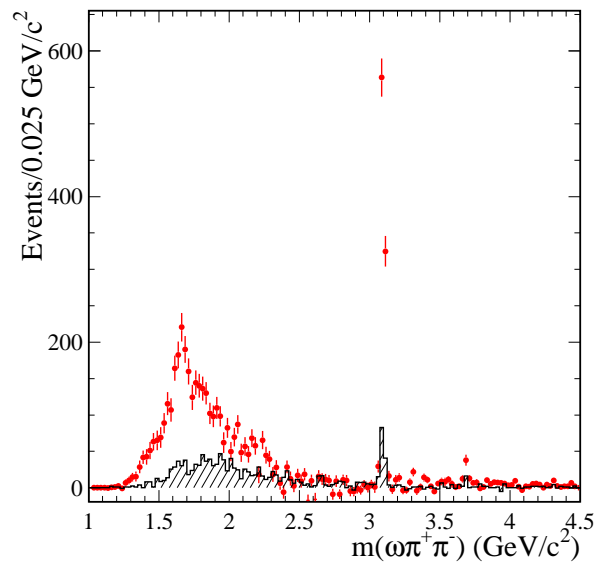}
\vspace{-0.4cm}
\caption{
  Invariant mass distribution for events in the $\omega\pipi$ signal
  region (points) and sidebands (hatched histogram).
  }
\label{nevomega}  
\end{center}
\end{figure}

\subsection{\bf\boldmath 
  The $\omega\pipi$ and $\omega f_0$ Intermediate States}
\label{Sec:omegapipi}
To extract the contribution of the $\omega\pipi$ intermediate state we
select the $\pipi\pi^0$ combination 
with mass closest to the $\omega$ mass. 
Events with this mass below 0.6~\gevcc are predominantly $\eta\pipi$
events, and we ignore them.
Subtracting the simulated $uds$ background and ISR-type background from 
the \chifourpipn control region,
as described in Sec.~\ref{sec:selection1},
we obtain the mass distribution shown in Fig.~\ref{3piommass} as points.
The histogram is for simulated $\omega\pipi$ events 
and describes the peak in the data well, but a background from 
non-$\omega\pipi$, non-$\eta\pipi$ events is still present. 

We define an $\omega$ signal region as $\pipi\pi^0$ mass in the range 745--825~\mevcc,
indicated by the inner vertical lines in Fig.~\ref{3piommass} and
containing 7693 events,
and two sidebands, 706--745 and 825--865~\mevcc, indicated by the
outer vertical lines.
Figure~\ref{nevomega} shows the invariant mass distributions for
events in the signal region (points) and sidebands (hatched histogram).
The sideband events are ISR $2(\pipi)\pi^0$ events but without an $\omega$ or
$\eta$.
They contribute mostly at higher energies including the $J/\psi$ peak.

\begin{table*}
\caption{Measurement of the $\ep\en\to\omega\pipi$ 
cross section (errors are statistical only).}
\label{omegapipi_tab}
\begin{ruledtabular}
\hspace{-1.8cm}
\begin{tabular}{ c c c c c c c c }
$E_{\rm c.m.}$ (GeV) & $\sigma$ (nb)  
& $E_{\rm c.m.}$ (GeV) & $\sigma$ (nb) 
& $E_{\rm c.m.}$ (GeV) & $\sigma$ (nb) 
& $E_{\rm c.m.}$ (GeV) & $\sigma$ (nb)  
\\
\hline

 1.1500 &  0.00 $\pm$  0.09 & 1.5000 &  1.06 $\pm$  0.25 & 1.8500 &  0.79 $\pm$  0.20 & 2.2000 &  0.11 $\pm$  0.14 \\
 1.1750 &  0.00 $\pm$  0.05 & 1.5250 &  1.33 $\pm$  0.24 & 1.8750 &  0.84 $\pm$  0.20 & 2.2250 &  0.52 $\pm$  0.14 \\
 1.2000 &  0.06 $\pm$  0.08 & 1.5500 &  1.67 $\pm$  0.27 & 1.9000 &  0.95 $\pm$  0.20 & 2.2500 &  0.35 $\pm$  0.13 \\
 1.2250 &  0.00 $\pm$  0.12 & 1.5750 &  1.30 $\pm$  0.27 & 1.9250 &  0.61 $\pm$  0.19 & 2.2750 &  0.27 $\pm$  0.12 \\
 1.2500 &  0.15 $\pm$  0.14 & 1.6000 &  2.10 $\pm$  0.28 & 1.9500 &  0.28 $\pm$  0.18 & 2.3000 & -0.04 $\pm$ 0.12~ \\
 1.2750 &  0.23 $\pm$  0.14 & 1.6250 &  2.21 $\pm$  0.29 & 1.9750 &  0.69 $\pm$  0.17 & 2.3250 &  0.11 $\pm$  0.11 \\
 1.3000 &  0.30 $\pm$  0.15 & 1.6500 &  2.80 $\pm$  0.30 & 2.0000 &  0.11 $\pm$  0.17 & 2.3500 & -0.03 $\pm$ 0.10~ \\
 1.3250 &  0.33 $\pm$  0.16 & 1.6750 &  2.19 $\pm$  0.28 & 2.0250 &  0.49 $\pm$  0.16 & 2.3750 &  0.00 $\pm$  0.10 \\
 1.3500 &  0.55 $\pm$  0.20 & 1.7000 &  1.99 $\pm$  0.26 & 2.0500 &  0.70 $\pm$  0.15 & 2.4000 &  0.05 $\pm$  0.10 \\
 1.3750 &  0.88 $\pm$  0.20 & 1.7250 &  1.38 $\pm$  0.25 & 2.0750 &  0.40 $\pm$  0.15 & 2.4250 &  0.02 $\pm$  0.10 \\
 1.4000 &  0.69 $\pm$  0.22 & 1.7500 &  1.51 $\pm$  0.24 & 2.1000 &  0.35 $\pm$  0.15 & 2.4500 & -0.04 $\pm$ 0.09~ \\
 1.4250 &  0.83 $\pm$  0.24 & 1.7750 &  1.45 $\pm$  0.23 & 2.1250 &  0.26 $\pm$  0.14 & 2.4750 &  0.05 $\pm$  0.10 \\
 1.4500 &  1.17 $\pm$  0.23 & 1.8000 &  1.18 $\pm$  0.23 & 2.1500 &  0.56 $\pm$  0.14 & 2.5000 &  0.02 $\pm$  0.10 \\
 1.4750 &  0.95 $\pm$  0.25 & 1.8250 &  1.19 $\pm$  0.21 & 2.1750 &  0.43 $\pm$  0.14 & 2.5250 &  0.15 $\pm$  0.09 \\

\end{tabular}
\end{ruledtabular}
\end{table*}

\begin{figure}[tbh]
\begin{center}
\includegraphics[width=0.9\linewidth]{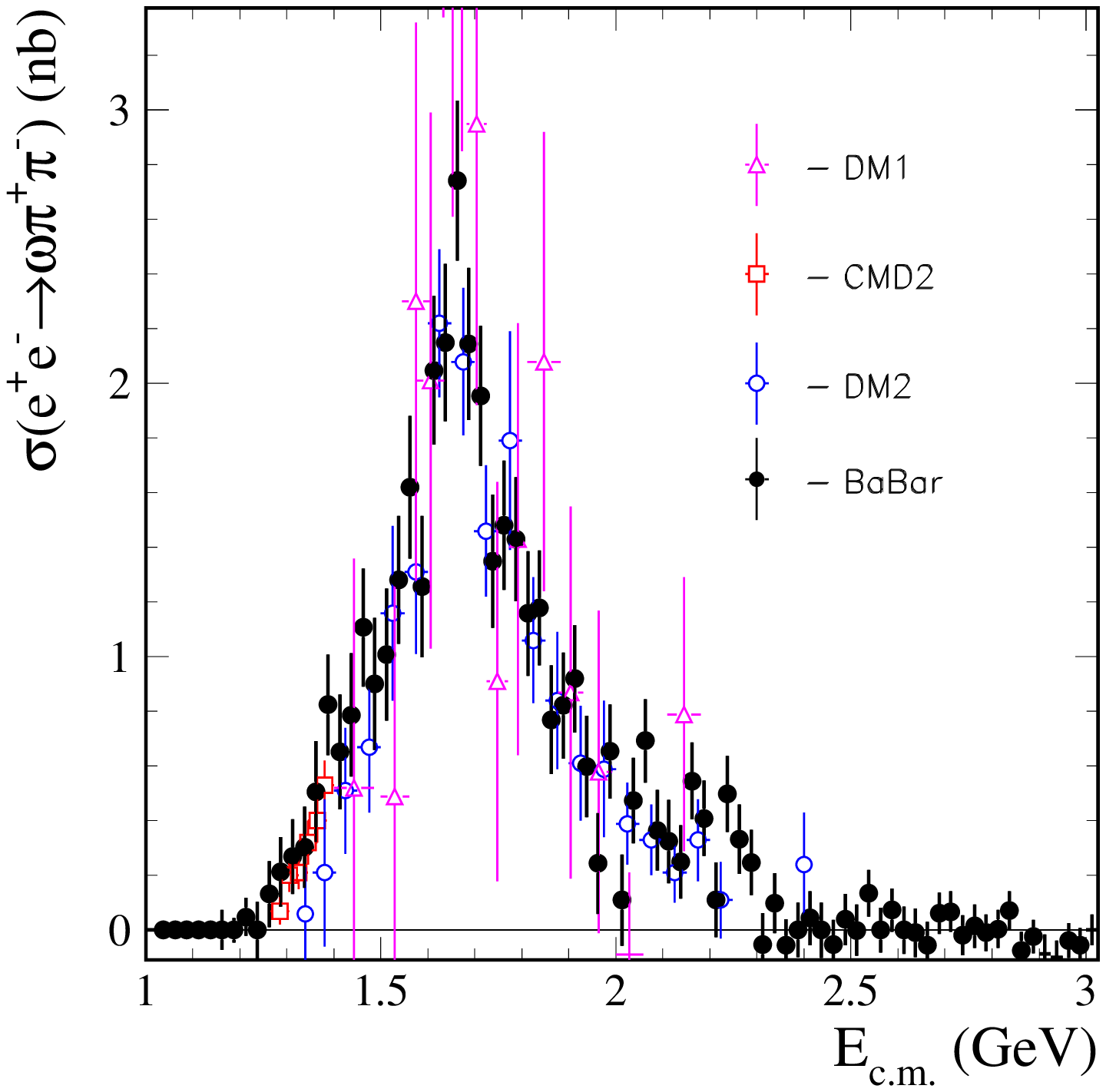}
\vspace{-0.4cm}
\caption{
  The $\epem\!\! \to\! \omega\pipi$ cross section as a function of c.m.\
  energy obtained via ISR at \babar. 
  The direct measurements from DM1, DM2 and CMD2 are also shown.
   Only statistical errors are shown.
  }
\label{xsomegapipi}  
\end{center}
\end{figure}
We evaluate the detection efficiency using the 
$\omega\pipi$ phase space simulation.
It is similar to that in Fig.~\ref{mc_acc1}, differing by few percent 
due to the additional selection criteria.
Subtracting the sideband background and dividing by the
corrected efficiency, ISR luminosity and the 89.1\% branching fraction 
of $\omega\! \to\! \pipi\pi^0$,
we obtain the $\epem\!\! \to\! \omega\pipi$ cross section shown
in Fig.~\ref{xsomegapipi}.
Also shown are the previous direct \epem measurements from the
DM2~\cite{omegadm2}, DM1~\cite{omegadm1} and CMD2~\cite{5picmd2} experiments. 
All measurements are consistent, 
and ours cover the widest energy range and are by far the most precise
above 1.4~\gev. 

The cross section is consistent with zero below 1.2~\gev, 
then rises to a peak value of about 2.5~\nb at about 1.65~\gev,
followed by a general decrease with increasing energy.
We list the cross section in Table~\ref{omegapipi_tab} for c.m.\ energies 
up to 2.4~\gev with statistical errors only.
The systematic uncertainties are the same as those discussed in
section~\ref{sec:xs4pipi0}, 
totalling about 8\% in this range.
Above 2.4~\gev the cross section is consistent with zero within the
current statistical errors, 
except for the $J/\psi$ peak, which is discussed below.

\begin{figure}[tbh]
\begin{center}
\includegraphics[width=0.48\linewidth]{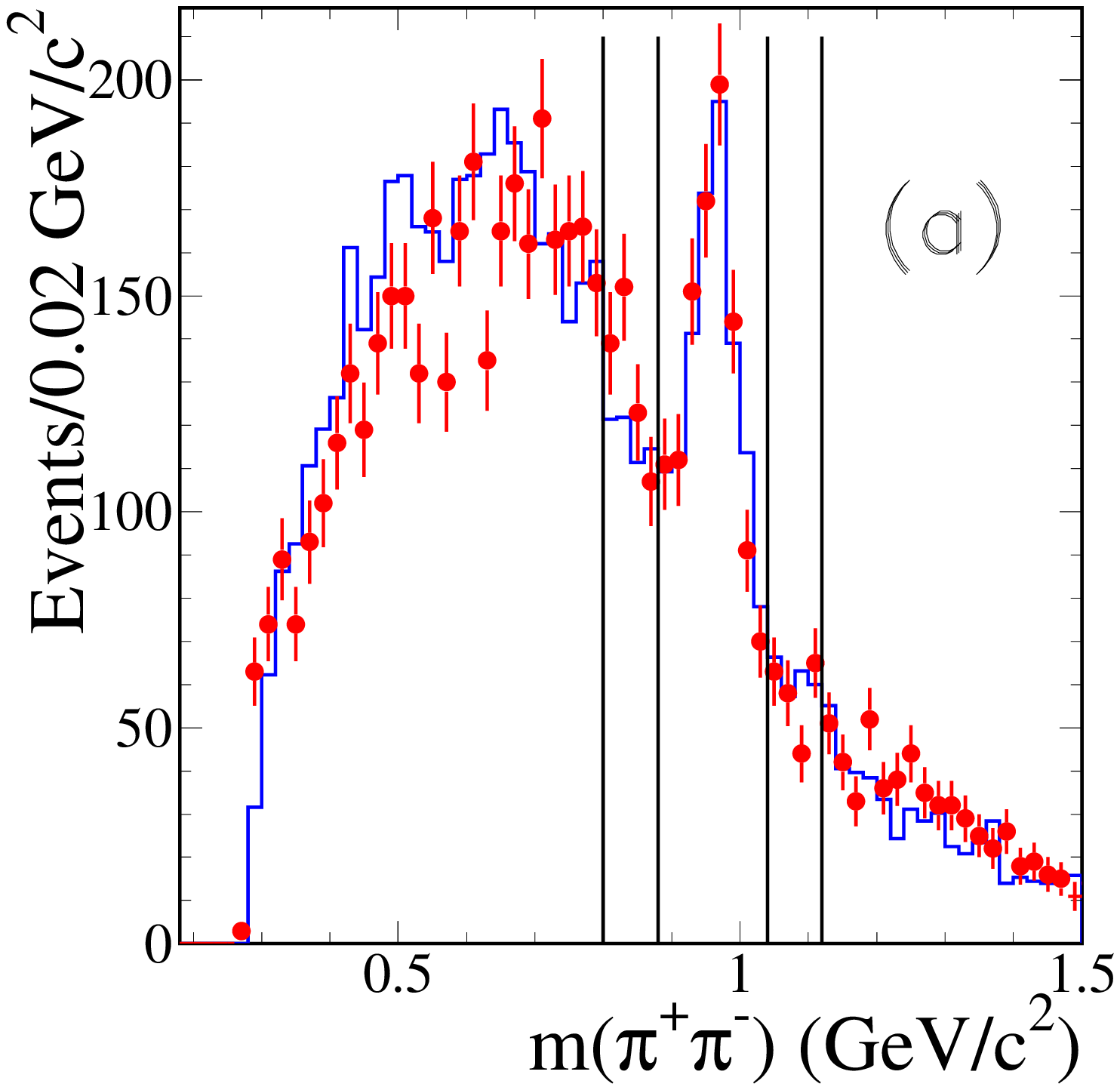}
\includegraphics[width=0.48\linewidth]{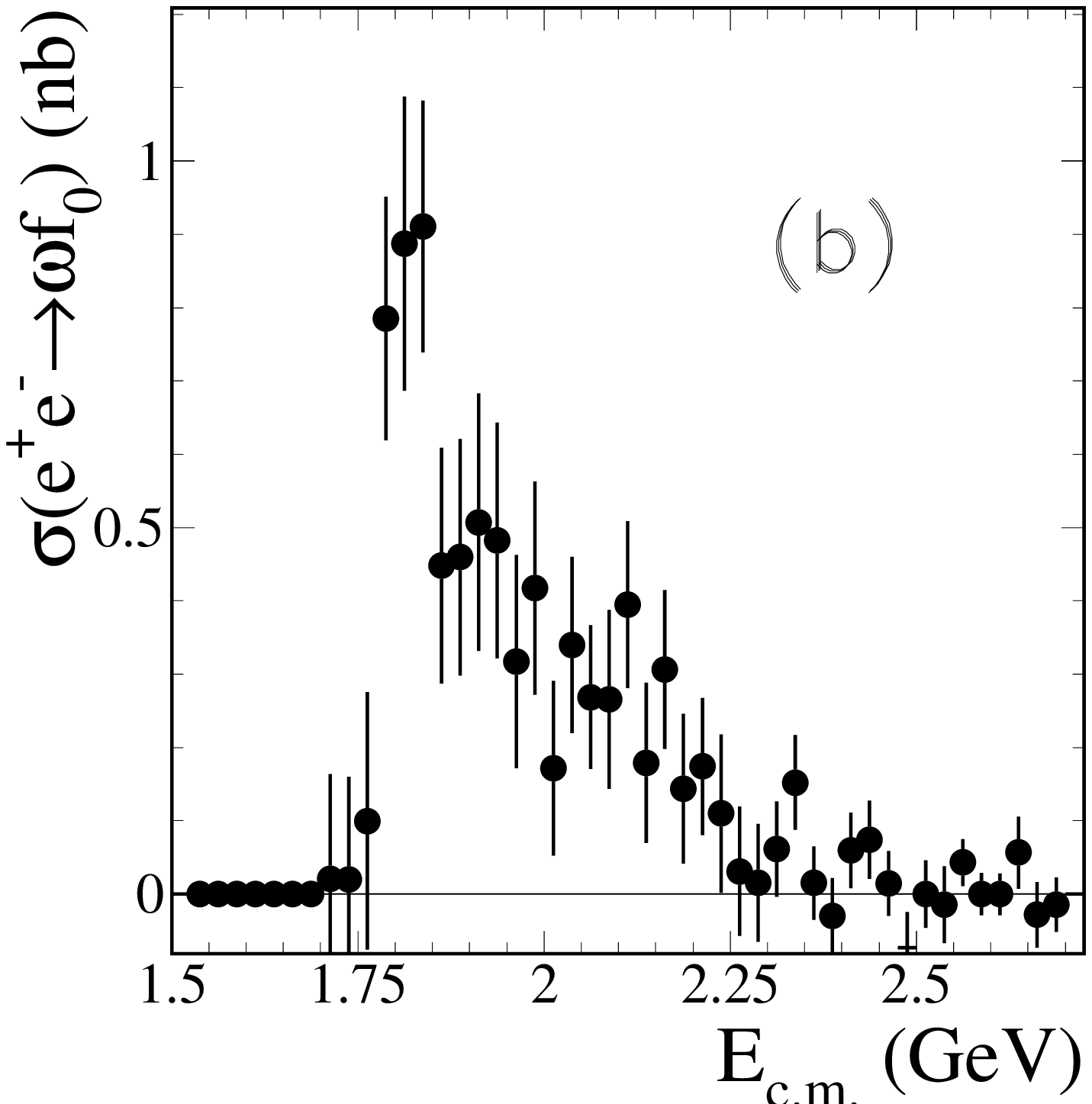}
\vspace{-0.2cm}
\caption{
  (a) Invariant mass distribution of the $\pipi$ pair not from the $\omega$
  in selected $\omega\pipi$ events in the data (points) and in simulated
  events (histogram).
  The vertical lines delimit the $f_0(980)$ signal region and sidebands.
  (b) The $\epem\!\! \to\! \omega f_0(980)$ cross section.
  }
\label{omf0}  
\end{center}
\end{figure}
\begin{table*}
\caption{Measurements of the $\ep\en\to\omega f_0(980)$ 
cross section (errors are statistical only).}
\label{omf0_tab}
\begin{ruledtabular}
\hspace{-1.8cm}
\begin{tabular}{ c c c c c c c c }
$E_{\rm c.m.}$ (GeV) & $\sigma$ (nb)  
& $E_{\rm c.m.}$ (GeV) & $\sigma$ (nb) 
& $E_{\rm c.m.}$ (GeV) & $\sigma$ (nb) 
& $E_{\rm c.m.}$ (GeV) & $\sigma$ (nb)  
\\
\hline

 1.7000 &  0.02 $\pm$  0.14 & 1.9000 &  0.51 $\pm$  0.18 & 2.1000 &  0.39 $\pm$  0.11 & 2.3000 &  0.06 $\pm$  0.07 \\
 1.7250 &  0.02 $\pm$  0.14 & 1.9250 &  0.48 $\pm$  0.16 & 2.1250 &  0.18 $\pm$  0.11 & 2.3250 &  0.15 $\pm$  0.06 \\
 1.7500 &  0.10 $\pm$  0.18 & 1.9500 &  0.32 $\pm$  0.15 & 2.1500 &  0.31 $\pm$  0.11 & 2.3500 &  0.02 $\pm$  0.05 \\
 1.7750 &  0.79 $\pm$  0.17 & 1.9750 &  0.42 $\pm$  0.15 & 2.1750 &  0.14 $\pm$  0.10 & 2.3750 &  0.03 $\pm$  0.05 \\
 1.8000 &  0.89 $\pm$  0.20 & 2.0000 &  0.17 $\pm$  0.12 & 2.2000 &  0.17 $\pm$  0.09 & 2.4000 &  0.06 $\pm$  0.05 \\
 1.8250 &  0.91 $\pm$  0.17 & 2.0250 &  0.34 $\pm$  0.12 & 2.2250 &  0.11 $\pm$  0.11 & 2.4250 &  0.07 $\pm$  0.05 \\
 1.8500 &  0.45 $\pm$  0.16 & 2.0500 &  0.27 $\pm$  0.10 & 2.2500 &  0.03 $\pm$  0.09 & 2.4500 &  0.01 $\pm$  0.04 \\
 1.8750 &  0.46 $\pm$  0.16 & 2.0750 &  0.27 $\pm$  0.12 & 2.2750 &  0.02 $\pm$  0.08 & 2.4750 &  0.01 $\pm$  0.05 \\

\end{tabular}
\end{ruledtabular}
\end{table*}

For events in the $\omega$ signal region with a five-pion mass below 
3.0~\gevcc,
we show the invariant mass distribution of the \pipi pair not from the
$\omega$ in Fig.~\ref{omf0}(a).
A peak is visible in the data at the $f_0(980)$ mass, and 
the histogram is for a simulation that includes $\omega\pipi$ phase
space and $\omega f_0(980)$ combined so as to describe the data.
We define an $\omega f_0(980)$ signal region by this $\pipi$ mass 
in the region 0.88--1.04~\gevcc, 
indicated by the inner vertical lines in Fig.~\ref{omf0}(a),
and sidebands 0.80--0.88 and 1.04--1.12~\gevcc,
indicated by the outer lines. 
Subtracting the sideband contribution from that in the signal region
and dividing by the corrected efficiency, ISR luminosity and
the 2/3 branching fraction of $f_0(980)\! \to\! \pipi$
(assuming 2$\pi$ decay mode dominance~\cite{PDG}),
we obtain the $\epem\!\! \to\! \omega f_0(980)$ cross section shown 
in Fig.~\ref{omf0}(b) and listed in Table~\ref{omf0_tab}.
This measurement of the cross section 
shows a very fast rise from threshold and a possible structure
at about 1.85~\gevcc, followed by a monotonic decrease with increasing
energy.

\begin{figure}[tbh]
\begin{center} 
\includegraphics[width=0.9\linewidth]{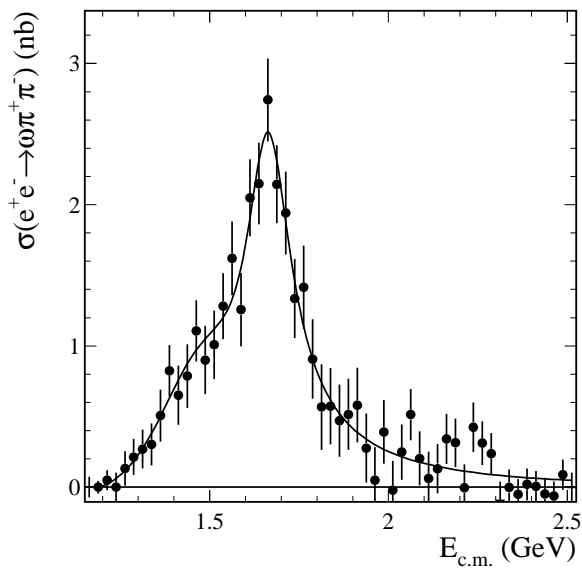}
\vspace{-0.4cm}
\caption{
  The $\epem\!\! \to\! \omega\pipi$ cross section excluding the
  $\omega f_0(980)$ contribution (points). 
  The curve shows the result of the fit of the $\omega(1420)$ and 
  $\omega(1650)$ resonances described in the text (Fit 3 in
  Table~\ref{omfit_tab}).
  }
\label{ompipixs}  
\end{center}
\end{figure}

We subtract the $\epem\!\! \to\! \omega f_0(980)$ cross section 
from the inclusive $\epem\!\! \to\! \omega\pipi$ cross section to
obtain the cross section shown in Fig.~\ref{ompipixs}.
A peak is visible, presumably from the $\omega(1650)$, and the
shoulder at lower masses can be attributed to the $\omega(1420)$.  
We fit this cross section as a function of $\Ecm = \sqrt{s}$ 
from threshold up to 2.4~\gev with a sum of vector resonances,
\begin{eqnarray}
\sigma(\Ecm)   &\! =\! & \frac{P(s)}{s}
    \left| \sum_{k=1}^{n} \frac{\sqrt{\sigma_{0k}} m_k^2\Gamma_k}
                 {m_k^2-s-i\sqrt{s}\Gamma_k}
                 \frac{e^{i\phi_k}}{\sqrt{P(m_k^2)}} \right|^2~              
\label{bwsum}
\\
\sigma_{0k} &\! =\! & \frac{12 \pi B_{ke}B_{kf} C}{m_k^2},~
\label{sigma0}
\\
 P(s)        &\! =\! & \sqrt{s - m_0^2} , \nonumber
\end{eqnarray}
where $\Gamma_k$ and $m_k$ are the full width and mass of the $k^{\rm th}$ resonance,
$B_{ke}$ and $B_{kf}$ are its branching fractions to \epem and the
final state $f\! =\! \omega\pipi$, respectively,
$P(s)$ is a simple approximation of the phase space with a
threshold cutoff at $m_0\! =\! 1.2~\gev$,
and $C\! =\! 3.893\cdot 10^5$ \nb~GeV$^2$ is a conversion constant. 
This formulation allows the extraction not only of the product 
$B_{ke}B_{kf}$ but also the peak cross section $\sigma_{0k}$ or the 
product $\Gamma_{ke}B_{vf}$.

\begin{table*}[tbh]
\caption{
  Summary of parameters obtained from the fits described in the text.
  The values without errors were fixed in that fit.
  }
\label{omfit_tab}
\begin{ruledtabular}
\begin{tabular}{l c c c c c } 
Fit &  1 & 2 & 3 & $3\pi$~\cite{isr3pi} & PDG~\cite{PDG}\\
\hline
$\sigma_{0w '}$ (\nb)     &    0.10$\pm$0.08   &  1.01$\pm$0.29
                                 &  0.64$\pm$0.34    
                                 &  --                     &   --   \\
$B_{ee}B_{w'f}\cdot 10^6$    & 0.013$\pm$0.010     &  0.13$\pm$0.04
                                 &  0.101$\pm$0.055    
                                 &  0.82$\pm$0.08    &   --   \\
$\Gamma_{ee}B_{w'f}$(\ev)   & 1.4$\pm$1.01      &  17.5$\pm$5.4
                                 &  37.8$\pm$12.1   
                                 &  369                   &   --   \\
$m_{w'}$(\gevcc)          &   1.381$\pm$0.032  &  1.382$\pm$0.023  
                                  & 1.463$\pm$0.070 
                                  &  1.350$\pm$0.030       &  1.40--1.45 \\
$\Gamma_{w'}$(\gev)    &  0.105$\pm$0.090        & 0.133$\pm$0.048   
                                  & 0.383$\pm$0.233 
                                  &   0.450$\pm$0.140      &  0.180--0.250 \\
$\phi_{w'}$ (rad.)          &   -1.93$\pm$0.73         &     $\pi$   
                                  &    -0.61$\pm$0.94          
                                  &      $\pi$      &   -- \\
$\sigma_{0w ''}$ (\nb)     &   2.14$\pm$0.18          & 2.47$\pm$0.18  
                                 &  1.03$\pm$0.54    
                                 &  --                    &   --   \\
$B_{ee}B_{w''f}\cdot 10^6$    &  0.41$\pm$0.03    &  0.47$\pm$0.04
                                 &  0.193$\pm$0.087    
                                 &  1.3$\pm$0.2                    &   --   \\
$\Gamma_{ee}B_{w''f}$(\ev)   & 96.5$\pm$10.9   &  103.5$\pm$8.3
                                 &  28.7$\pm$7.7    
                                 &  286                   &   --   \\
$m_{w''}$(\gevcc)          &  1.673$\pm$0.011        &  1.667$\pm$0.013   
                                  &  1.661$\pm$0.032 
                                  &  1.660$\pm$0.011       & 1.670$\pm$0.030 \\
$\Gamma_{w''}$(\gev)    &  0.236$\pm$0.029         &  0.222$\pm$0.025   
                                  & 0.148$\pm$0.037 
                                  &  0.220$\pm$0.040       &  0.315$\pm$0.035 \\
$\phi_{w''}$ (rad.)          &   0      &       0
                                  &   0.02$\pm$0.71         
                                  &      0     &   -- \\
$\sigma_{0w }$ (\nb)     & 0                     &  102$\pm$67
                                 &  147$\pm$140    
                                 &   PDG                    &   --   \\
\chisq /n.d.f.           &    36.2/48      &     34.9/48
                         &    32.2/46      
                         &    --      &     --     \\
\end{tabular}
\end{ruledtabular}
\end{table*}                           

We consider the resonances, $k\! =\! \omega(782)$ ($\omega$), 
$\omega(1420)$ ($\omega'$), and $\omega(1650)$ ($\omega''$),
where $\omega(782)$ is below threshold and is used as a convenient 
``coherent background'' with parameters fixed to PDG values~\cite{PDG}
and with $\phi_{\omega(782)}$ set to 0.
We perform three fits, the results of which are listed in 
Table~\ref{omfit_tab} and compared with results of a similar fit 
from our study of the ISR $\pipi\piz$ process~\cite{isr3pi} and 
with current PDG values~\cite{PDG}.
In the first fit, we set the contribution from $\omega$
to zero and set $\phi_{\omega''}\! =\! 0$.
The fitted cross section is dominated by $\omega''$, and the $\omega'$ 
has a relatively narrow width.

Next we float the contribution from $\omega$ but fix the relative 
phases to the values used in our ISR $\pipi\pi^0$ study~\cite{isr3pi},
$\phi_{\omega'}\! =\! \pi$ and $\phi_{\omega''}\! =\! 0$.
The resulting contribution from $\omega'$ is almost 10 times higher 
due to destructive interference,
but the masses and widths are similar to those from the first fit.
In particular the $\omega'$ width is lower than that found in 
Ref.~\cite{isr3pi}.
The fitted peak $\omega$ cross section corresponds to a large 
$\omega\pipi$ branching fraction of about 7\%, 
but this is driven by the data above 2~\gev and should be considered a
measure of the coherent background.

An interference with other unaccounted vector mesons could 
produce deviations from the assumed values of the phases.
To demonstrate this
we float the $\omega$ level and both relative phases,
and show the result as the curve in Fig.~\ref{ompipixs}.
The coherent background is larger and both the $\omega'$ and
$\omega''$ peak cross sections are much lower than in the second fit.
Both masses are consistent with the other fits, but the $\omega'$ 
 ($\omega''$) width is larger (smaller) and statistically consistent
with our ISR $\pipi\piz$ study.

A better understanding of the background, 
including any structure above 2~\gev and any contribution from excited
$\rho$ or $\phi$ states,
is needed in order to make precise measurements of the excited $\omega$ 
resonance parameters.
Taking the results from the second fit and using the differences from
the other fits to estimate systematic errors, we obtain:
\begin{eqnarray*}
      m_{\omega(1420)} & = & 1.38\;\: \pm 0.02\;\: \pm 0.07\;\:~\gevcc, \\
 \Gamma_{\omega(1420)} & = & 0.13\;\: \pm 0.05\;\: \pm 0.10\;\:~\gev,   \\
      m_{\omega(1650)} & = & 1.667    \pm 0.013    \pm 0.006~\gevcc, \\
 \Gamma_{\omega(1650)} & = & 0.222    \pm 0.025    \pm 0.020~\gev.
\end{eqnarray*}
The $\omega(1650)$ width is significantly different from the PDG 
value~\cite{PDG} based on the DM2 results~\cite{omegadm2, 5picmd2}, 
but consistent with our measurement in ISR $\pipi\piz$ events~\cite{isr3pi}.
Note that the structure, observed in our study of ISR $\omega(782)\eta$ 
events~\cite{isr6pi} and described by a resonance with
$m\! =\! 1.645 \pm 0.008$~\gevcc and $\Gamma\! =\! 0.114\pm0.014$~\gev
can also be interpreted as $\omega(1650)$.

\begin{figure}[tbh]
\begin{center}
\includegraphics[width=0.9\linewidth]{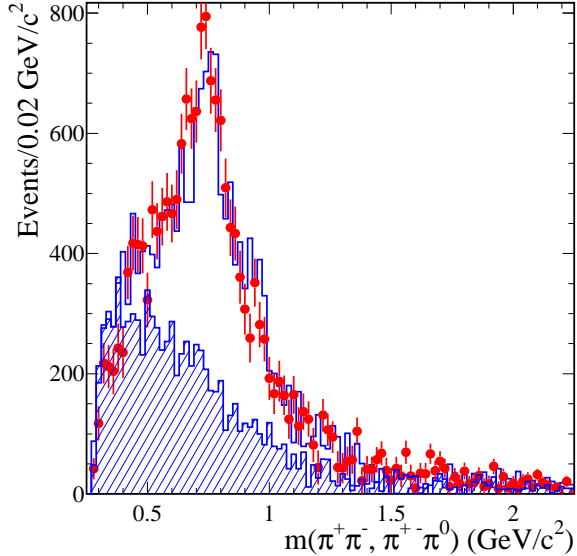}
\vspace{-0.4cm}
\caption{
  The background-subtracted invariant mass distributions for all \pipi pairs 
  (open histogram, four entries per event),
  $\pi^\pm\piz$ pairs (points, four entries per event)
  and $\pip\pip$ and $\pim\pim$ pairs
  (hatched histogram, two entries per event),
  for events with no $\eta$ or $\omega\! \to\! \pipi\piz$
  candidates.
  }
\label{rho_sel}  
\end{center}
\end{figure}

\begin{figure}[tbh]
\begin{center}
\includegraphics[width=0.48\linewidth]{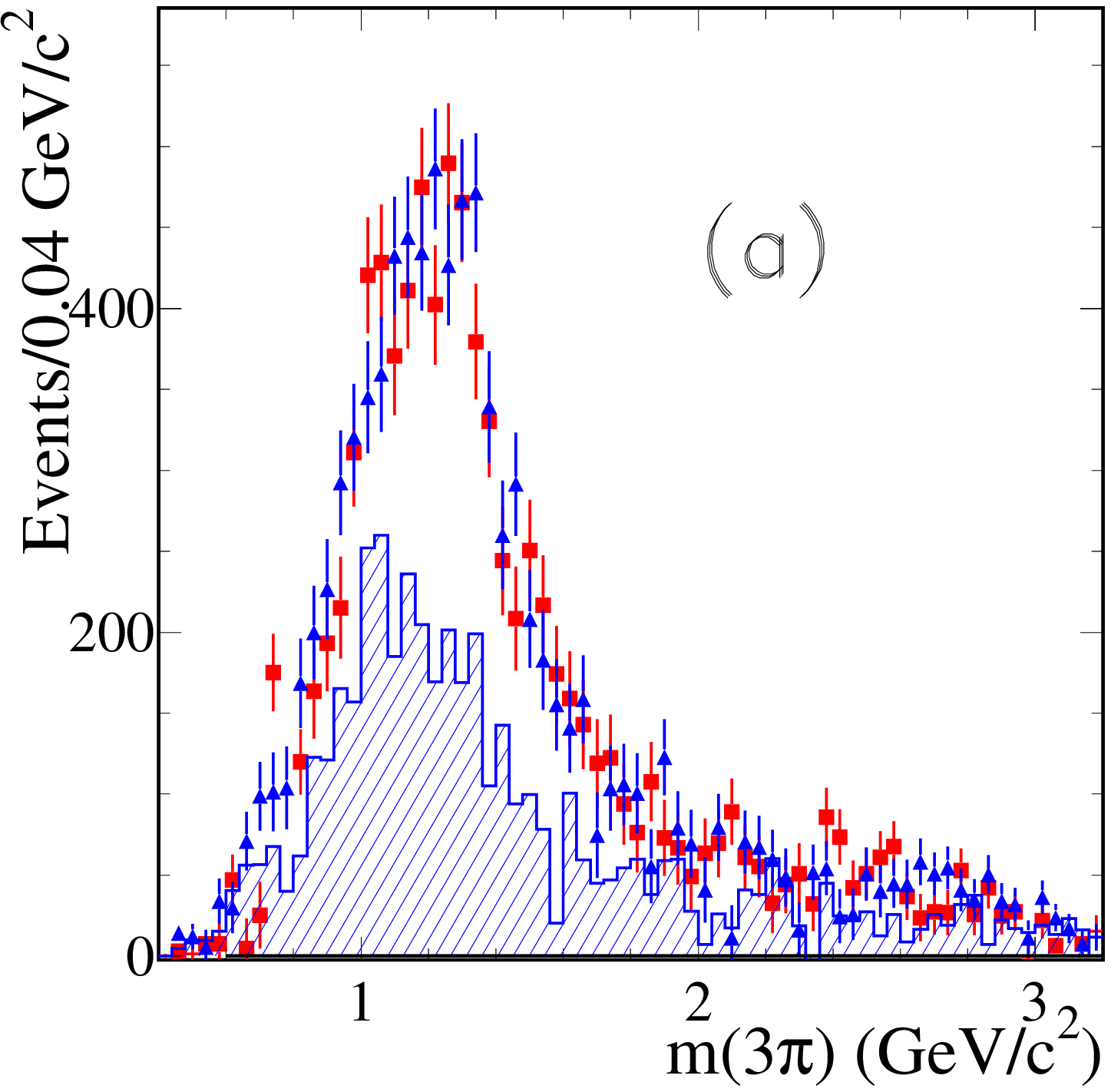}
\includegraphics[width=0.48\linewidth]{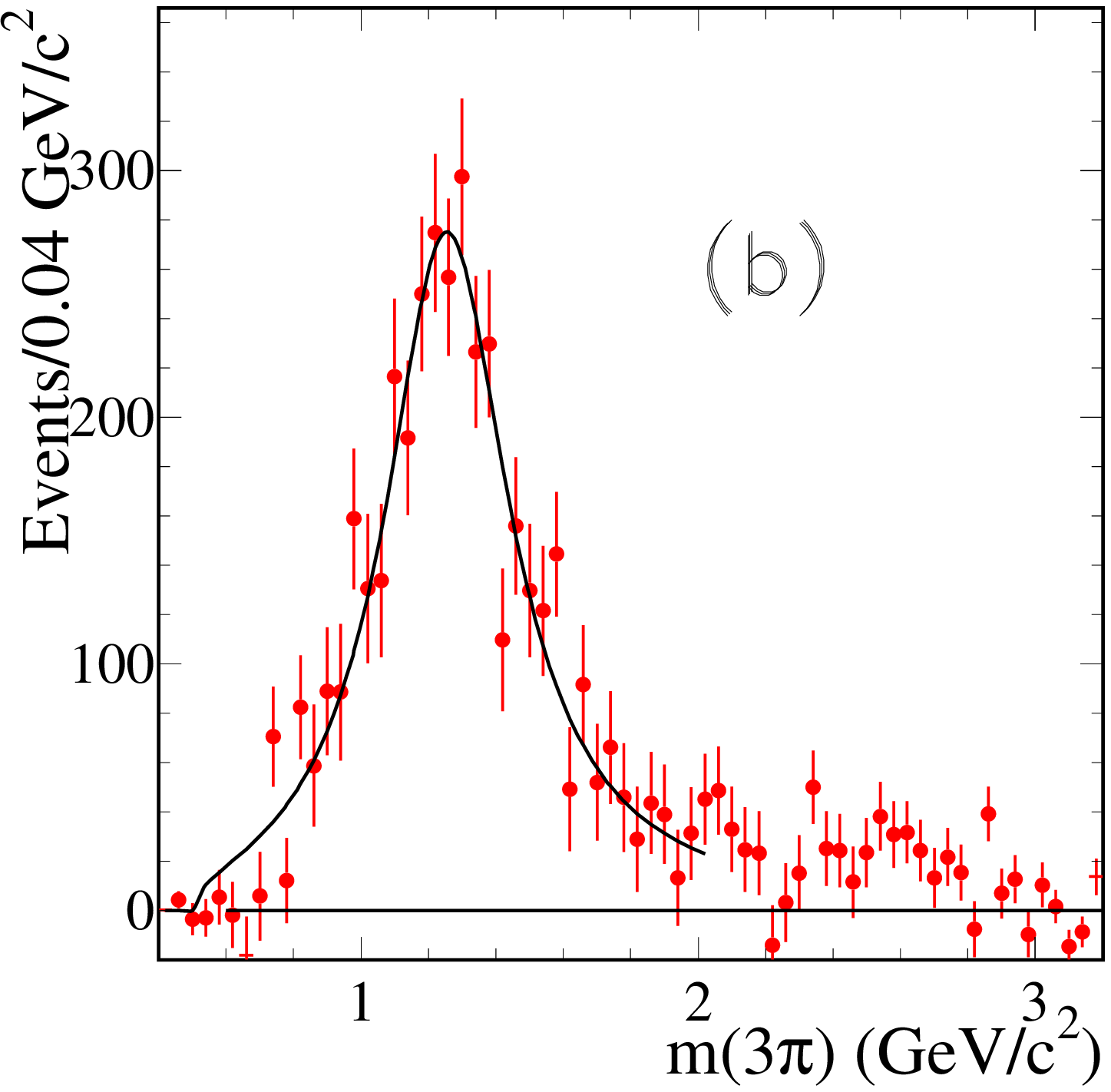}
\vspace{-0.4cm}
\caption{
  (a) The invariant mass distributions for $\pipi\piz$ (triangles) and 
  $\pipi\pi^\pm$ (squares) combinations for events in which the other \pipi or
  $\pi^{\pm}\piz$ pair is in the $\rho$ signal region.
  The hatched histogram is an estimate of the combinatorial background.
  (b) The average of the triangles and squares in (a) minus the combinatorial
  background.
  The line is the result of a Breit-Wigner fit.
  }
\label{3pirho_sel}  
\end{center}
\end{figure}

\subsection{\bf\boldmath The $\rho(770)3\pi$ 
  Intermediate States}

To study events containing a charged or neutral $\rho(770)$ we first
exclude any event in which a $\pipi\piz$ combination has invariant
mass within 25~\mevcc of the $\eta$ mass or within 40~\mevcc of the $\omega$
mass. 
For this study we also exclude events with a five-pion mass within
50~\mevcc of the $J/\psi$ mass.
Figure~\ref{rho_sel} shows the invariant mass distributions for all 
four $\pipi$ pairs and all four $\pi^\pm\pi^0$ pairs in the remaining events.
The ISR and non-ISR backgrounds are subtracted using the procedures 
described above.
These two distributions are quite similar and show strong 
$\rho(770)$ peaks.
The hatched histogram in Fig.~\ref{rho_sel} shows the mass distribution for 
the $\pip\pip$ and $\pim\pim$ pairs (two entries per event),
which gives an estimate of the combinatorial background.
The difference between these distributions is consistent with an
average of two $\rho$ per event: 
one $\rho$ is charged and the other neutral,
since the yields are consistent and $\epem\!\! \to\! \rho^0\rho^0\pi^0$ 
is forbidden by C-parity.
This suggests one or more quasi-two-body intermediate states,
$X^{\pm ,0}\rho^{\mp ,0}$, 
where $X$ could be $a_1(1260)$, $\pi(1300)$ or $a_2(1370)$,
which have $I=1$ and a dominant $\rho\pi$ decay. 

\begin{figure}[t]
\begin{center}
\includegraphics[width=0.49\linewidth]{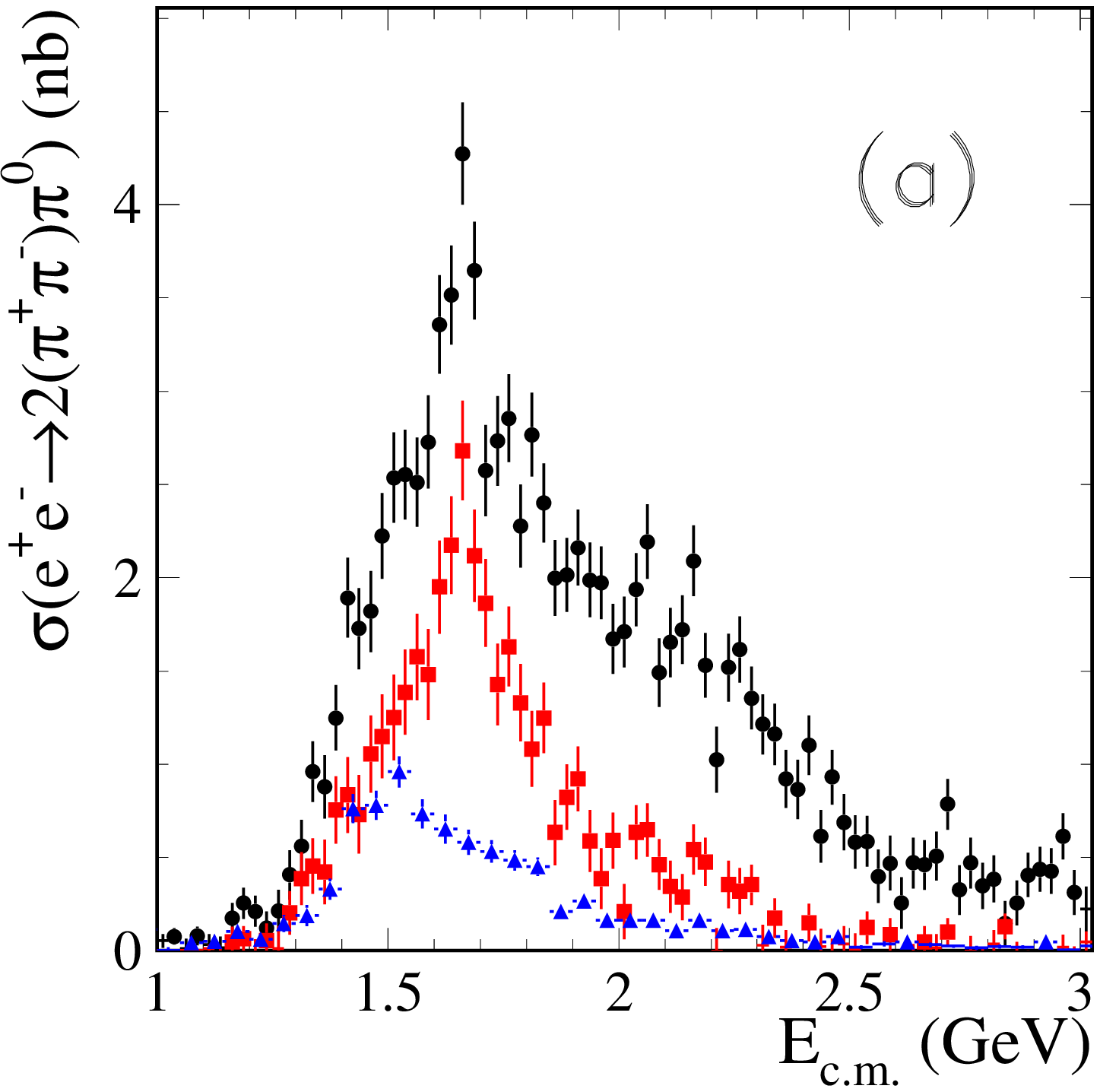}
\includegraphics[width=0.49\linewidth]{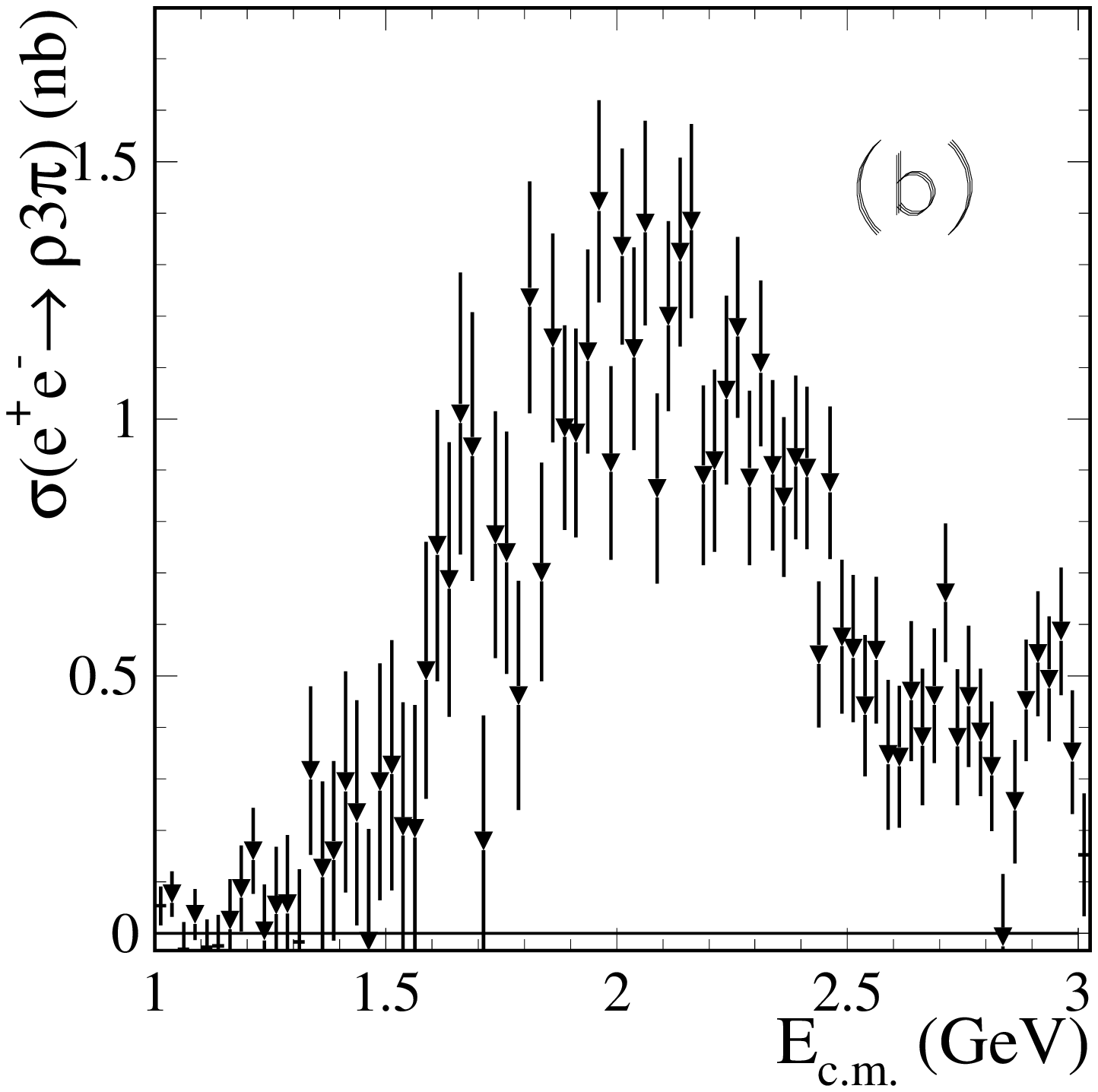}
\vspace{-0.4cm}
\caption{
  (a) The total $\epem\!\! \to\! \fourpipn$ cross section (circles) and 
  the contributions from $\omega\pipi$ (squares) and $\eta\pipi$ (triangles).
  (b) The cross section obtained as the 
  difference between the total and the latter two contributions, which
  is dominated by the $\epem\!\! \to\! \rho(770) X$ process.
  }
\label{xsrhox1}  
\end{center}
\end{figure}

We now select events that contain a $\pi^\pm\pi^0$ or $\pipi$ pair
with mass within 150~\mevcc of the $\rho$ mass.
Figure~\ref{3pirho_sel}(a) shows the mass distributions for the other
charged (squares) and neutral (triangles) three-pion combinations (up
to four total entries per event).
These two distributions are consistent and the hatched histogram is an 
estimate of the combinatorial background from doubly charged 
($\pip\pip\piz$ and $\pim\pim\piz$) combinations.
By averaging the charged and neutral distributions and subtracting the 
combinatorial background we obtain the distribution shown in 
Fig.~\ref{3pirho_sel}(b),
which is consistent with a resonant structure.
Fitting a single Breit-Wigner function gives 
\begin{eqnarray*}
      m(X) & = & 1.243 \pm 0.012 \pm 0.020~\gevcc; \\
 \Gamma(X) & = & 0.410 \pm 0.031 \pm 0.030~\gev .
\end{eqnarray*}
The first errors are statistical and the second systematic, 
dominated by the background subtraction procedure.
These values are inconsistent with the $a_2(1320)$ resonance,
but consistent, within large uncertainties, with the
$\pi(1300)$ and  $a_1(1260)$~\cite{PDG}.
An angular analysis could distinguish between these possibilities, 
but requires substantially higher statistics due to the large 
combinatorial background.

Since the events that do not contain an $\eta$ or $\omega$ 
appear 
to be predominantly $X\rho(770)\! \to\! \rho^0\rho^{\pm}\pi^{\mp}$ events, 
where $X$ is consistent with a single resonance,
we obtain an $\epem\!\! \to\! X\rho(770)$ cross section as the
difference between the total $\epem\!\! \to\! \fourpipn$ cross section 
(Fig.~\ref{4pipi0_ee_babar}) and the 
$\epem\!\! \to\! \omega\pipi$ and $\epem\!\! \to\! \eta\pipi$ cross sections
(Figs.~\ref{xsomegapipi} and~\ref{xsetapipi} with branching fraction 
corrections removed).
We show these three cross sections in Fig.~\ref{xsrhox1}(a) for
energies up to 3~\gev,
and the difference in Fig.~\ref{xsrhox1}(b);
it shows no sharp structure.
Above 3~\gev, the contributions from $\eta\pipi$ and $\omega\pipi$ are
consistent with zero, 
so the cross section is as in Fig.~\ref{4pipi0_ee_babar}, except for
the $J/\psi$ and $\psi(2S)$ peaks, which can have a different substructure
(see Sec.~\ref{sec:charmonium}).

\begin{figure}[tbh]
\begin{center}
\includegraphics[width=0.9\linewidth]{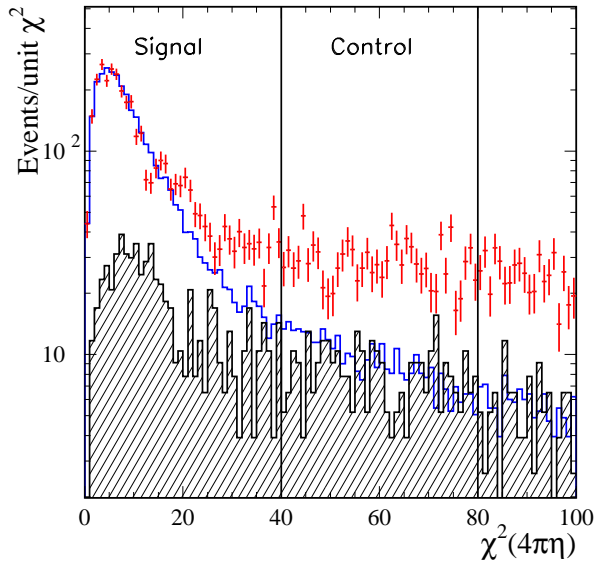}
\vspace{-0.4cm}
\caption{
   Distribution of \chisq from the 5C fit for \fourpieta candidates in
   the data (points).
   The open and hatched histograms are the distribution for simulated
   signal events and non-ISR background, respectively,
   normalized as described in the text.
   }
\label{4pieta_chi2_all}
\end{center}
\end{figure}

\section{\bf\boldmath The \fourpieta final state}
\subsection{Final selection and background}

To suppress \KKppeta background, we require that no more than one
track in the event is identified as a kaon,
and we also fit under all four possible \KKppeta hypotheses and
require $\chiKKppeta\! >\! 30$.
We suppress any $K^\pm\KS\pi^\mp\eta$ background
by requiring all tracks to extrapolate within 2.5 mm of the beam axis.
The \chifourpieta distribution for the remaining events is shown as
points in Fig.~\ref{4pieta_chi2_all}, 
and the distribution for simulated \fourpipn events (open histogram)
is normalized to the data in the region $\chifourpieta\! <\! 10$.
We do not simulate \fourpieta events, 
but we expect the efficiency and resolution for $\eta$ to be
similar to that for \piz.
The hatched histogram represents the non-ISR background contribution 
obtained from the JETSET simulation.
It is dominated by $\fourpi\piz\eta$ events,
and we use the same normalization factor as for the \fourpipn events  
described in Sec.~\ref{sec:eff1}.
We define a signal region, $\chifourpieta\! <\! 40$, 
containing 4272 events, 
and a control region for the estimation of other backgrounds, 
$40\! <\!\chifourpieta\! <\! 80$, containing 1485 events.

\begin{figure}[tbh]
\begin{center}
\includegraphics[width=0.9\linewidth]{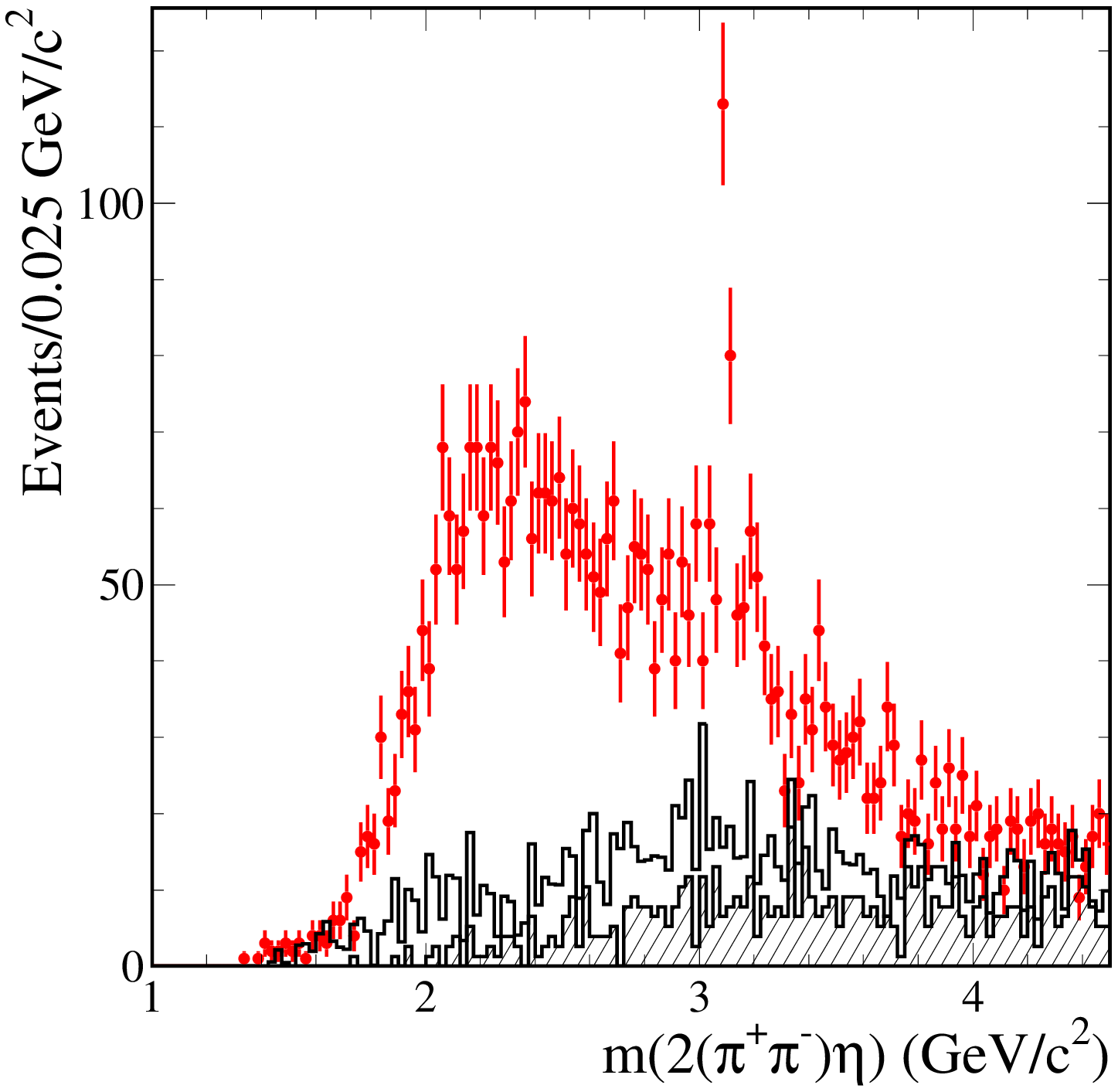}
\vspace{-0.4cm}
\caption{
  Invariant mass distribution for selected \fourpieta events in the
  data (points).
  The hatched and open histograms represent, cumulatively, the
  non-ISR background and the background from the control region
  of Fig.~\ref{4pieta_chi2_all}.
  }
\label{4pieta_babar}
\end{center}
%
%
\begin{center}
\includegraphics[width=0.9\linewidth]{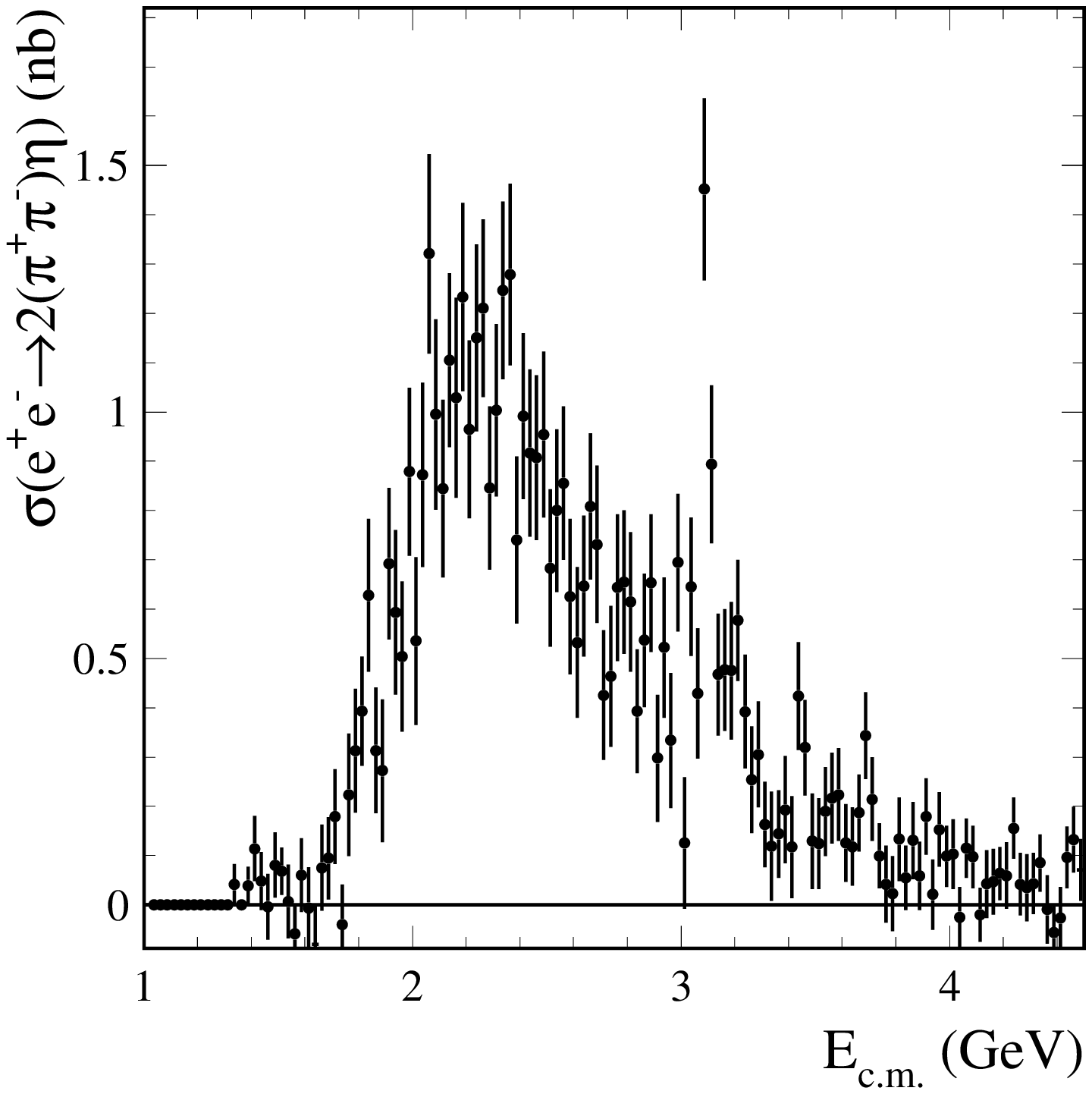}
\vspace{-0.4cm}
\caption{
  The $\epem\!\! \to\! \fourpieta$ cross section as a function of
  c.m.\ energy measured with ISR data.
  Only statistical errors are shown.
  }
\label{4pieta_ee_babar}
\end{center}
\end{figure} 

Figure~\ref{4pieta_babar} shows the \fourpieta invariant mass
distribution from threshold up to 4.5~\gevcc for events in the signal region.
A $J/\psi$ signal is visible.
The hatched histogram represents the non-ISR background,
and the open histogram represents the sum of all backgrounds, where
the ISR-type background is estimated from the control region.
Both backgrounds are relatively small at low mass, about 20\% altogether,
but they account for 50--80\% of the observed data in the
3.0--4.5~\gevcc region.  
We subtract this total background in each bin to obtain a number of
signal events.

\begin{table*}
\caption{Measurements of the $\ep\en\to 2(\pipi)\eta$ 
cross section (errors are statistical only).}
\label{4pieta_tab}
\begin{ruledtabular}
\hspace{-1.8cm}
\begin{tabular}{ c c c c c c c c }
$E_{\rm c.m.}$ (GeV) & $\sigma$ (nb)  
& $E_{\rm c.m.}$ (GeV) & $\sigma$ (nb) 
& $E_{\rm c.m.}$ (GeV) & $\sigma$ (nb) 
& $E_{\rm c.m.}$ (GeV) & $\sigma$ (nb)  
\\
\hline

 1.3125 &  0.00 $\pm$  0.00 & 2.1125 &  0.84 $\pm$  0.18 & 2.9125 &  0.30 $\pm$  0.13 & 3.7125 &  0.21 $\pm$  0.08 \\
 1.3375 &  0.04 $\pm$  0.04 & 2.1375 &  1.10 $\pm$  0.18 & 2.9375 &  0.52 $\pm$  0.14 & 3.7375 &  0.10 $\pm$  0.07 \\
 1.3625 &  0.00 $\pm$  0.00 & 2.1625 &  1.03 $\pm$  0.20 & 2.9625 &  0.33 $\pm$  0.14 & 3.7625 &  0.04 $\pm$  0.08 \\
 1.3875 &  0.04 $\pm$  0.04 & 2.1875 &  1.23 $\pm$  0.19 & 2.9875 &  0.69 $\pm$  0.14 & 3.7875 &  0.02 $\pm$  0.08 \\
 1.4125 &  0.11 $\pm$  0.07 & 2.2125 &  0.96 $\pm$  0.18 & 3.0125 &  0.13 $\pm$  0.13 & 3.8125 &  0.13 $\pm$  0.08 \\
 1.4375 &  0.05 $\pm$  0.06 & 2.2375 &  1.15 $\pm$  0.19 & 3.0375 &  0.65 $\pm$  0.14 & 3.8375 &  0.06 $\pm$  0.07 \\
 1.4625 &  0.00 $\pm$  0.07 & 2.2625 &  1.21 $\pm$  0.18 & 3.0625 &  0.43 $\pm$  0.13 & 3.8625 &  0.13 $\pm$  0.08 \\
 1.4875 &  0.08 $\pm$  0.07 & 2.2875 &  0.85 $\pm$  0.17 & 3.0875 &  1.45 $\pm$  0.19 & 3.8875 &  0.06 $\pm$  0.07 \\
 1.5125 &  0.07 $\pm$  0.05 & 2.3125 &  1.00 $\pm$  0.17 & 3.1125 &  0.89 $\pm$  0.16 & 3.9125 &  0.18 $\pm$  0.08 \\
 1.5375 &  0.01 $\pm$  0.07 & 2.3375 &  1.25 $\pm$  0.18 & 3.1375 &  0.47 $\pm$  0.12 & 3.9375 &  0.02 $\pm$  0.07 \\
 1.5625 & -0.06 $\pm$ 0.06~ & 2.3625 &  1.28 $\pm$  0.18 & 3.1625 &  0.48 $\pm$  0.12 & 3.9625 &  0.15 $\pm$  0.08 \\
 1.5875 &  0.06 $\pm$  0.07 & 2.3875 &  0.74 $\pm$  0.17 & 3.1875 &  0.48 $\pm$  0.14 & 3.9875 &  0.10 $\pm$  0.06 \\
 1.6125 & -0.01 $\pm$ 0.08~ & 2.4125 &  0.99 $\pm$  0.17 & 3.2125 &  0.58 $\pm$  0.12 & 4.0125 &  0.10 $\pm$  0.07 \\
 1.6375 & -0.08 $\pm$ 0.08~ & 2.4375 &  0.92 $\pm$  0.17 & 3.2375 &  0.39 $\pm$  0.12 & 4.0375 & -0.02 $\pm$ 0.06~ \\
 1.6625 &  0.08 $\pm$  0.09 & 2.4625 &  0.91 $\pm$  0.17 & 3.2625 &  0.25 $\pm$  0.11 & 4.0625 &  0.11 $\pm$  0.06 \\
 1.6875 &  0.09 $\pm$  0.08 & 2.4875 &  0.95 $\pm$  0.17 & 3.2875 &  0.31 $\pm$  0.11 & 4.0875 &  0.10 $\pm$  0.06 \\
 1.7125 &  0.18 $\pm$  0.10 & 2.5125 &  0.68 $\pm$  0.16 & 3.3125 &  0.16 $\pm$  0.09 & 4.1125 & -0.02 $\pm$ 0.05~ \\
 1.7375 & -0.04 $\pm$ 0.08~ & 2.5375 &  0.80 $\pm$  0.17 & 3.3375 &  0.12 $\pm$  0.11 & 4.1375 &  0.04 $\pm$  0.07 \\
 1.7625 &  0.22 $\pm$  0.12 & 2.5625 &  0.86 $\pm$  0.16 & 3.3625 &  0.14 $\pm$  0.09 & 4.1625 &  0.05 $\pm$  0.07 \\
 1.7875 &  0.31 $\pm$  0.13 & 2.5875 &  0.63 $\pm$  0.16 & 3.3875 &  0.19 $\pm$  0.11 & 4.1875 &  0.06 $\pm$  0.05 \\
 1.8125 &  0.39 $\pm$  0.11 & 2.6125 &  0.53 $\pm$  0.15 & 3.4125 &  0.12 $\pm$  0.10 & 4.2125 &  0.06 $\pm$  0.07 \\
 1.8375 &  0.63 $\pm$  0.16 & 2.6375 &  0.65 $\pm$  0.14 & 3.4375 &  0.42 $\pm$  0.11 & 4.2375 &  0.16 $\pm$  0.06 \\
 1.8625 &  0.31 $\pm$  0.13 & 2.6625 &  0.81 $\pm$  0.15 & 3.4625 &  0.32 $\pm$  0.10 & 4.2625 &  0.04 $\pm$  0.06 \\
 1.8875 &  0.27 $\pm$  0.14 & 2.6875 &  0.73 $\pm$  0.16 & 3.4875 &  0.13 $\pm$  0.10 & 4.2875 &  0.03 $\pm$  0.07 \\
 1.9125 &  0.69 $\pm$  0.15 & 2.7125 &  0.43 $\pm$  0.13 & 3.5125 &  0.12 $\pm$  0.09 & 4.3125 &  0.04 $\pm$  0.06 \\
 1.9375 &  0.59 $\pm$  0.17 & 2.7375 &  0.46 $\pm$  0.14 & 3.5375 &  0.19 $\pm$  0.09 & 4.3375 &  0.09 $\pm$  0.06 \\
 1.9625 &  0.50 $\pm$  0.15 & 2.7625 &  0.64 $\pm$  0.15 & 3.5625 &  0.22 $\pm$  0.09 & 4.3625 & -0.01 $\pm$ 0.07~ \\
 1.9875 &  0.88 $\pm$  0.17 & 2.7875 &  0.65 $\pm$  0.15 & 3.5875 &  0.22 $\pm$  0.09 & 4.3875 & -0.06 $\pm$ 0.06~ \\
 2.0125 &  0.54 $\pm$  0.17 & 2.8125 &  0.61 $\pm$  0.14 & 3.6125 &  0.13 $\pm$  0.08 & 4.4125 & -0.03 $\pm$  0.06 \\
 2.0375 &  0.87 $\pm$  0.19 & 2.8375 &  0.39 $\pm$  0.13 & 3.6375 &  0.12 $\pm$  0.08 & 4.4375 &  0.10 $\pm$  0.06 \\
 2.0625 &  1.32 $\pm$  0.20 & 2.8625 &  0.54 $\pm$  0.14 & 3.6625 &  0.19 $\pm$  0.08 & 4.4625 &  0.13 $\pm$  0.07 \\
 2.0875 &  0.99 $\pm$  0.19 & 2.8875 &  0.65 $\pm$  0.14 & 3.6875 &  0.34 $\pm$  0.09 & 4.4875 &  0.07 $\pm$  0.06 \\

\end{tabular}
\end{ruledtabular}
\end{table*}

\subsection{\boldmath Cross section for $\epem\to \fourpieta$}
\label{4pieta}

We calculate the cross section for the $\epem\!\! \to\! \fourpieta$
process as described in Sec.~\ref{sec:xs4pipi0}, 
by dividing the number of events in each \fourpieta mass bin by the 
corrected detection efficiency and differential luminosity.
The angular acceptance is uniform in all of our simulations,
and this has been demonstrated in our previous studies of four- and
six-pion final states~\cite{isr4pi, isr6pi}.
We therefore use the same detection efficiency as for the \fourpipn process,
shown in Fig.~\ref{mc_acc1}, 
divided by the $\eta\! \to\! \gamma\gamma$ branching fraction of 
39.28\% ~\cite{PDG},
and with the systematic error increased to 5\%. 
We use the same corrections and uncertainties for the \chisq cut, 
tracking efficiency and $\eta$-finding efficiency.

We show the cross section as a function of energy in 
Fig.~\ref{4pieta_ee_babar} with statistical errors only, 
and provide a list of our results in Table~\ref{4pieta_tab}.
This is the first measurement of this cross section,
which shows a peak value of about 1.2~\nb at about 2.2~\gev, 
followed by a monotonic decrease toward higher energies, 
broken only by a peak at the $J/\psi$ mass, 
discussed in Sec.~\ref{sec:charmonium}.
Again, the energy resolution is much smaller than the bin width and we
apply no correction.
The overall systematic error is about 10\% for energies below 3~\gev, 
rising to 30-50\% in the 3--4.5~\gev region.

\begin{figure}[tbh]
\begin{center}
\includegraphics[width=0.48\linewidth]{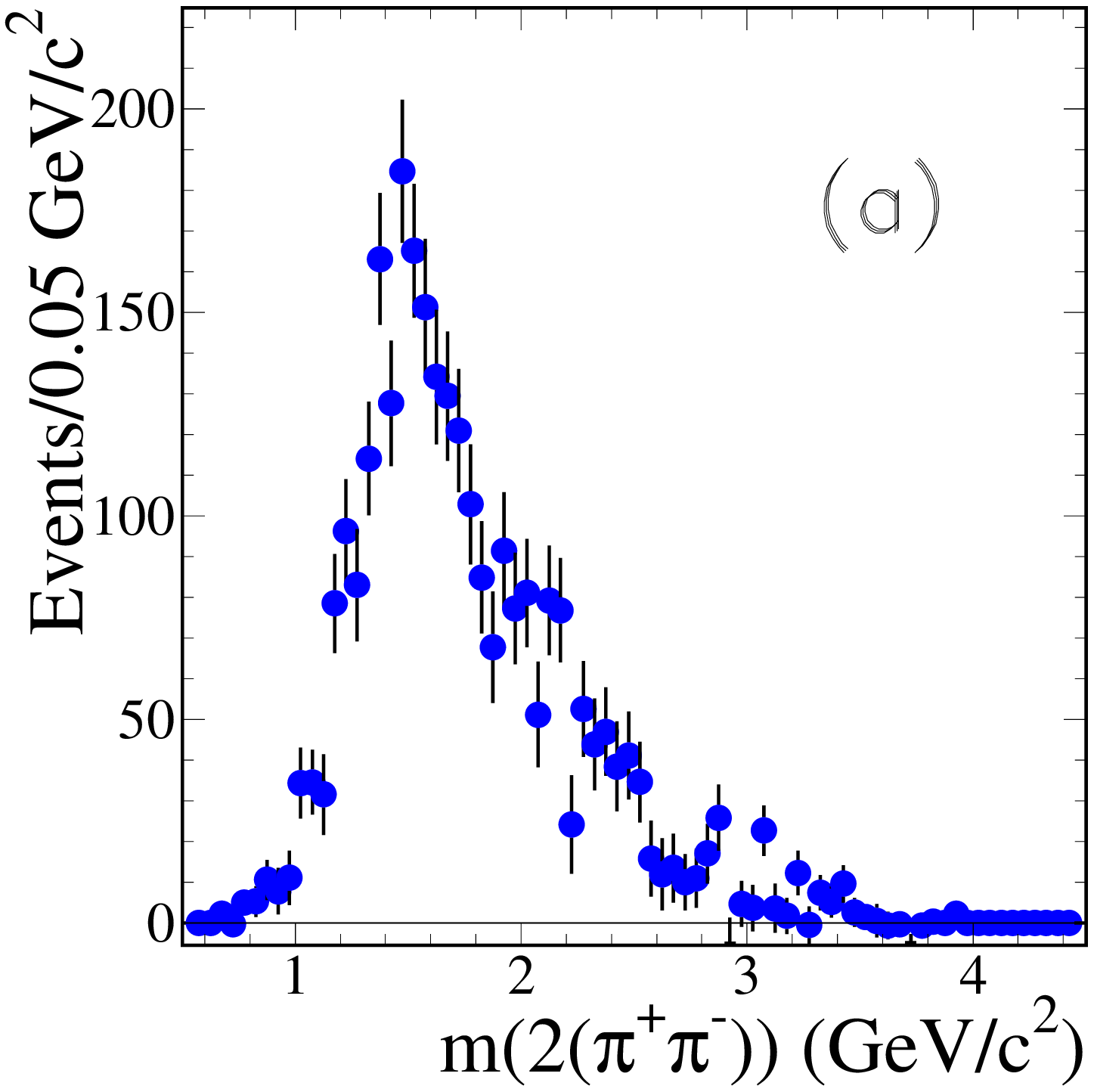}
\includegraphics[width=0.48\linewidth]{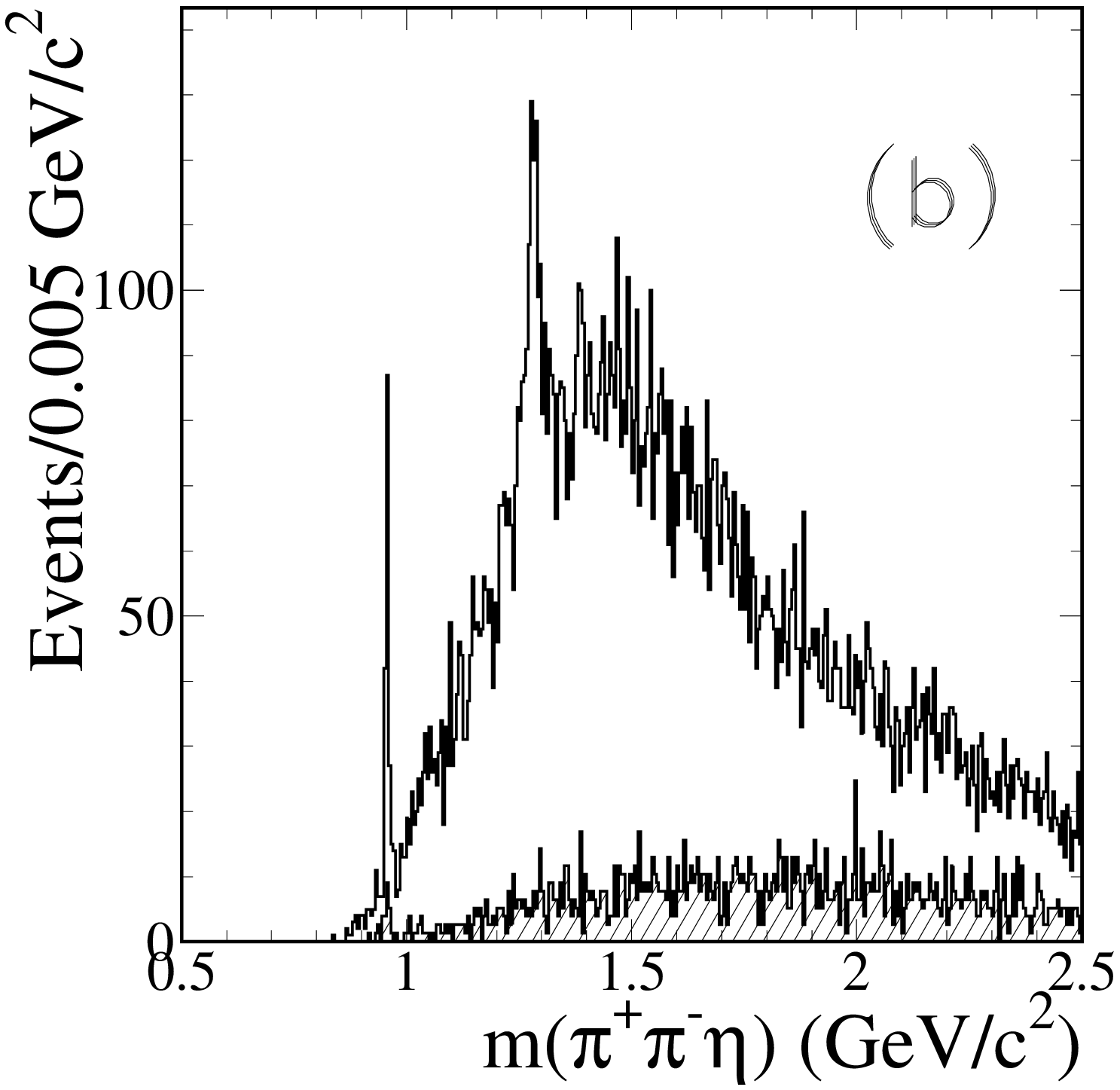}
\vspace{-0.2cm}
\caption{
   (a) Background-subtracted invariant mass distribution for the four
   pions recoiling against the $\eta$ in $2(\pipi)\eta$ events.
   (b) Invariant mass distribution for all $\pipi\eta$ combination 
   (four entries per event) in selected $2(\pipi)\eta$ candidates
   (open histogram) and the estimated non-ISR background (hatched).
   }
\label{2pivs3pieta}  
\end{center}
\end{figure}

\subsection{\boldmath Substructure in the \fourpieta Final State}

We might expect a rich internal structure in the $2(\pipi)\eta$ final state.
Figure ~\ref{2pivs3pieta}(a) shows the invariant mass distribution of
the $\pipi\pipi$ system recoiling against the $\eta$, after subtraction 
of the ISR and non-ISR backgrounds.
The concentration around 1.5~\gevcc is consistent with final state
$\eta\rho(1450)$ being one of the dominant channels in the
$2(\pipi)\eta$ process.
Figure~\ref{2pivs3pieta}(b) shows the mass distribution for all
neutral $\pipi\eta$ combinations (four entries per event).
Signals from the $\eta'(958)$ and 
a peak at 1.3~\gevcc are evident.
There are two candidates decaying to $\eta\pipi$ and
allowed by quantum numbers for the later: $f_1(1285)$ and $\eta(1295)$~\cite{PDG}.
For events with an entry in one of these peaks,
the mass of the remaining \pipi pair is concentrated in the
$\rho(770)$ region, 
indicating that these events are predominantly from the
$\eta'(958)\rho$ and $\eta(1295)\rho$ ($f_1(1285)\rho$). 
The process $\epem\to f_1(1285)\rho(770)$ seems to
be prefered, because $f_1(1285)$ has the decay to
$\gamma\rho(770)$, but $\eta(1295)$ decays to $\eta\pi\pi$ with pions
in S-wave~\cite{PDG} (and not well studied).
We now study these events in detail.

\begin{figure}[tbh]
\begin{center}
\includegraphics[width=0.48\linewidth]{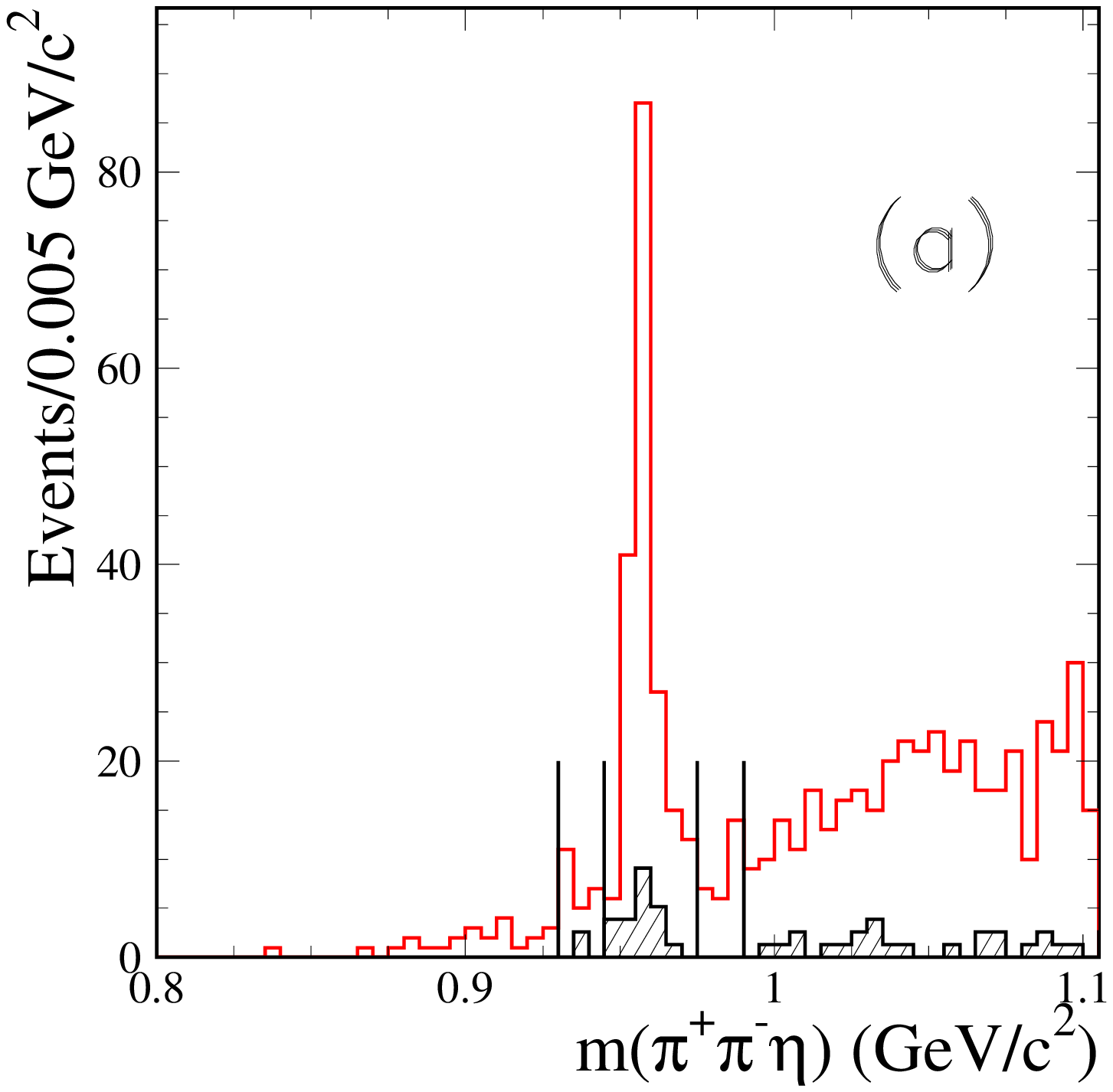}
\includegraphics[width=0.48\linewidth]{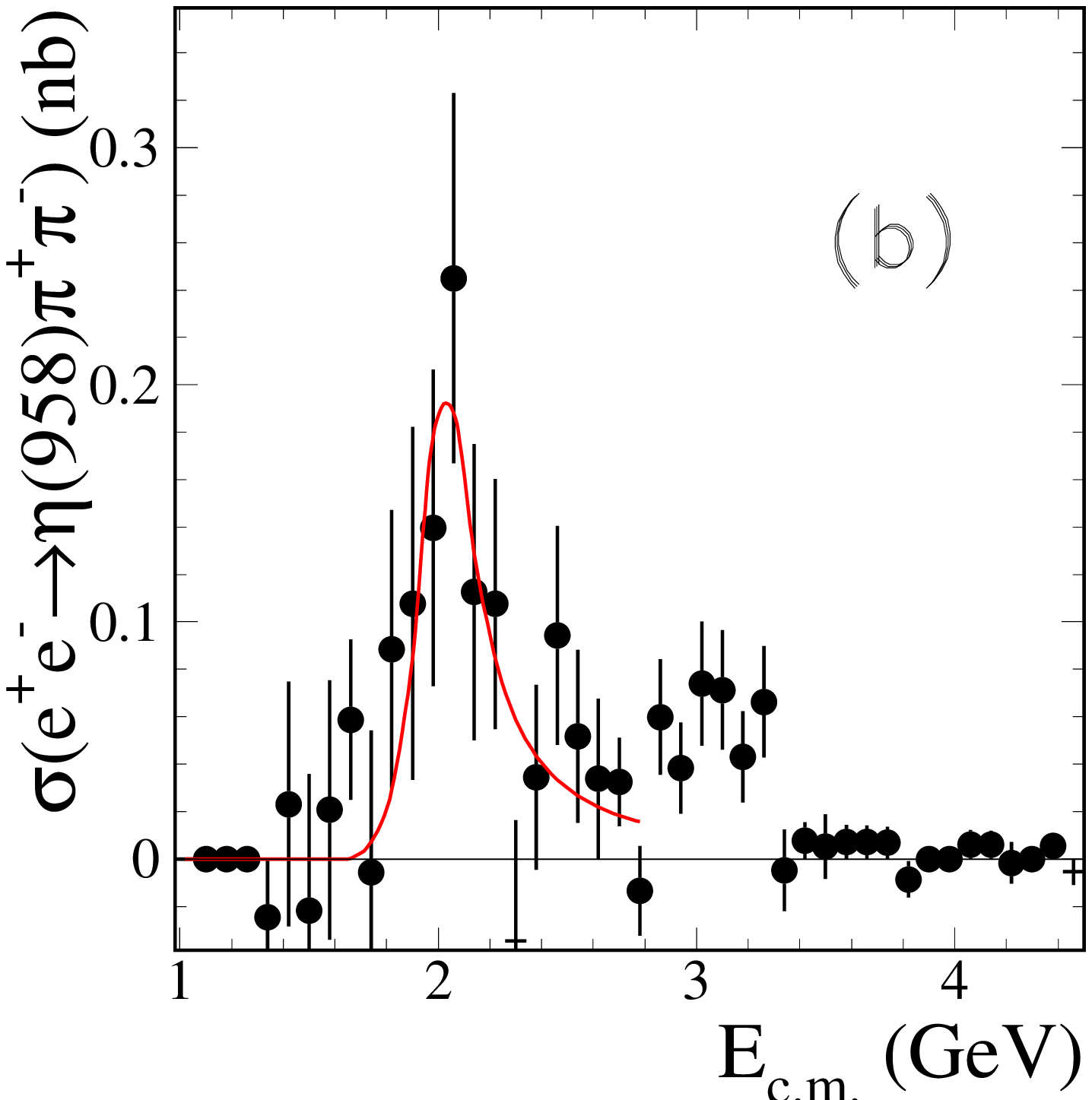}
\vspace{-0.2cm}
\caption{
  (a) Distribution of the $\pipi\eta$ mass closest to the $\eta'$ mass
  in the data (open histogram), along with the estimated non-ISR
  background contribution (hatched).
  The vertical lines indicate the $\eta'(958)$ signal and side band regions.
  (b) The $\epem\!\! \to\! \eta'(958)\pipi$ cross section obtained via ISR.
  The line is the result of the Breit-Wigner fit described in the text. 
  }
\label{2pietaprimmass}  
\end{center}
\end{figure}

\subsection{\boldmath The $\eta'(958)\pipi$ Intermediate State}

To extract the contribution from the $\eta'(958)\pipi$ channel,
we select the $\pipi\eta$ combination with mass closest to the $\eta'(958)$
mass.
Figure~\ref{2pietaprimmass}(a) shows the distribution of this mass
below 1.1~\gevcc.
A clean $\eta'(958)$ signal is visible.
Also shown is the estimated contribution from non-ISR background,
which is small but also shows an $\eta'$ peak.

\begin{table*}
\caption{Measurements of the $\ep\en\to\eta'(958)\pipi$ 
cross section (errors are statistical only).}
\label{etaprimpipi_tab}
\begin{ruledtabular}
\hspace{-1.8cm}
\begin{tabular}{ c c c c c c c c }
$E_{\rm c.m.}$ (GeV) & $\sigma$ (nb)  
& $E_{\rm c.m.}$ (GeV) & $\sigma$ (nb) 
& $E_{\rm c.m.}$ (GeV) & $\sigma$ (nb) 
& $E_{\rm c.m.}$ (GeV) & $\sigma$ (nb)  
\\
\hline

   1.58 &  0.02 $\pm$  0.05 &   2.06 &  0.24 $\pm$  0.08 &   2.54 &  0.05 $\pm$  0.04 &   3.02 &  0.07 $\pm$  0.03 \\
   1.66 &  0.06 $\pm$  0.03 &   2.14 &  0.10 $\pm$  0.06 &   2.62 &  0.03 $\pm$  0.03 &   3.10 &  0.07 $\pm$  0.03 \\
   1.74 &  0.01 $\pm$  0.06 &   2.22 &  0.11 $\pm$  0.05 &   2.70 &  0.03 $\pm$  0.02 &   3.18 &  0.04 $\pm$  0.02 \\
   1.82 &  0.07 $\pm$  0.06 &   2.30 & -0.05 $\pm$ 0.05~ &   2.78 & -0.01 $\pm$ 0.02~ &   3.26 &  0.07 $\pm$  0.02 \\
   1.90 &  0.11 $\pm$  0.07 &   2.38 &  0.03 $\pm$  0.04 &   2.86 &  0.06 $\pm$  0.02 &   3.34 &  0.00 $\pm$  0.02 \\
   1.98 &  0.16 $\pm$  0.06 &   2.46 &  0.09 $\pm$  0.05 &   2.94 &  0.04 $\pm$  0.02 &   3.42 &  0.01 $\pm$  0.01 \\

\end{tabular}
\end{ruledtabular}
\end{table*}

We obtain a cross section in a manner similar to that described in
Sec.~\ref{sec:xs4pipi0}.
We first subtract the non-ISR background in each mass bin, 
then subtract the events in two sidebands, 930--945 and 975--990~\mevcc,
from those in an $\eta'(958)$ signal region, 945--975~\mevcc,
obtaining a total of $120\pm14$ $\eta'(958)\pipi$ events.
Repeating this procedure in bins of the \fourpieta invariant mass and
dividing by the efficiency, ISR luminosity, $\eta\! \to\! \gamma\gamma$ 
branching fraction, and the $\eta'(958)\! \to\! \eta\pipi$ branching
fraction of 0.445~\cite{PDG},
we obtain the $\epem\!\! \to\! \eta'(958)\pipi$ cross section shown in 
Fig.~\ref{2pietaprimmass}(b) and listed in Table~\ref{etaprimpipi_tab}.
It shows a resonance-like behavior at around 2.1~\gev and a sharp drop
at 3.3~\gev. 
Fitting a single Breit-Wigner function,
Eq.~\ref{bwsum} with $m_0 = 1.5~\gev$ to describe the phase space,
we obtain:
\begin{eqnarray*}
  \sigma_0 & = & 0.18 \pm 0.07~\nb, \\
     m_x   & = & 1.99 \pm 0.08~\gevcc,   \\
  \Gamma_x & = & 0.31 \pm 0.14~\gev~.
\end{eqnarray*}
This might be the $\rho(2150)$, 
the next radial excitation of the $\rho$ family, 
reported previously and listed in the detailed PDG tables~\cite{PDG}.
The structure around the $J/\psi$ region can not be explained by 
the $J/\psi\to\eta'(958)\pipi$ decay, 
but could be a background from other $J/\psi$ decay modes with a missing 
$\pi^0$ or undetected radiative photon(s).

\subsection{\bf\boldmath The $f_1 (1285)\pipi$ Intermediate State}

\begin{figure}[tbh]
\begin{center}
\includegraphics[width=0.48\linewidth]{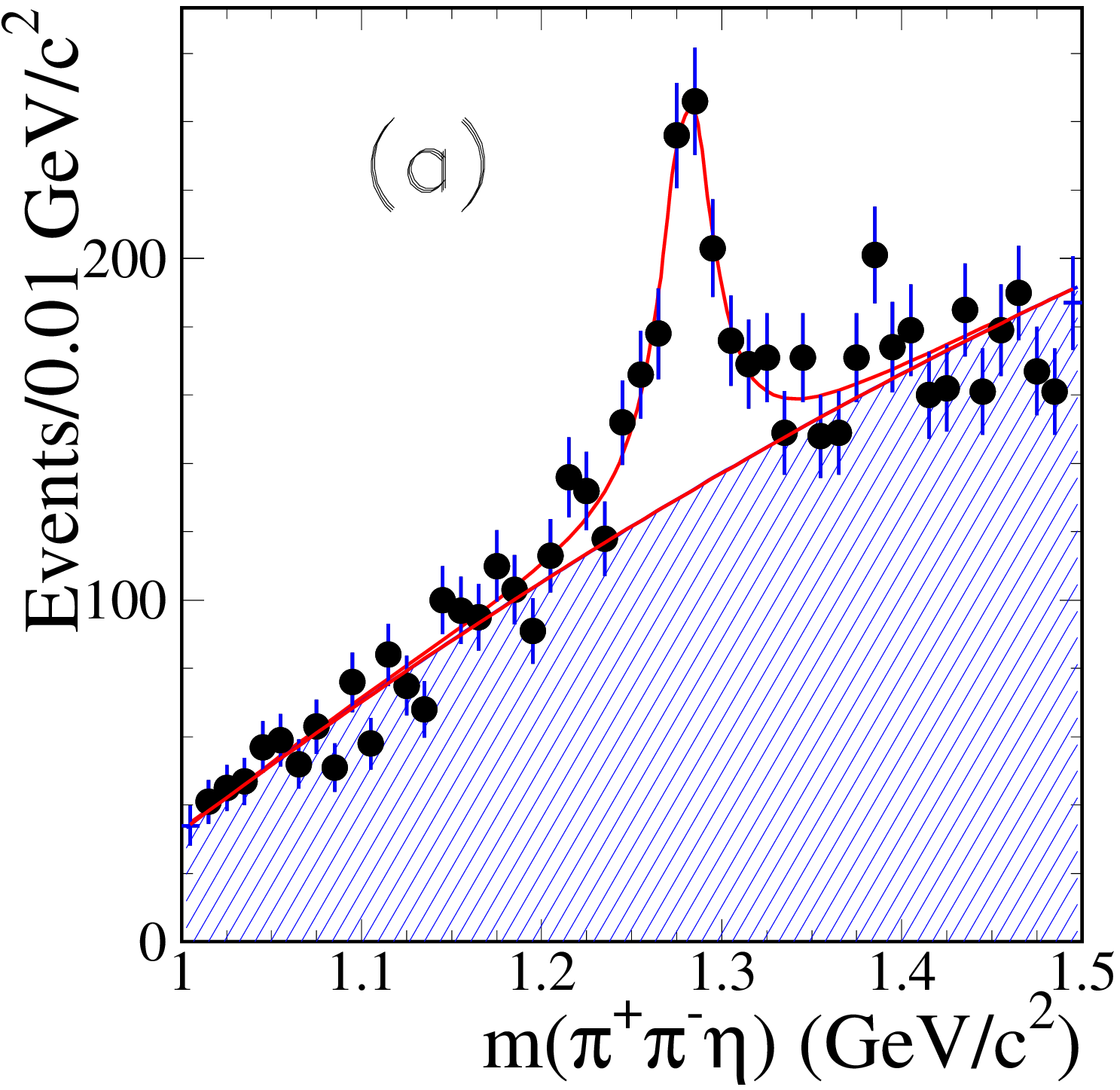}
\includegraphics[width=0.48\linewidth]{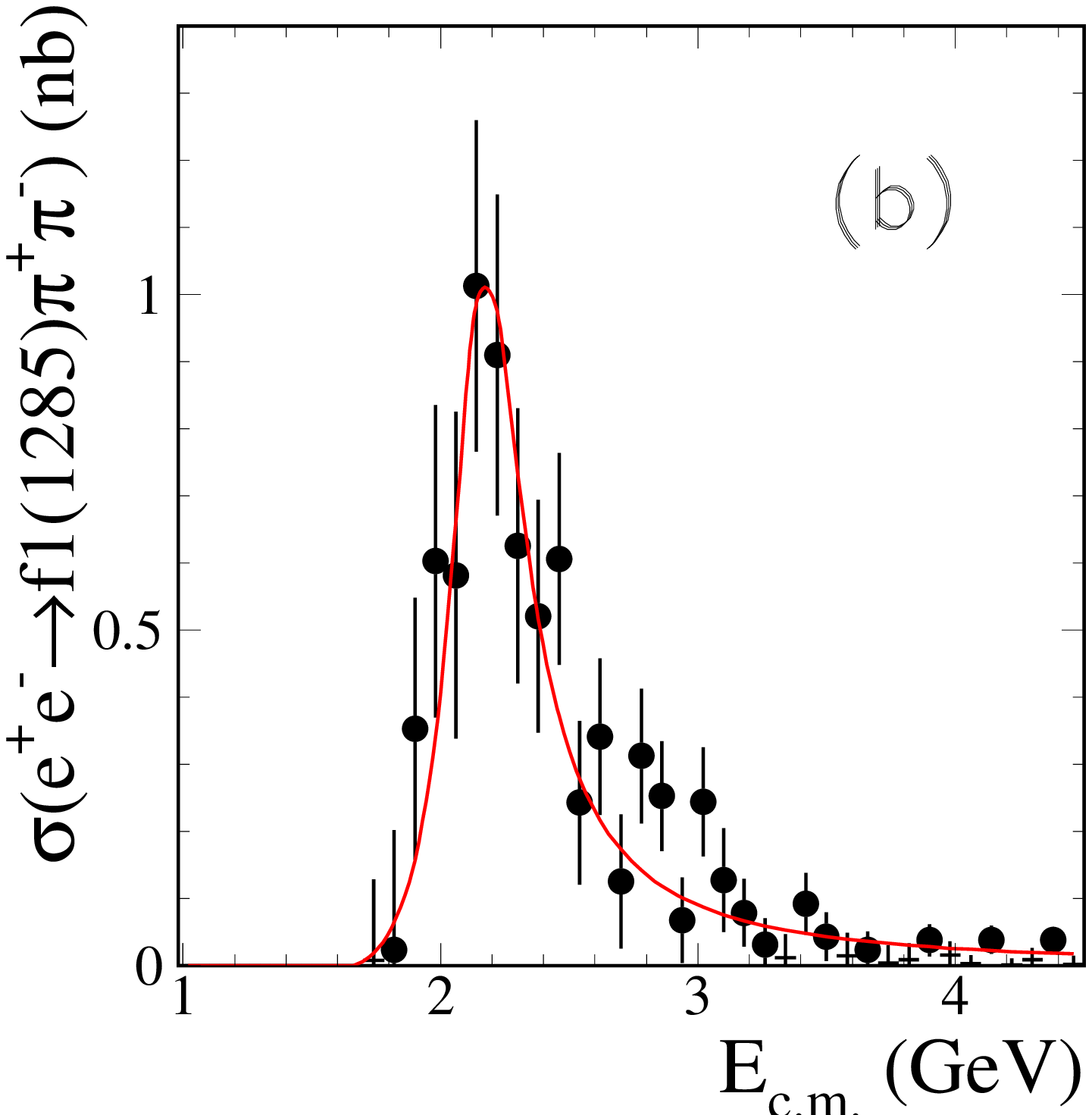}
\vspace{-0.2cm}
\caption{
  (a) Distribution of the $\pipi\eta$ invariant mass in the
  1--1.5~\gevcc region for the data (points).
  The line represents the result of the fit described in the text,
  with the shaded region representing the background component.
  (b) The $\epem\!\! \to\! f_1 (1285)\pipi$ cross section obtained via ISR.
  The line is the result of the Breit-Wigner fit described in the text.
  }
\label{eta1295}  
\end{center}
\end{figure}

Figure~\ref{eta1295}(a) shows an expanded view of the $\pipi\eta$
invariant mass distribution [Fig.~\ref{2pivs3pieta}(b)] in the region
around 1.3~\gevcc.
We fit this distribution with a Breit-Wigner signal function plus
a second order polynomial to describe the combinatorial background,
obtaining $649\pm85$ events in the peak.
The fitted mass and width, $1.281\pm0.002\pm0.001~\gevcc$ and
$0.035\pm0.006\pm0.004~\gev$, are compatible with the $f_1(1285)$
parameters~\cite{PDG} and are not in agreement with those listed for
the $\eta(1295)$. We conclude that contribution from the
$\eta(1295)\pipi$ is small and more data is needed for detailed
study. 
A similar fit to the $\eta'$ peak gives a mass shifted from the PDG
value by $-0.9\pm0.4$~\mev, from which we estimate a 1~\mev systematic
error on the $f_1 (1285)$ mass.
The systematic error on the width is estimated by varying the shape of
the combinatorial background.

\begin{table*}
\caption{Measurements of the $\ep\en\to f_1(1285)\pipi$ 
cross section (errors are statistical only).}
\label{eta1295pipi_tab}
\begin{ruledtabular}
\hspace{-1.8cm}
\begin{tabular}{ c c c c c c c c }
$E_{\rm c.m.}$ (GeV) & $\sigma$ (nb)  
& $E_{\rm c.m.}$ (GeV) & $\sigma$ (nb) 
& $E_{\rm c.m.}$ (GeV) & $\sigma$ (nb) 
& $E_{\rm c.m.}$ (GeV) & $\sigma$ (nb)  
\\
\hline

   1.66 &  0.00 $\pm$  0.00 &   2.14 &  0.99 $\pm$  0.24 &   2.62 &  0.32 $\pm$  0.12 &   3.10 &  0.11 $\pm$  0.08 \\
   1.74 &  0.01 $\pm$  0.11 &   2.22 &  0.89 $\pm$  0.24 &   2.70 &  0.13 $\pm$  0.10 &   3.18 &  0.08 $\pm$  0.05 \\
   1.82 &  0.02 $\pm$  0.18 &   2.30 &  0.64 $\pm$  0.21 &   2.78 &  0.31 $\pm$  0.10 &   3.26 &  0.03 $\pm$  0.04 \\
   1.90 &  0.35 $\pm$  0.20 &   2.38 &  0.54 $\pm$  0.17 &   2.86 &  0.25 $\pm$  0.08 &   3.34 &  0.02 $\pm$  0.03 \\
   1.98 &  0.61 $\pm$  0.23 &   2.46 &  0.61 $\pm$  0.16 &   2.94 &  0.07 $\pm$  0.06 &   3.42 &  0.09 $\pm$  0.05 \\
   2.06 &  0.59 $\pm$  0.24 &   2.54 &  0.25 $\pm$  0.12 &   3.02 &  0.24 $\pm$  0.08 &   3.50 &  0.05 $\pm$  0.04 \\

\end{tabular}
\end{ruledtabular}
\end{table*}

We extract the number of $f_1 (1285)\pipi$ events in 80~\mevcc bins of
the \fourpieta mass using similar fits with the $f_1 (1285)$ mass and
width fixed to the above values.
Due to the uncertainty in the background shape we assign an additional 
10\% systematic error on the number of signal events.
Dividing the fitted number of events by the efficiency, ISR luminosity
and $\eta\! \to\! \gamma\gamma$ and $f_1 (1285)\! \to\! \eta\pipi$
branching fractions yields the cross section shown in
Fig.~\ref{eta1295}(b) and listed in Table~\ref{eta1295pipi_tab}.
There is no evidence of the $J/\psi$ decay into this mode, and
the cross section again demonstrates resonance-like behavior at around
2.1~\gev. 
Fitting with a single Breit-Wigner function, 
Eq.~\ref{bwsum} with $m_0 = 1.8~\gev$,
we obtain:
\begin{eqnarray*}
  \sigma_0 & = & 1.00 \pm 0.18 \pm 0.15~\nb, \\
     m_x   & = & 2.15 \pm 0.04 \pm 0.05~\gevcc,    \\
  \Gamma_x & = & 0.35 \pm 0.04 \pm 0.05~\gev.
\end{eqnarray*}
The mass and width are consistent with those obtained above for the
$\eta'(958)\pipi$ channel, 
and with those listed in the PDG tables~\cite{PDG} for the $\rho(2150)$,
but the cross section is substantially larger. 
\begin{figure}[tbh]
\includegraphics[width=0.9\linewidth]{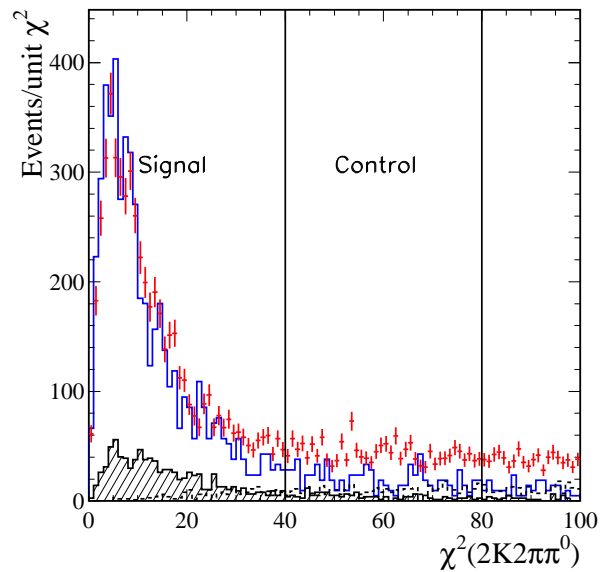}
\vspace{-0.4cm}
\caption{
  Distribution of \chisq from the 5C fit for \KKpppn candidates
  in the data (points).
  The open histogram is the distribution for simulated signal events, 
  normalized as described in the text.
  The hatched and dashed histograms are the backgrounds from non-ISR
  events and ISR $\KpKm\pipi$ events, respectively, 
  estimated as described in the text
  }
\label{2k3pi_chi2_all}
\end{figure}
\section{\boldmath The \KKpppn  Final State}
\subsection{Final Selection and Backgrounds}
To suppress ISR \fourpipn background, 
we fit each event under that hypothesis and require 
$\chifourpipn\! >\! 30$.
The \chiKKpppn distribution for the remaining events is shown as
points in Fig.~\ref{2k3pi_chi2_all} 
and the distribution for simulated \KKpppn events (open histogram)
is normalized to the data in the region $\chiKKpppn\! <\! 20$.
The hatched histogram represents the non-ISR background,
which is dominated by $\Kp\Km\pipi\piz\piz$ events and checked
against data as in Sec.~\ref{sec:selection1}.
Since the statistics are low and the scale factor is consistent, 
we use the same value as for the \fourpipn final state.
The largest remaining background is from ISR $\Kp\Km\pipi$ events.
This contribution, 
estimated from the simulation using the measured cross
section~\cite{isr2K2pi},
is shown as the dashed histogram in Fig.~\ref{2k3pi_chi2_all}.
All other backgrounds are either negligible or distributed uniformly 
in \chiKKpppn.
We define a signal region, $\chiKKpppn\! <\! 40$, 
containing 5565 events
and a control region, $40\! <\! \chiKKpppn\! <\! 80$,
containing 1758 events.

\begin{figure}[tbh]
\includegraphics[width=0.9\linewidth]{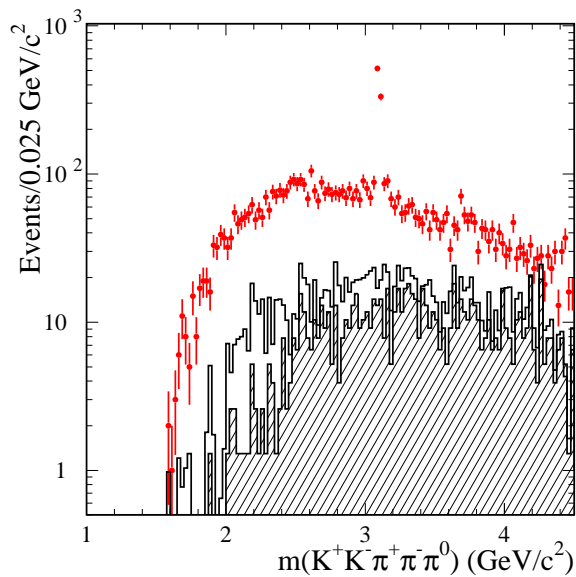}
\vspace{-0.4cm}
\caption{
  Invariant mass distribution for selected \KKpppn candidates in the
  data (points).  The hatched and open histograms represent, cumulatively, 
  the non-ISR background and the ISR background from the control
  region of Fig.~\ref{2k3pi_chi2_all}.
  }
\label{2k3pi_babar}
\end{figure}

Figure~\ref{2k3pi_babar} shows the \KKpppn invariant mass distribution 
from threshold up to 4.5~\gevcc for events in the signal region.
The hatched histogram represents the non-ISR background, and the open
histogram represents the sum of all backgrounds, where the ISR-type
background is estimated from the control region.
The total background is about 15\% at low mass,
but accounts for a large fraction of the selected events above 
about 3.5~\gevcc.
We subtract the sum of backgrounds from the number of selected events 
in each mass bin to obtain a number of signal events.
Considering uncertainties in the cross sections for the background processes,
the normalization of events in the control region and the simulation
statistics, 
we estimate a systematic uncertainty on the signal yield
of less than 5\% in the 1.6--3.0~\gevcc region, 
but increasing to 10\% in the region above 3~\gevcc.

\subsection{Selection Efficiency}

\begin{figure}[t]
\includegraphics[width=0.9\linewidth]{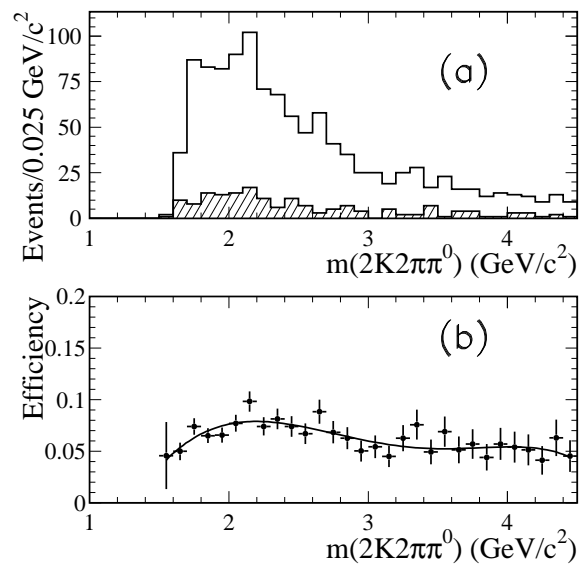}
\vspace{-0.4cm}
\caption{
  (a) Invariant mass distribution for simulated \KKpppn events in the
  signal (open) and control (hatched) regions of Fig.~\ref{2k3pi_chi2_all}.
  (b) Net reconstruction and selection efficiency as a function of mass
  obtained from this simulation;
  the curve represents a third order polynomial fit.
  }
\label{mc_acc3}
\end{figure}

The detection efficiency is determined in the same manner as in 
Sec.~\ref{sec:eff1}. 
Figure~\ref{mc_acc3}(a) shows the simulated \KKpppn invariant mass
distributions in the signal and control regions from the phase space
simulation.
We divide the number of reconstructed events in each
mass interval by the number generated in that interval to
obtain the efficiency shown as the points in Fig.~\ref{mc_acc3}(b);
the curve represents a third order polynomial fit to the points, 
which we use to calculate the cross section.
Simulations assuming dominance of the $\phi\pipi\piz$ and
$\eta\Kp\Km$ channels give consistent results, 
and we apply a 5\% systematic uncertainty for possible model 
dependence, as in Sec.~\ref{sec:eff1}.

We correct for track- and $\pi^0$-finding efficiencies, 
and the shape of the \chiKKpppn distribution as in Sec.~\ref{sec:eff1}.
We measure the kaon identification efficiency using 
$\epem \!\!\to\! \phi(1020)\gamma \!\to\! \Kp\Km\gamma$ events,
as described in Ref.~\cite{isr2K2pi},
and apply a correction of $+(2.0\pm2.0)$\% to the efficiency.
The total efficiency correction is $+$8\% and we estimate a
systematic error of 10\% for masses below 3.0~\gevcc, increasing
to 30\% at 4.5~\gevcc.

\begin{figure}[tbh]
\includegraphics[width=0.9\linewidth]{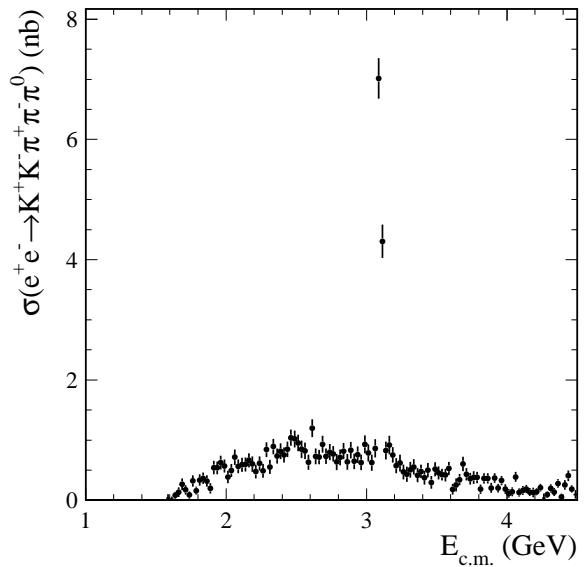}
\vspace{-0.5cm}
\caption{
  The $\epem \!\!\to\! \KKpppn$ cross section as a function of \epem
  c.m.\ energy measured with ISR data.
  The errors are statistical only.
  }
\label{2k3pi_ee_babar}
\end{figure} 

\subsection{\boldmath Cross Section for $\epem \to \KKpppn$}
\label{sec:2k3pixs}

We calculate the cross section for the $\epem\!\! \to\! \KKpppn$ process 
by dividing the number of events in each \KKpppn mass bin by the
corrected efficiency and differential luminosity.
We show the first measurement of this cross section
in Fig.~\ref{2k3pi_ee_babar}, with statistical errors only,
and list the results in Table~\ref{2k3pi_tab}.
Again, the energy resolution is much smaller than the bin width and
we apply no correction.
The cross section rises to a peak value near 1~\nb at 2.5~\gev,
followed by a slow decrease with increasing energy.
The only statistically significant structure is the $J/\psi$ peak.

\begin{table*}
\caption{Measurements of the $\ep\en\to K^+ K^-\pipi\pi^0$ 
cross section (errors are statistical only).}
\label{2k3pi_tab}
\begin{ruledtabular}
\hspace{-1.8cm}
\begin{tabular}{ c c c c c c c c }
$E_{\rm c.m.}$ (GeV) & $\sigma$ (nb)  
& $E_{\rm c.m.}$ (GeV) & $\sigma$ (nb) 
& $E_{\rm c.m.}$ (GeV) & $\sigma$ (nb) 
& $E_{\rm c.m.}$ (GeV) & $\sigma$ (nb)  
\\
\hline

 1.6125 &  0.02 $\pm$  0.04 & 2.3375 &  0.89 $\pm$  0.13 & 3.0625 &  0.86 $\pm$  0.15 & 3.7875 &  0.38 $\pm$  0.10 \\
 1.6375 &  0.08 $\pm$  0.06 & 2.3625 &  0.73 $\pm$  0.13 & 3.0875 &  7.01 $\pm$  0.34 & 3.8125 &  0.18 $\pm$  0.08 \\
 1.6625 &  0.13 $\pm$  0.08 & 2.3875 &  0.82 $\pm$  0.13 & 3.1125 &  4.31 $\pm$  0.28 & 3.8375 &  0.36 $\pm$  0.09 \\
 1.6875 &  0.27 $\pm$  0.10 & 2.4125 &  0.75 $\pm$  0.13 & 3.1375 &  0.83 $\pm$  0.15 & 3.8625 &  0.36 $\pm$  0.08 \\
 1.7125 &  0.17 $\pm$  0.08 & 2.4375 &  0.85 $\pm$  0.13 & 3.1625 &  0.92 $\pm$  0.15 & 3.8875 &  0.20 $\pm$  0.08 \\
 1.7375 &  0.09 $\pm$  0.06 & 2.4625 &  1.04 $\pm$  0.13 & 3.1875 &  0.75 $\pm$  0.13 & 3.9125 &  0.37 $\pm$  0.08 \\
 1.7625 &  0.32 $\pm$  0.10 & 2.4875 &  1.02 $\pm$  0.14 & 3.2125 &  0.57 $\pm$  0.13 & 3.9375 &  0.20 $\pm$  0.07 \\
 1.7875 &  0.16 $\pm$  0.07 & 2.5125 &  0.95 $\pm$  0.14 & 3.2375 &  0.62 $\pm$  0.14 & 3.9625 &  0.33 $\pm$  0.08 \\
 1.8125 &  0.34 $\pm$  0.09 & 2.5375 &  0.85 $\pm$  0.15 & 3.2625 &  0.47 $\pm$  0.12 & 3.9875 &  0.18 $\pm$  0.08 \\
 1.8375 &  0.36 $\pm$  0.09 & 2.5625 &  0.82 $\pm$  0.14 & 3.2875 &  0.43 $\pm$  0.13 & 4.0125 &  0.11 $\pm$  0.07 \\
 1.8625 &  0.32 $\pm$  0.09 & 2.5875 &  0.63 $\pm$  0.12 & 3.3125 &  0.50 $\pm$  0.13 & 4.0375 &  0.14 $\pm$  0.07 \\
 1.8875 &  0.20 $\pm$  0.09 & 2.6125 &  1.20 $\pm$  0.15 & 3.3375 &  0.55 $\pm$  0.13 & 4.0625 &  0.39 $\pm$  0.08 \\
 1.9125 &  0.54 $\pm$  0.11 & 2.6375 &  0.73 $\pm$  0.13 & 3.3625 &  0.42 $\pm$  0.12 & 4.0875 &  0.13 $\pm$  0.07 \\
 1.9375 &  0.54 $\pm$  0.11 & 2.6625 &  0.71 $\pm$  0.12 & 3.3875 &  0.48 $\pm$  0.11 & 4.1125 &  0.16 $\pm$  0.07 \\
 1.9625 &  0.63 $\pm$  0.11 & 2.6875 &  0.93 $\pm$  0.14 & 3.4125 &  0.37 $\pm$  0.11 & 4.1375 &  0.19 $\pm$  0.07 \\
 1.9875 &  0.57 $\pm$  0.11 & 2.7125 &  0.73 $\pm$  0.13 & 3.4375 &  0.50 $\pm$  0.12 & 4.1625 &  0.13 $\pm$  0.07 \\
 2.0125 &  0.39 $\pm$  0.10 & 2.7375 &  0.80 $\pm$  0.13 & 3.4625 &  0.29 $\pm$  0.11 & 4.1875 &  0.13 $\pm$  0.08 \\
 2.0375 &  0.49 $\pm$  0.11 & 2.7625 &  0.77 $\pm$  0.13 & 3.4875 &  0.52 $\pm$  0.12 & 4.2125 &  0.14 $\pm$  0.06 \\
 2.0625 &  0.71 $\pm$  0.13 & 2.7875 &  0.64 $\pm$  0.14 & 3.5125 &  0.45 $\pm$  0.11 & 4.2375 &  0.21 $\pm$  0.06 \\
 2.0875 &  0.56 $\pm$  0.11 & 2.8125 &  0.71 $\pm$  0.13 & 3.5375 &  0.42 $\pm$  0.10 & 4.2625 &  0.04 $\pm$  0.08 \\
 2.1125 &  0.59 $\pm$  0.12 & 2.8375 &  0.82 $\pm$  0.13 & 3.5625 &  0.42 $\pm$  0.10 & 4.2875 &  0.10 $\pm$  0.06 \\
 2.1375 &  0.59 $\pm$  0.12 & 2.8625 &  0.64 $\pm$  0.13 & 3.5875 &  0.53 $\pm$  0.11 & 4.3125 &  0.20 $\pm$  0.07 \\
 2.1625 &  0.66 $\pm$  0.12 & 2.8875 &  0.83 $\pm$  0.14 & 3.6125 &  0.18 $\pm$  0.09 & 4.3375 &  0.14 $\pm$  0.07 \\
 2.1875 &  0.60 $\pm$  0.13 & 2.9125 &  0.65 $\pm$  0.13 & 3.6375 &  0.25 $\pm$  0.11 & 4.3625 &  0.28 $\pm$  0.07 \\
 2.2125 &  0.48 $\pm$  0.11 & 2.9375 &  0.76 $\pm$  0.14 & 3.6625 &  0.34 $\pm$  0.10 & 4.3875 &  0.05 $\pm$  0.06 \\
 2.2375 &  0.61 $\pm$  0.12 & 2.9625 &  0.63 $\pm$  0.13 & 3.6875 &  0.60 $\pm$  0.12 & 4.4125 &  0.25 $\pm$  0.08 \\
 2.2625 &  0.49 $\pm$  0.11 & 2.9875 &  0.93 $\pm$  0.15 & 3.7125 &  0.43 $\pm$  0.10 & 4.4375 &  0.41 $\pm$  0.09 \\
 2.2875 &  0.84 $\pm$  0.13 & 3.0125 &  0.78 $\pm$  0.14 & 3.7375 &  0.36 $\pm$  0.10 & 4.4625 &  0.19 $\pm$  0.06 \\
 2.3125 &  0.55 $\pm$  0.12 & 3.0375 &  0.62 $\pm$  0.13 & 3.7625 &  0.38 $\pm$  0.10 & 4.4875 &  0.10 $\pm$  0.07 \\

\end{tabular}
\end{ruledtabular}
\end{table*}
 
\begin{table*}
\caption{Measurements of the $\ep\en\to\phi(1020)\eta$ 
cross section (errors are statistical only).}
\label{phieta_tab}
\begin{ruledtabular}
\hspace{-1.8cm}
\begin{tabular}{ c c c c c c c c }
$E_{\rm c.m.}$ (GeV) & $\sigma$ (nb)  
& $E_{\rm c.m.}$ (GeV) & $\sigma$ (nb) 
& $E_{\rm c.m.}$ (GeV) & $\sigma$ (nb) 
& $E_{\rm c.m.}$ (GeV) & $\sigma$ (nb)  
\\
\hline

   1.56 &  0.00 $\pm$  0.41 &   1.84 &  0.65 $\pm$  0.29 &   2.12 &  0.63 $\pm$  0.26 &   2.40 &  0.10 $\pm$  0.10 \\
   1.60 &  0.57 $\pm$  0.33 &   1.88 &  0.63 $\pm$  0.28 &   2.16 &  0.31 $\pm$  0.18 &   2.44 &  0.19 $\pm$  0.13 \\
   1.64 &  0.70 $\pm$  0.35 &   1.92 &  0.85 $\pm$  0.32 &   2.20 &  0.30 $\pm$  0.18 &   2.48 &  0.00 $\pm$  0.19 \\
   1.68 &  2.28 $\pm$  0.61 &   1.96 &  0.58 $\pm$  0.26 &   2.24 &  0.00 $\pm$  0.20 &   2.52 &  0.28 $\pm$  0.16 \\
   1.72 &  1.53 $\pm$  0.48 &   2.00 &  0.57 $\pm$  0.25 &   2.28 &  0.20 $\pm$  0.14 &   2.56 &  0.00 $\pm$  0.19 \\
   1.76 &  2.02 $\pm$  0.54 &   2.04 &  0.22 $\pm$  0.16 &   2.32 &  0.10 $\pm$  0.10 &   2.60 &  0.19 $\pm$  0.13 \\
   1.80 &  1.23 $\pm$  0.41 &   2.08 &  0.54 $\pm$  0.24 &   2.36 &  0.29 $\pm$  0.17 &   2.64 &  0.09 $\pm$  0.09 \\

\end{tabular}
\end{ruledtabular}
\end{table*}
 
\begin{table*}
\caption{Measurements of the $\ep\en\to\omega(782) K^+ K^-$ 
cross section (errors are statistical only).}
\label{omega2k_tab}
\begin{ruledtabular}
\hspace{-1.8cm}
\begin{tabular}{ c c c c c c c c }
$E_{\rm c.m.}$ (GeV) & $\sigma$ (nb)  
& $E_{\rm c.m.}$ (GeV) & $\sigma$ (nb) 
& $E_{\rm c.m.}$ (GeV) & $\sigma$ (nb) 
& $E_{\rm c.m.}$ (GeV) & $\sigma$ (nb)  
\\
\hline

   1.80 &  0.09 $\pm$  0.04 &   2.48 &  0.31 $\pm$  0.07 &   3.16 &  0.07 $\pm$  0.04 &   3.84 &  0.04 $\pm$  0.02 \\
   1.84 &  0.33 $\pm$  0.08 &   2.52 &  0.27 $\pm$  0.06 &   3.20 &  0.12 $\pm$  0.04 &   3.88 &  0.01 $\pm$  0.01 \\
   1.88 &  0.25 $\pm$  0.06 &   2.56 &  0.18 $\pm$  0.05 &   3.24 &  0.17 $\pm$  0.05 &   3.92 &  0.01 $\pm$  0.01 \\
   1.92 &  0.52 $\pm$  0.09 &   2.60 &  0.20 $\pm$  0.05 &   3.28 &  0.06 $\pm$  0.04 &   3.96 &  0.01 $\pm$  0.01 \\
   1.96 &  0.60 $\pm$  0.10 &   2.64 &  0.20 $\pm$  0.05 &   3.32 &  0.07 $\pm$  0.04 &   4.00 &  0.02 $\pm$  0.01 \\
   2.00 &  0.45 $\pm$  0.09 &   2.68 &  0.18 $\pm$  0.06 &   3.36 &  0.06 $\pm$  0.03 &   4.04 &  0.02 $\pm$  0.01 \\
   2.04 &  0.43 $\pm$  0.09 &   2.72 &  0.10 $\pm$  0.04 &   3.40 &  0.01 $\pm$  0.02 &   4.08 &  0.00 $\pm$  0.01 \\
   2.08 &  0.61 $\pm$  0.10 &   2.76 &  0.16 $\pm$  0.05 &   3.44 &  0.07 $\pm$  0.03 &   4.12 &  0.00 $\pm$  0.01 \\
   2.12 &  0.37 $\pm$  0.08 &   2.80 &  0.18 $\pm$  0.05 &   3.48 &  0.04 $\pm$  0.02 &   4.16 &  0.00 $\pm$  0.01 \\
   2.16 &  0.28 $\pm$  0.07 &   2.84 &  0.12 $\pm$  0.05 &   3.52 &  0.07 $\pm$  0.03 &   4.20 &  0.02 $\pm$  0.01 \\
   2.20 &  0.39 $\pm$  0.08 &   2.88 &  0.12 $\pm$  0.05 &   3.56 &  0.08 $\pm$  0.03 &   4.24 &  0.00 $\pm$  0.01 \\
   2.24 &  0.36 $\pm$  0.08 &   2.92 &  0.17 $\pm$  0.05 &   3.60 &  0.07 $\pm$  0.03 &   4.28 &  0.00 $\pm$  0.01 \\
   2.28 &  0.17 $\pm$  0.06 &   2.96 &  0.17 $\pm$  0.05 &   3.64 &  0.05 $\pm$  0.03 &   4.32 &  0.04 $\pm$  0.02 \\
   2.32 &  0.22 $\pm$  0.06 &   3.00 &  0.16 $\pm$  0.05 &   3.68 &  0.14 $\pm$  0.04 &   4.36 &  0.02 $\pm$  0.01 \\
   2.36 &  0.34 $\pm$  0.07 &   3.04 &  0.05 $\pm$  0.04 &   3.72 &  0.09 $\pm$  0.03 &   4.40 &  0.02 $\pm$  0.01 \\
   2.40 &  0.23 $\pm$  0.06 &   3.08 &  0.51 $\pm$  0.09 &   3.76 &  0.02 $\pm$  0.02 &   4.44 &  0.00 $\pm$  0.01 \\
   2.44 &  0.19 $\pm$  0.06 &   3.12 &  0.24 $\pm$  0.06 &   3.80 &  0.00 $\pm$  0.02 &   4.48 &  0.00 $\pm$  0.01 \\

\end{tabular}
\end{ruledtabular}
\end{table*}

\begin{figure}[tbh]
\begin{center}
\includegraphics[width=0.45\linewidth]{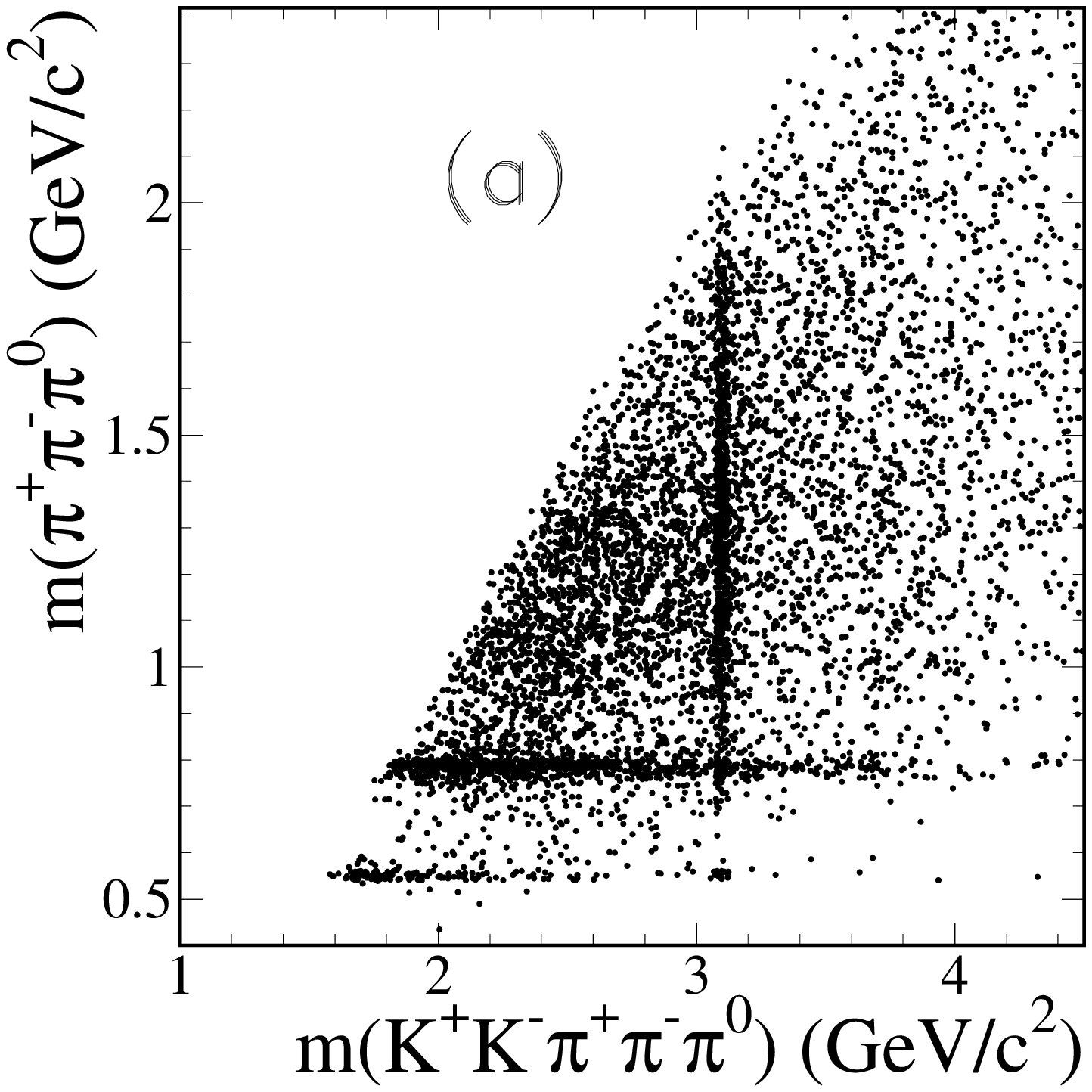}
\includegraphics[width=0.45\linewidth]{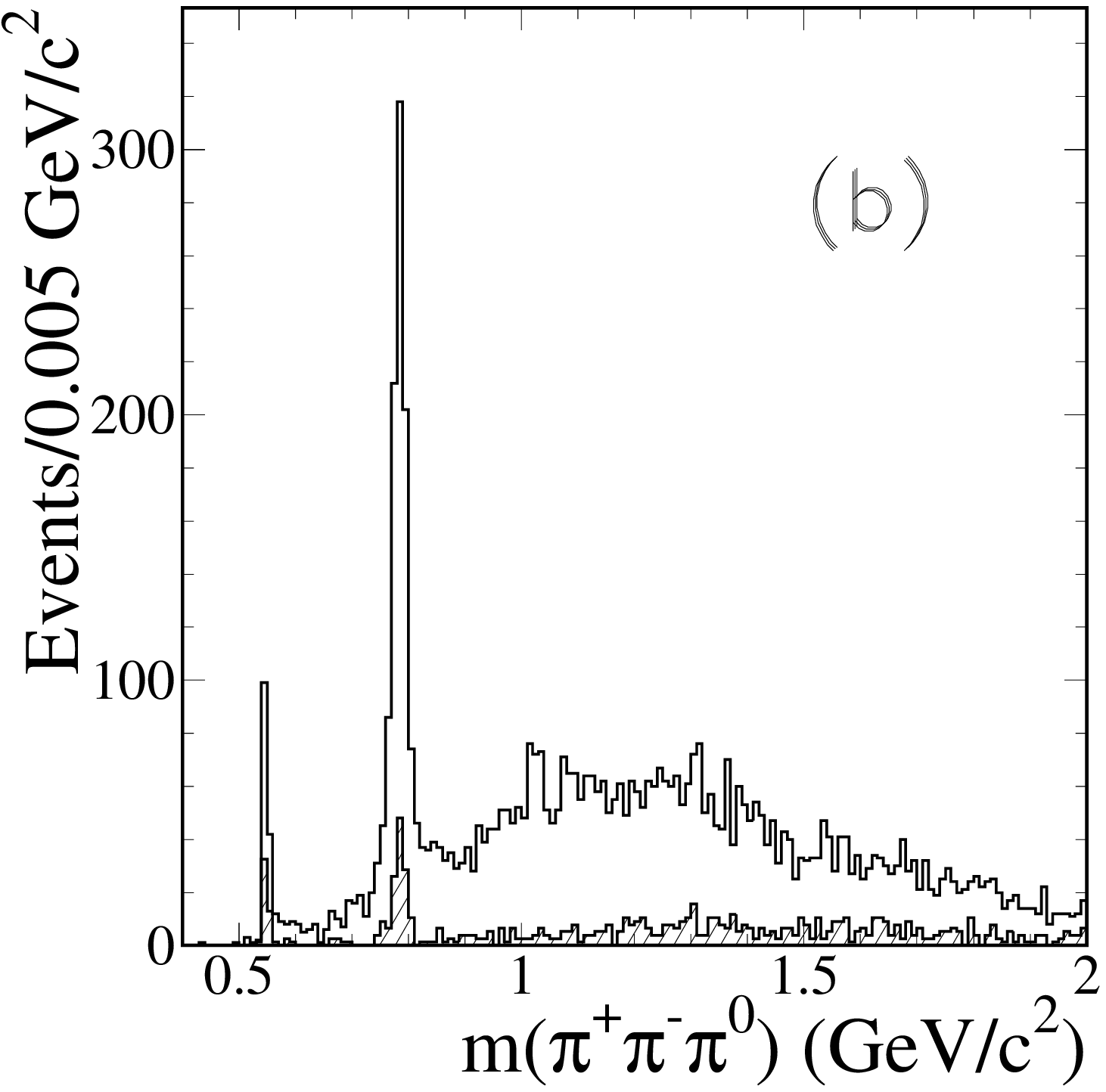}
\vspace{-0.2cm}
\caption{
  (a) The $\pipi\piz$ mass versus two-kaon-three-pion mass, and
  (b) the $\pipi\piz$ mass projection for selected \KKpppn candidates. 
  The hatched histogram represents the estimated non-ISR background.
  }
\label{3pivs2k3pi}  
\end{center}
\end{figure}

\begin{figure}[tbh]
\begin{center}
\includegraphics[width=0.45\linewidth]{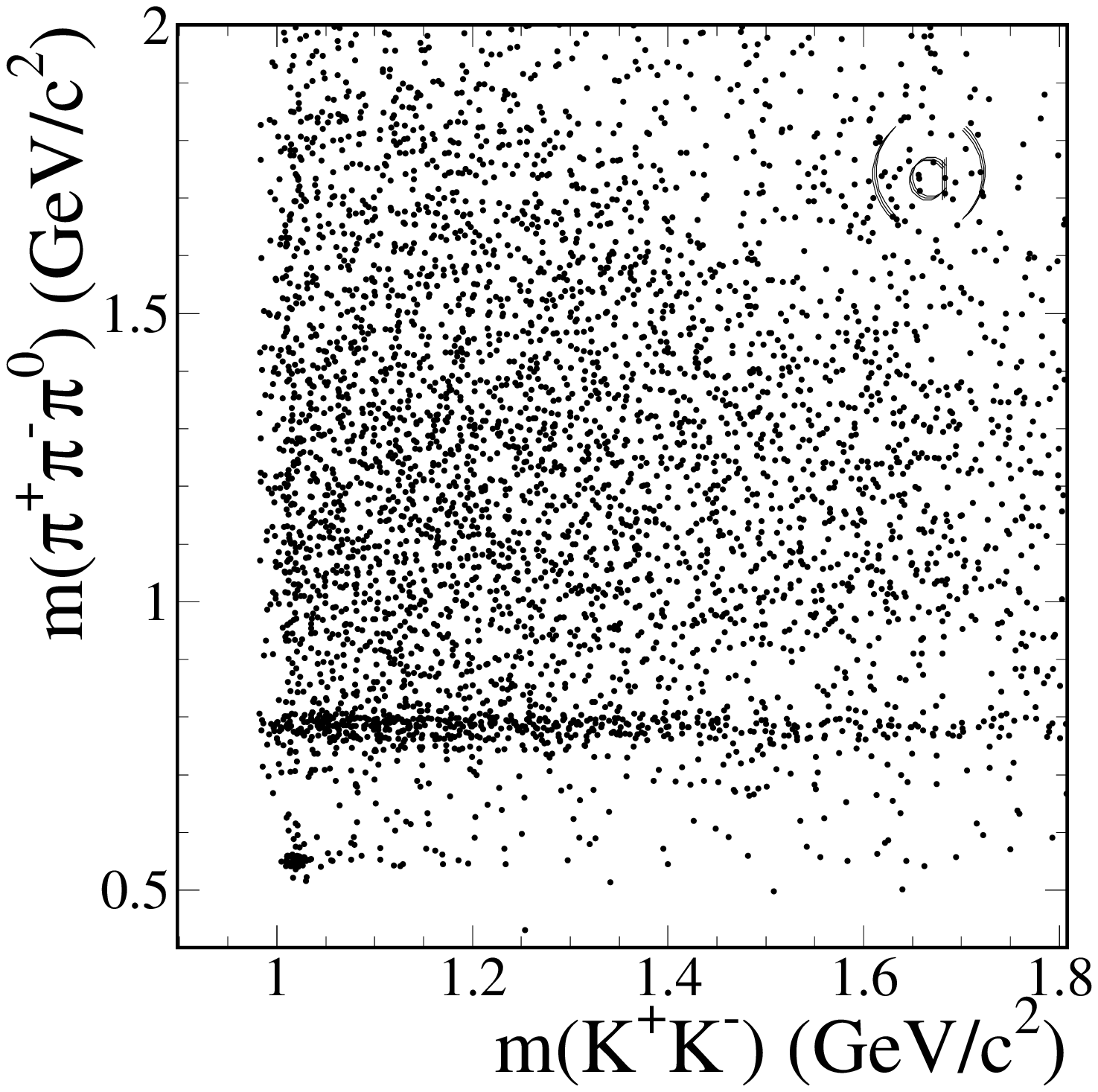}
\includegraphics[width=0.45\linewidth]{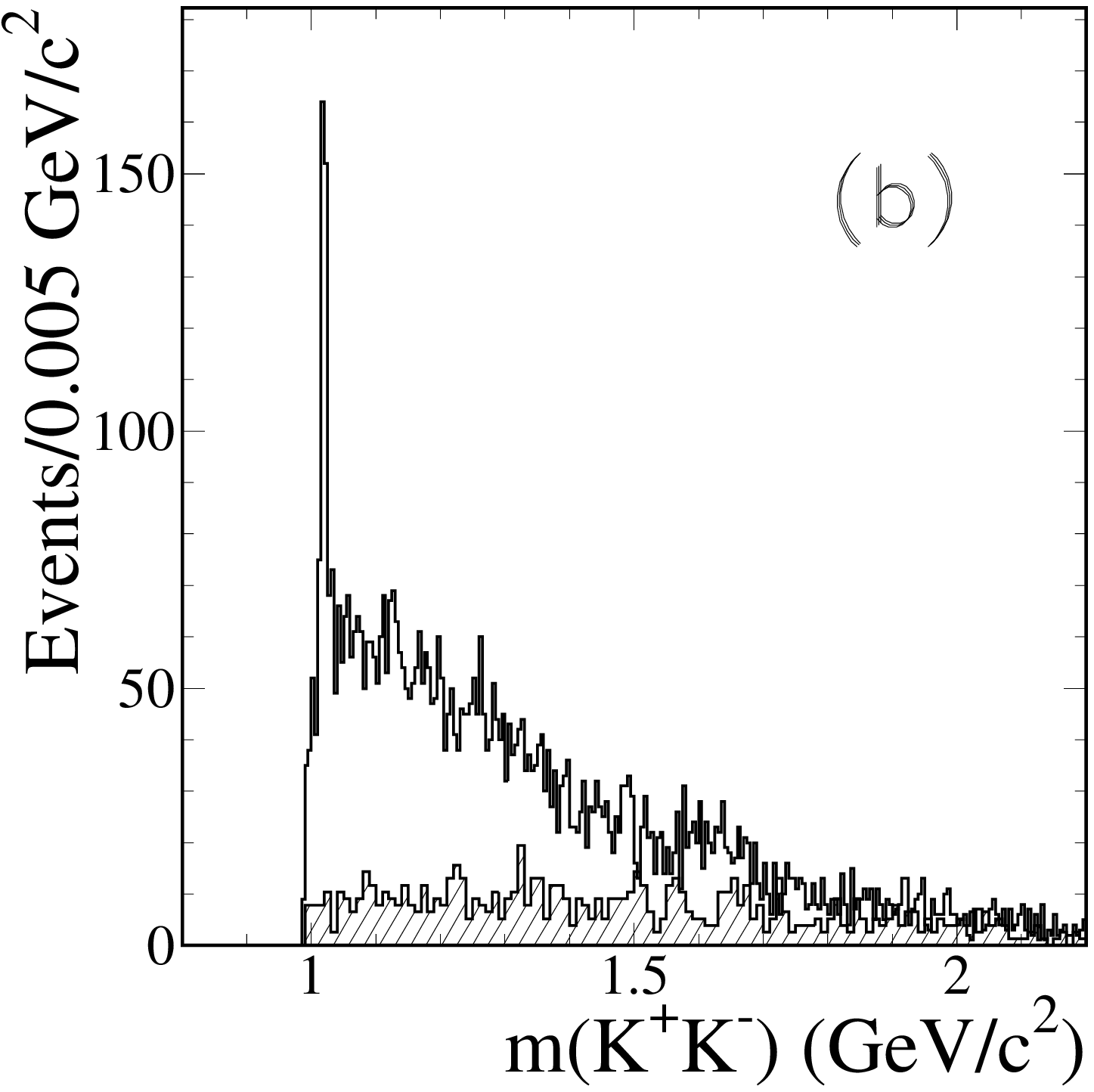}
\vspace{-0.2cm}
\caption{
  (a) The $\pipi\piz$ mass versus the $\Kp\Km$ mass and
  (b) the $\Kp\Km$ mass projection for selected \KKpppn candidates.
  The hatched histogram represents the estimated non-ISR background.
  }
\label{3pivs2k}  
\end{center}
\end{figure}

\begin{figure}[tbh]
\begin{center}
\includegraphics[width=0.48\linewidth]{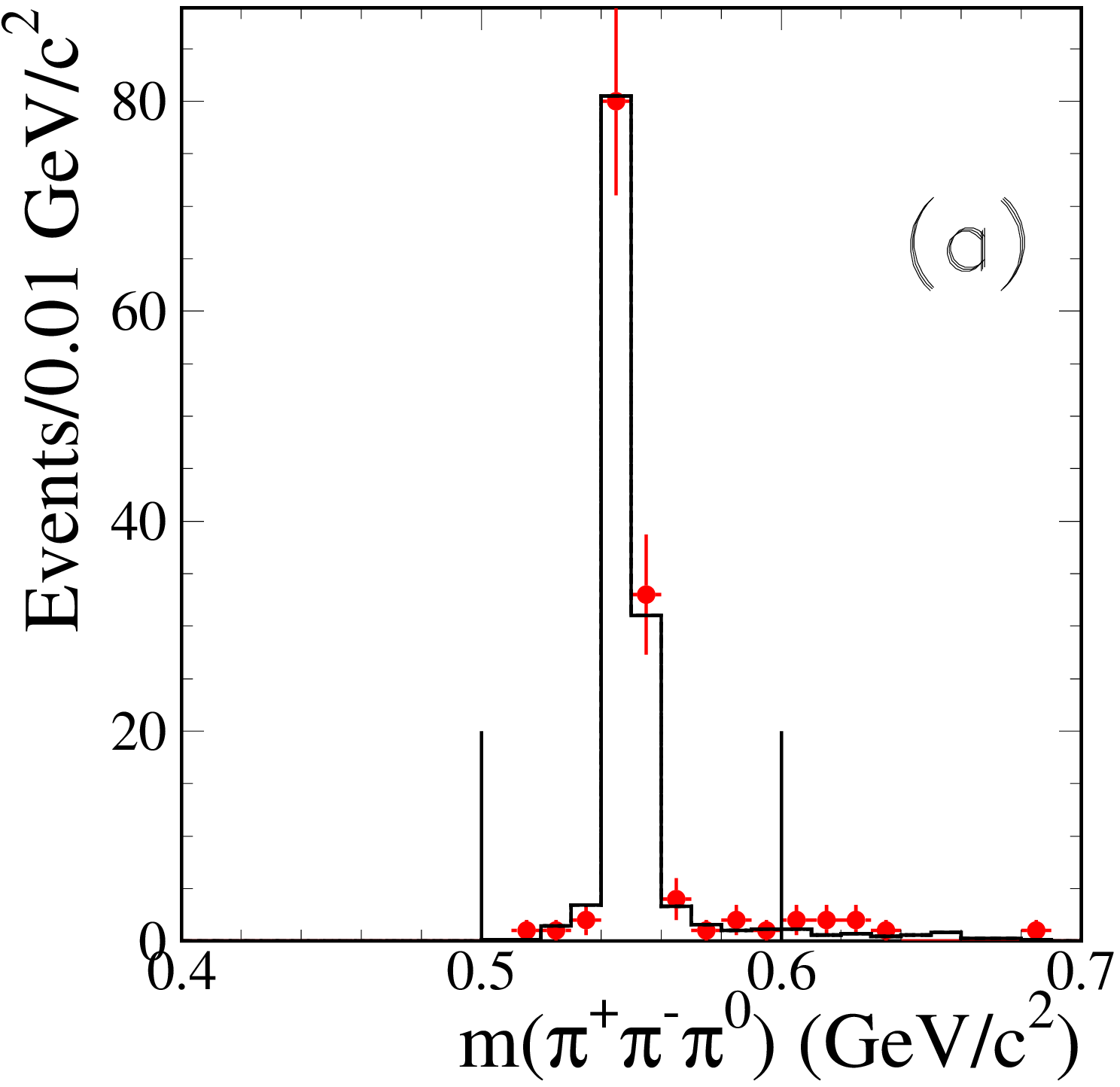}
\includegraphics[width=0.48\linewidth]{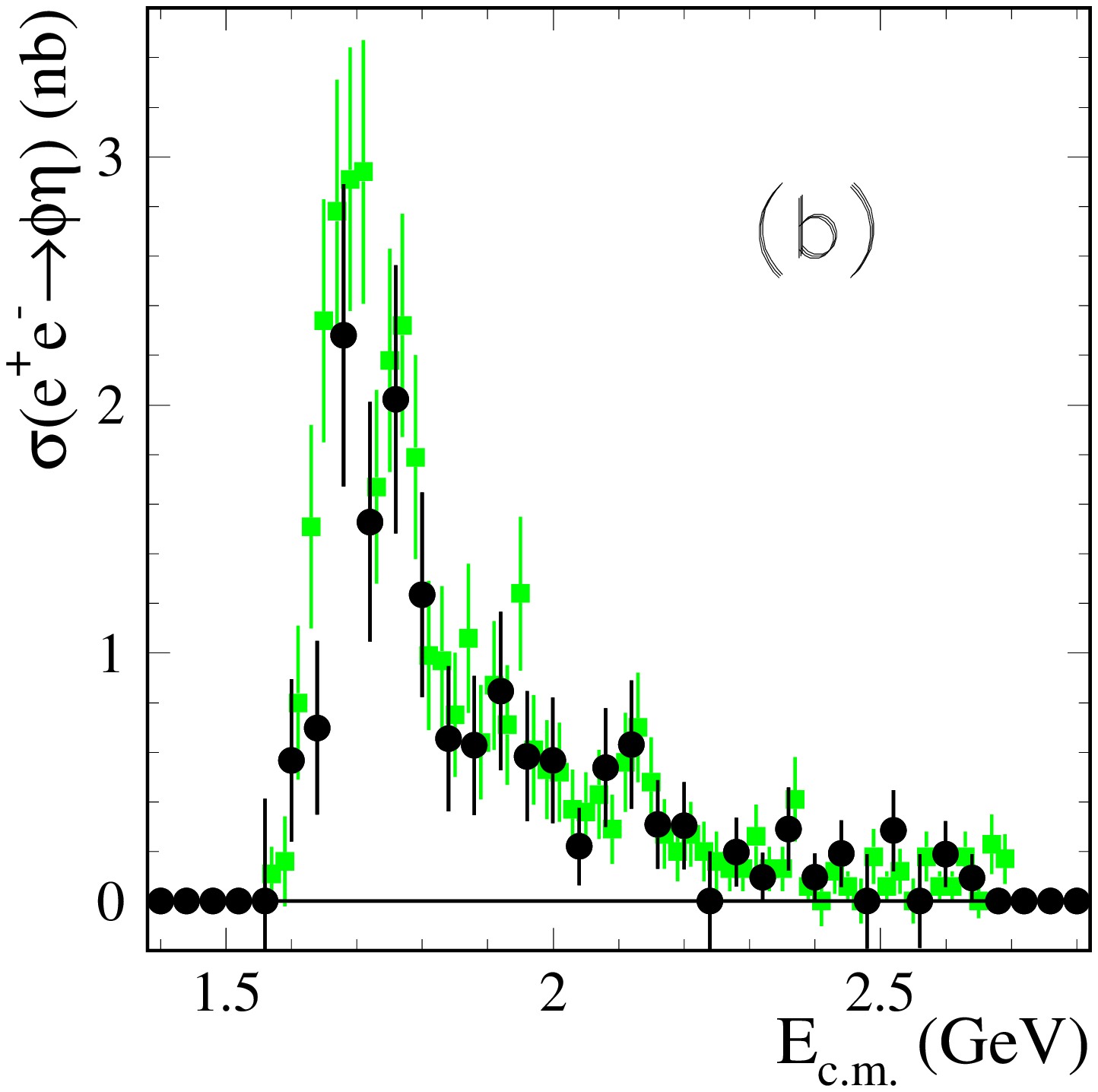}
\vspace{-0.2cm}
\caption{
  (a) The $\pipi\pi^0$ mass distribution in the region below 700~\mevcc
  for events with a $\Kp\Km$ mass within 15~\mevcc of the $\phi$ mass
  (points).
  The histogram is the distribution for the ISR $\phi\eta$ simulation.
  (b) The $\epem\!\! \to\! \phi\eta$ cross section measured here in the
  \KKpppn final state (large dots), compared with the previous \babar\
  measurement in the ISR $\Kp\Km\gamma\gamma$ final state~\cite{isrphieta}.
  }
\label{2ketamass}  
\end{center}
\end{figure}

\subsection{\boldmath Substructure in the \KKpppn Final State}
\label{sec:2k3piplots}

Figure~\ref{3pivs2k3pi}(a) shows a scatter plot of the $\pipi\piz$ mass 
versus the \KKpppn mass in the selected \KKpppn events.
There are horizontal bands corresponding to the
$\eta$ signal and $\omega(782)$ resonance.
The $\pipi\piz$ mass projection, Fig.~\ref{3pivs2k3pi}(b),
shows $\eta$ and $\omega$ peaks, as well as a small signal for the
$\phi(1020)$.
The non-ISR background contribution is shown as the hatched histogram and also
contains $\eta$ and $\omega$ signals.
Figure~\ref{3pivs2k}(a) shows a scatter plot of the $\pipi\pi^0$ mass 
versus the $\Kp\Km$ mass in the event. 
A vertical band corresponding to the $\phi(1020)$ is visible, 
and almost all $\eta$ are produced through the $\phi\eta$ channel,
whereas the $\omega(782)$ band is spread out across the plot.
The $\Kp\Km$ mass projection in Fig.~\ref{3pivs2k}(b) shows a
$\phi(1020)$ signal,
but the non-ISR background distribution has no structure.

\subsection{\bf\boldmath The $\phi\eta$ Intermediate State}
\label{sec:phieta}

Requiring the $\Kp\Km$ mass to be within $\pm$15~\mevcc of the nominal
$\phi(1020)$ mass and plotting the mass of the recoiling $\pipi\piz$
system below 700~\mevcc,
we obtain the distribution shown in Fig.~\ref{2ketamass}(a).
The distribution from the ISR $\phi\eta$ simulation is also shown,
and this channel can account for all entries below 700~\mevcc.
Counting events with a three-pion mass in the 0.5--0.6~\gevcc
region in bins of the \KKpppn mass,
and dividing by the corrected efficiency [Fig.~\ref{mc_acc3}(b)],
differential luminosity and $\phi\! \to\! \Kp\Km$ and 
$\eta\! \to\! \pipi\piz$ branching fractions,
we obtain the
$\epem\!\! \to\! \phi\eta$ cross section shown in Fig.~\ref{2ketamass}(b)
and listed in Table~\ref{phieta_tab}.  

The cross section shows a rise from threshold to a peak value of about
2~\nb at around 1.7~\gev, followed by a monotonic decrease with
increasing energy.
This measurement is consistent with the more precise \babar\
measurement in the $\Kp\Km\gamma\gamma$ final state~\cite{isrphieta},
which is also shown in Fig.~\ref{2ketamass}(b)

\begin{figure}[tbh]
\begin{center}
\includegraphics[width=0.45\linewidth]{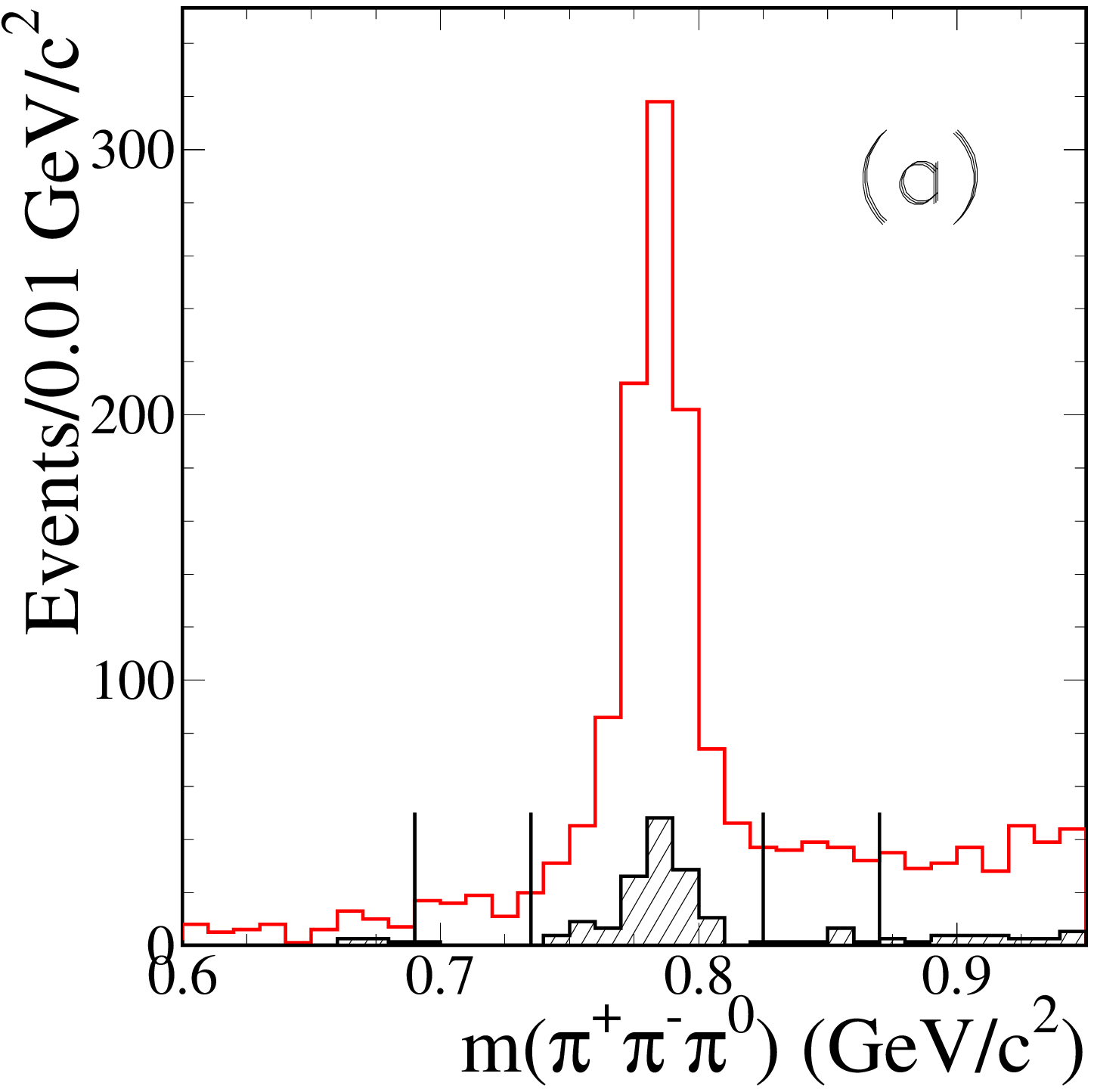}
\includegraphics[width=0.45\linewidth]{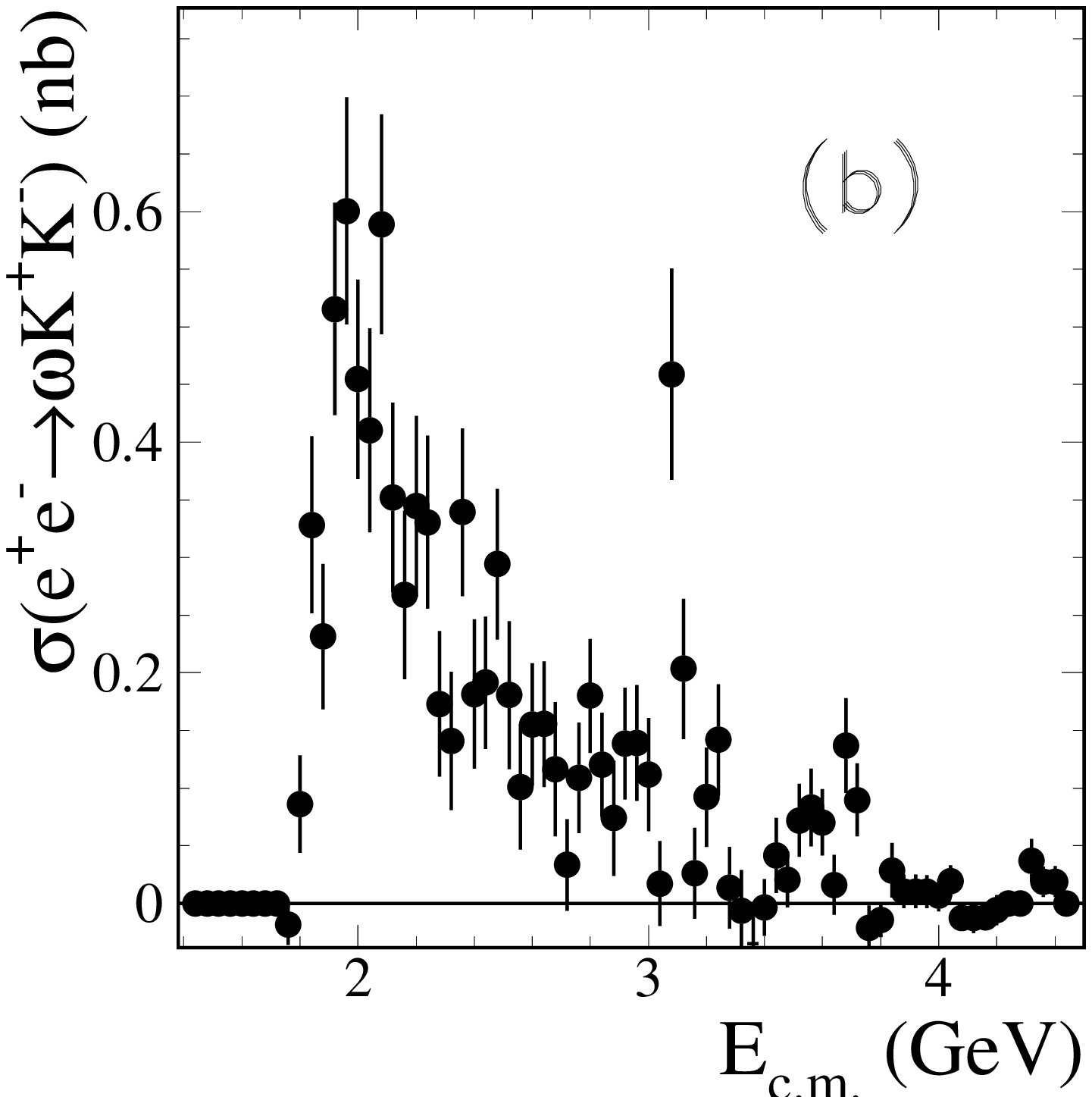}
\vspace{-0.2cm}
\caption{
   (a) Expanded view of the $\pipi\pi^0$ invariant mass distribution 
   [Fig.~\ref{3pivs2k3pi}(b)] in the region near the $\omega(782)$ for
   all selected events in the data (open histogram) and the estimated
   non-ISR background (hatched).
   The vertical lines delimit the $\omega$ signal region and sidebands.
   (b) The $\epem\!\! \to\! \omega(782)\Kp\Km$ cross section obtained 
   via ISR. 
   }
\label{omega2k}  
\end{center}
\end{figure}

\subsection{\bf\boldmath The $\omega(782)\Kp\Km$ Intermediate State}
\label{sec:omegakk}

Figure~\ref{omega2k}(a) shows the $\pipi\piz$ mass distribution
in the region near the $\omega(782)$ mass for all selected \KKpppn
events in the data and the estimated non-ISR contribution.
We first subtract the non-ISR background in each mass bin, 
then subtract the events in two sidebands, 690--735 and 825-870~\mevcc, 
from those in an $\omega$ signal region, 735--825~\mevcc.
The mass distribution of the recoiling $\Kp\Km$ pair after this
background subtraction shows no resonant structure.
Dividing by the corrected efficiency, differential luminosity and 
$\omega\! \to\! \pipi\piz$ branching fraction,
we obtain the first measurement of the $\epem\!\! \to\! \omega(782)\Kp\Km$
cross section, shown in Fig.~\ref{omega2k}(b) and listed in
Table~\ref{omega2k_tab}.  

The cross section rises from threshold to a peak value of about
0.55~\nb at about 2~\gev, 
then decreases with increasing energy except for peaks at the
$J/\psi$ and $\psi(2S)$ masses.
The events in the latter peak are partly due to the 
$\psi(2S)\! \to\! J/\psi\pipi$, $J/\psi\! \to\! \Kp\Km\piz$ decay.

\begin{figure}[tbh]
\begin{center}
\includegraphics[width=0.9\linewidth]{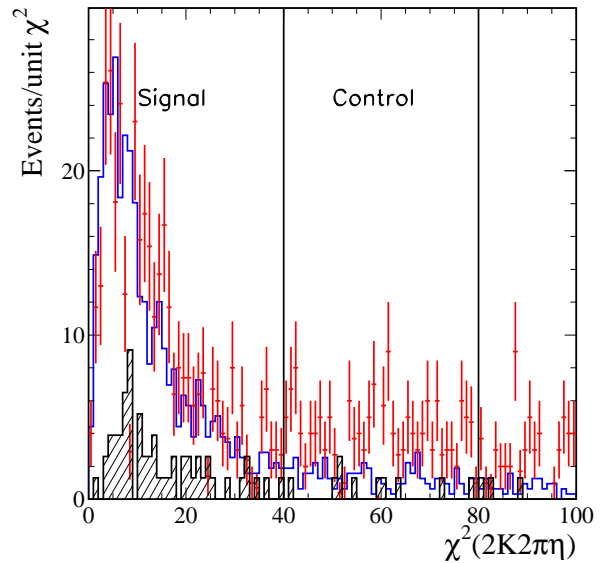}
\vspace{-0.4cm}
\caption{
   Distribution of \chisq from the 5C fit for \KKppeta candidates in
   the data (points).  
   The open histogram is the distribution for simulated signal events,
   and the hatched histogram is the estimated background from non-ISR
   events.
   }
\label{2k2pieta_chi2_all}
\end{center}
\end{figure}

\section{\bf\boldmath The \KKppeta Final State}
\subsection{Final selection and backgrounds}

To suppress ISR \fourpieta background, 
we fit each event under that hypothesis and require 
$\chifourpieta\! >\! 30$.
The \chiKKppeta distribution for the remaining events is shown as
points in Fig.~\ref{2k2pieta_chi2_all}, 
and the distribution for simulated ISR $\phi\eta\! \to\! \KKpppn$ 
events (open histogram) is normalized to the data in the region 
$\chiKKpppn\! <\! 20$.
We do not simulate \KKppeta events, 
but we expect the resolution and efficiency to be indistinguishable
from the generated mode.
The hatched histogram represents the non-ISR background,
which is dominated by $\KKpppn\eta$ events and is evaluated
from the simulation using the same scale factor as for the \KKpppn
final state.
We define a signal region, $\chiKKppeta\! <\! 40$, 
containing 375 events
and a control region, $40\! <\! \chiKKppeta\! <\! 80$,
containing 162 events.
We subtract the non-ISR background and the ISR-type background,
estimated from the control region, to obtain a number of signal events.

\begin{figure}[tbh]
\begin{center}
\includegraphics[width=0.9\linewidth]{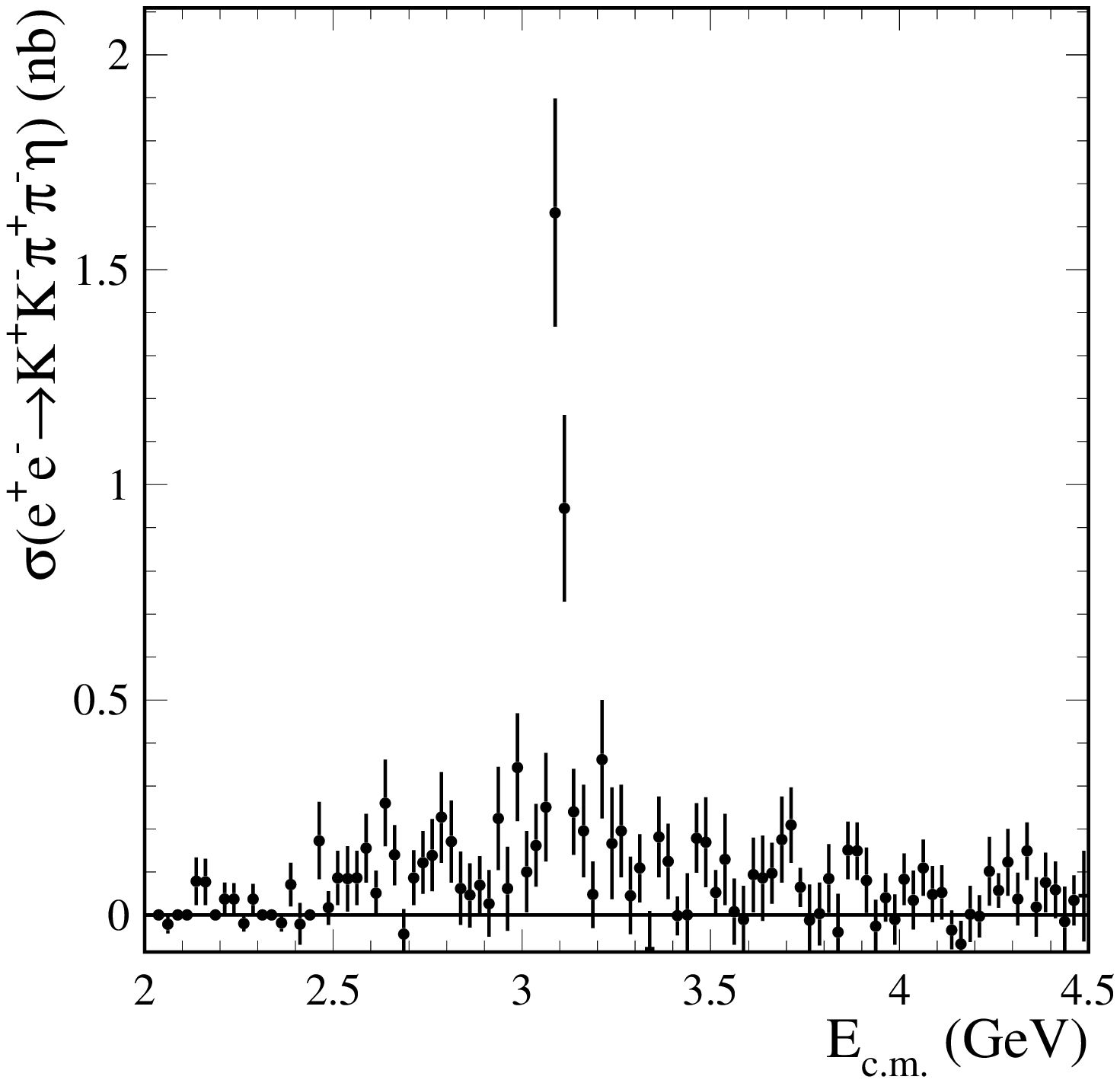}
\vspace{-0.5cm}
\caption{
   The $\epem\!\! \to\! \KKppeta$ cross section as a function of c.m.\
   energy measured with ISR data.
   Only statistical errors are shown.
   }
\label{2k2pieta_ee_babar}
\end{center}
\end{figure} 
\begin{figure}[tbh]
\begin{center}
\includegraphics[width=0.48\linewidth]{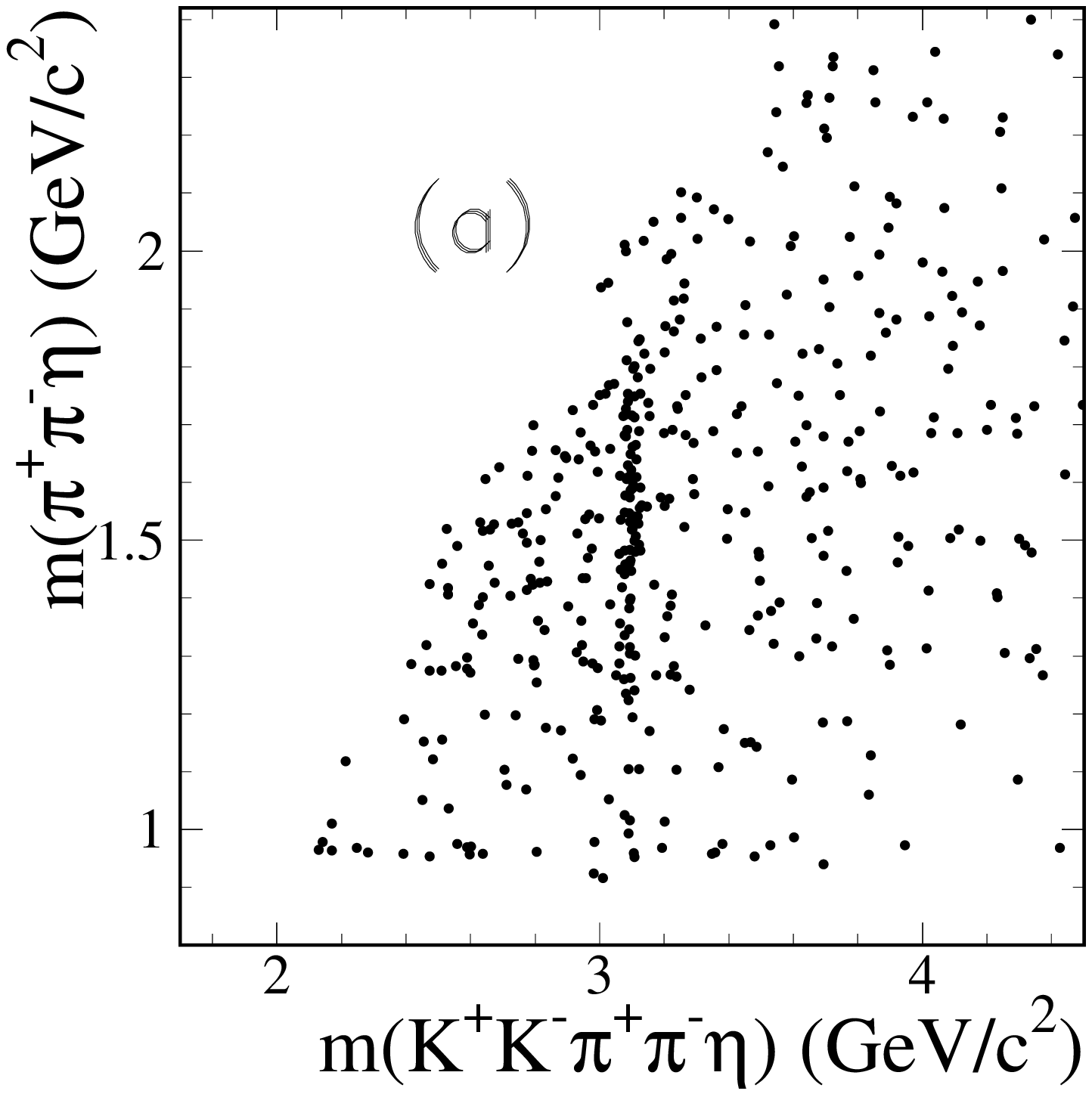}
\includegraphics[width=0.48\linewidth]{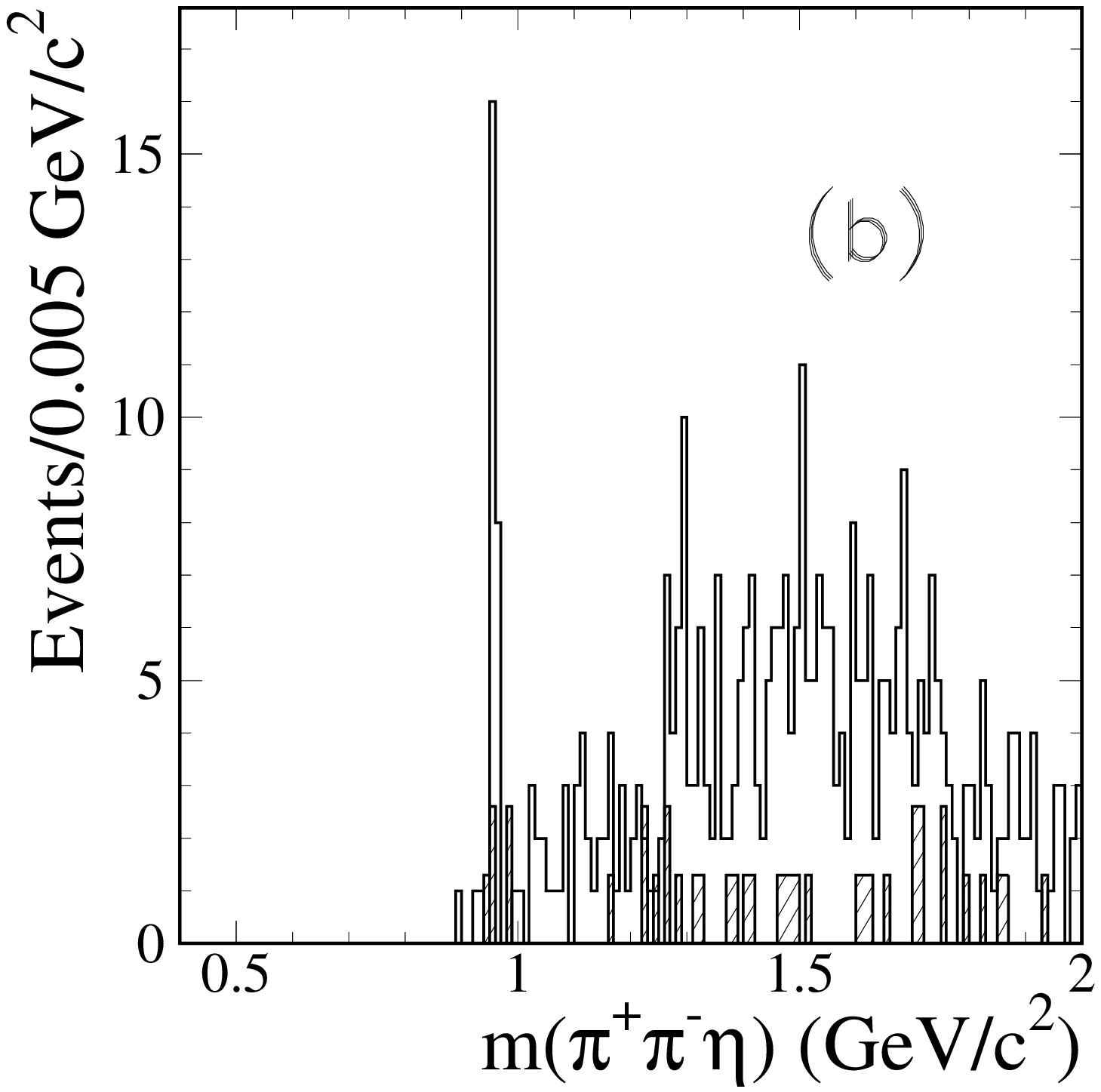}
\vspace{-0.2cm}
\caption{
   (a) The $\pipi\eta$ mass versus the \KKppeta mass, and
   (b) the $\pipi\eta$ mass projection for selected \KKppeta candidates.
   The hatched histogram represents the estimated non-ISR background.
   }
\label{2pietavs2k2pieta}  
\end{center}
\end{figure}
\begin{figure}[tbh]
\begin{center}
\includegraphics[width=0.48\linewidth]{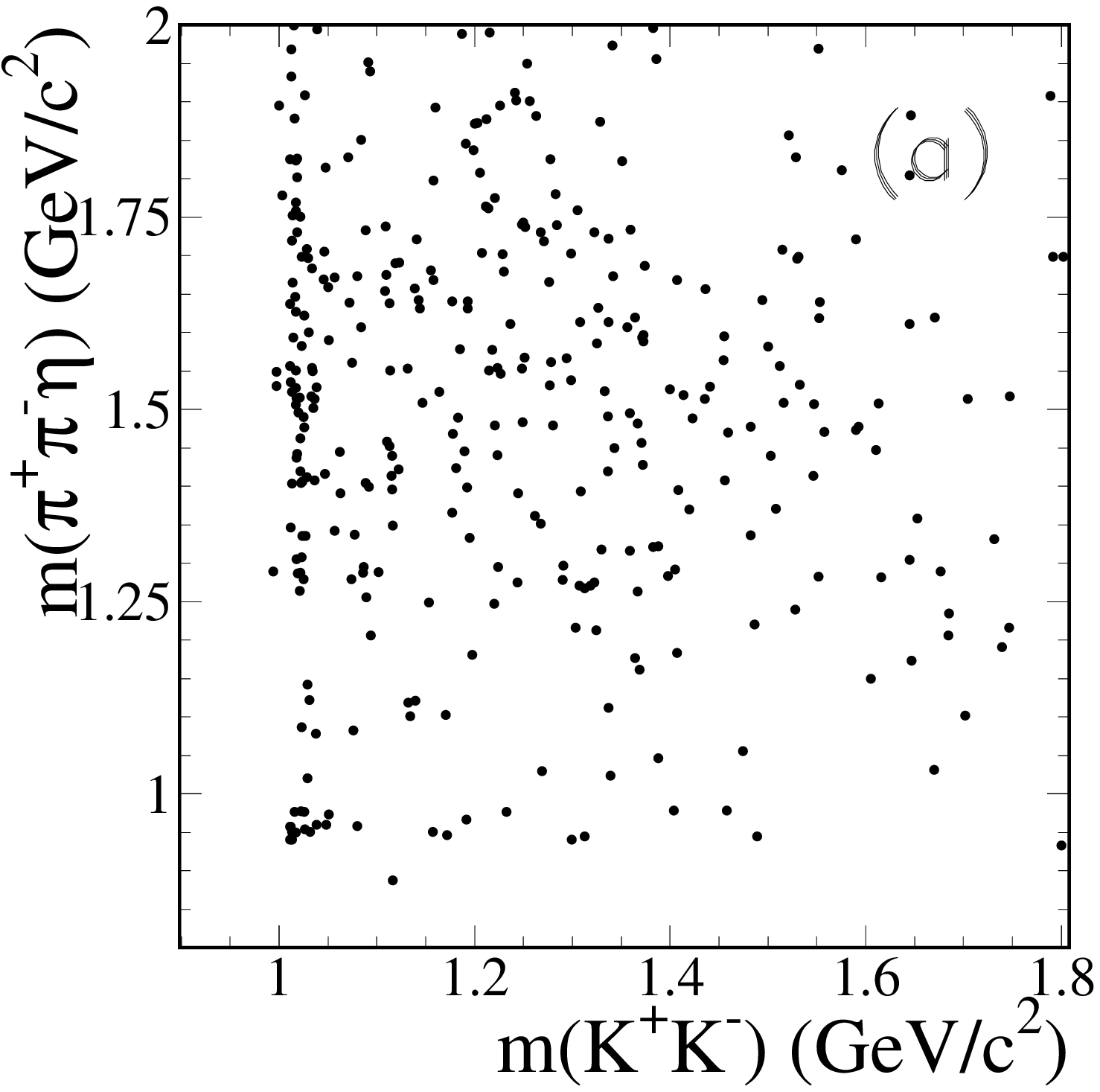}
\includegraphics[width=0.48\linewidth]{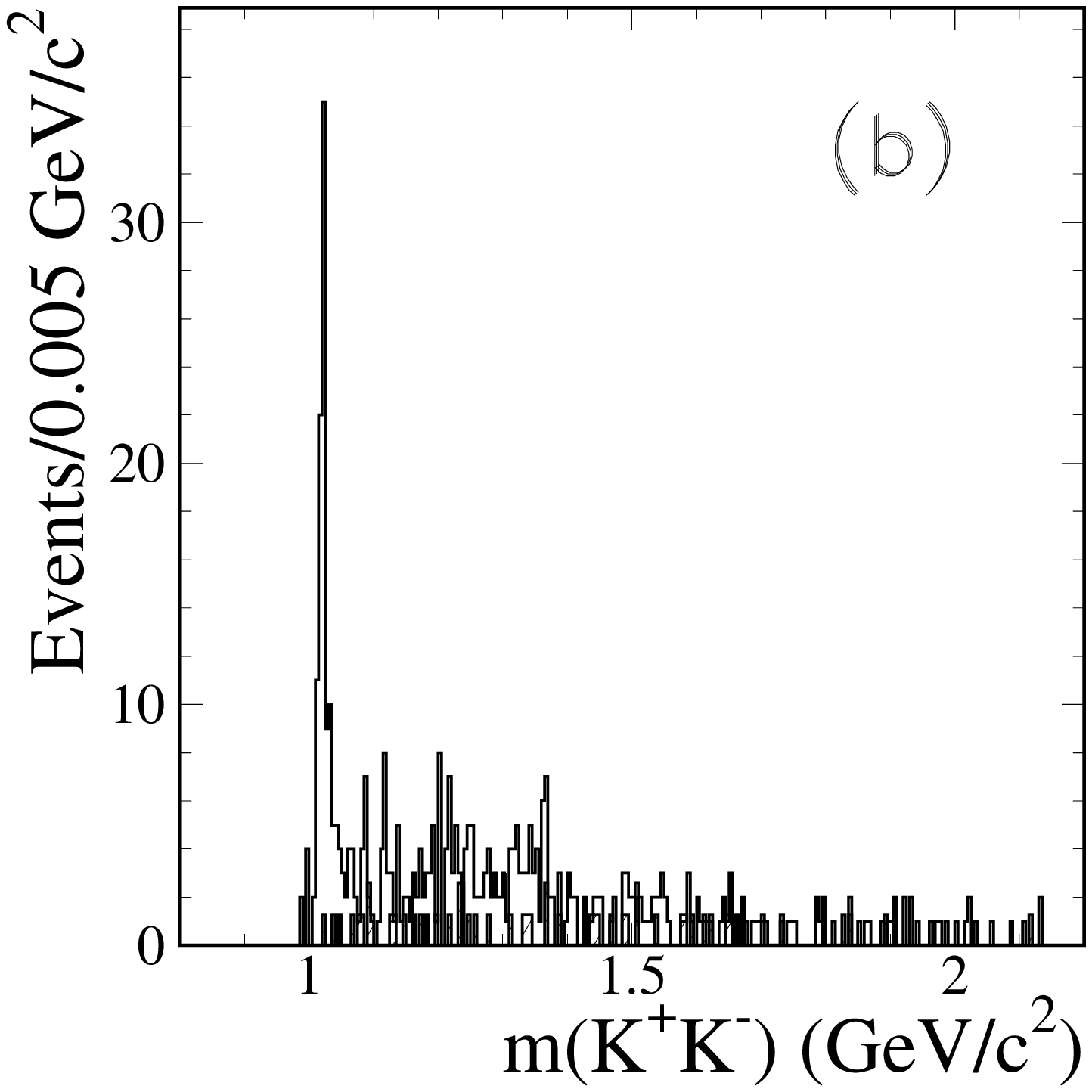}
\vspace{-0.2cm}
\caption{
   (a) The $\pipi\eta$ mass versus the $\Kp\Km$ mass and
   (b) the $\Kp\Km$ mass projection for selected \KKppeta candidates.
   The hatched histogram represents the estimated non-ISR background.
   }
\label{2pietavs2k}  
\end{center}
\end{figure}
\subsection{\boldmath Cross Section for \KKppeta}
\label{2k2pieta}
We calculate the cross section for the process $\epem\!\! \to\! \KKppeta$
as described in Sec.~\ref{sec:xs4pipi0}, by dividing the number
of background-subtracted events in each \KKppeta mass bin by the
corrected detection efficiency and differential luminosity.
Since the model dependence of the acceptance is small,
we use the efficiency for \KKpppn events shown in Fig.~\ref{mc_acc3}(b),
divided by the $\eta\! \to\! \gamma\gamma$ branching fraction
and with an increased systematic error of 5\%. 

We show the cross section as a function of energy in 
Fig.~\ref{2k2pieta_ee_babar} with statistical errors only.
This is the first measurement of this cross section, 
which shows a rise from threshold to a maximum value of about 0.2~\nb
at about 2.8~\gev,
followed by a monotonic decrease with increasing energy, 
except for a prominent peak at the $J/\psi$ mass.
The systematic errors are similar to those for the other modes
presented here, totalling about 10\% at all energies.
\subsection{\boldmath Substructure in the \KKppeta Final State}
\label{sec:2k2pietaplots}
Figure~\ref{2pietavs2k2pieta}(a) shows a scatter plot of the
$\pipi\eta$ mass versus the \KKppeta mass
and Fig.~\ref{2pietavs2k2pieta}(b) shows the $\pipi\eta$ mass projection.
A horizontal band and peak, respectively, corresponding to the $\eta'(958)$  
are visible.
The non-ISR background, 
shown as the shaded histogram in Fig.~\ref{2pietavs2k2pieta}(b), 
is small, but may include a few $\eta'$.
Figure~\ref{2pietavs2k}(a) shows a scatter plot of the $\pipi\eta$
mass versus the $\Kp\Km$ mass in the event. 
A vertical band corresponding to the $\phi(1020)$ is visible,
and almost all $\eta'(958)$ are produced through the $\phi\eta'$ channel.
The $\Kp\Km$ mass projection in Fig.~\ref{2pietavs2k}(b) shows a
$\phi(1020)$ signal, 
but the non-ISR background distribution has no resonant structure.
Due to the low statistics, we do not study the mass dependence of
these channels.
\begin{table*}
\caption{Measurements of the $\ep\en\to K^+ K^-\pipi\eta$ 
cross section (errors are statistical only).}
\label{2k2pieta_tab}
\begin{ruledtabular}
\hspace{-1.8cm}
\begin{tabular}{ c c c c c c c c }
$E_{\rm c.m.}$ (GeV) & $\sigma$ (nb)  
& $E_{\rm c.m.}$ (GeV) & $\sigma$ (nb) 
& $E_{\rm c.m.}$ (GeV) & $\sigma$ (nb) 
& $E_{\rm c.m.}$ (GeV) & $\sigma$ (nb)  
\\
\hline

 2.1125 &  0.00 $\pm$  0.00 & 2.7125 &  0.09 $\pm$  0.06 & 3.3125 &  0.11 $\pm$  0.08 & 3.9125 &  0.08 $\pm$  0.08 \\
 2.1375 &  0.08 $\pm$  0.06 & 2.7375 &  0.12 $\pm$  0.07 & 3.3375 & -0.08 $\pm$  0.09~ & 3.9375 & -0.03 $\pm$  0.06~ \\
 2.1625 &  0.08 $\pm$  0.05 & 2.7625 &  0.14 $\pm$  0.08 & 3.3625 &  0.18 $\pm$  0.09 & 3.9625 &  0.04 $\pm$  0.06 \\
 2.1875 &  0.00 $\pm$  0.00 & 2.7875 &  0.23 $\pm$  0.11 & 3.3875 &  0.12 $\pm$  0.09 & 3.9875 & -0.01 $\pm$  0.06~ \\
 2.2125 &  0.04 $\pm$  0.04 & 2.8125 &  0.17 $\pm$  0.09 & 3.4125 &  0.00 $\pm$  0.05 & 4.0125 &  0.08 $\pm$  0.06 \\
 2.2375 &  0.04 $\pm$  0.04 & 2.8375 &  0.06 $\pm$  0.08 & 3.4375 &  0.00 $\pm$  0.10 & 4.0375 &  0.03 $\pm$  0.07 \\
 2.2625 & -0.02 $\pm$  0.02~ & 2.8625 &  0.05 $\pm$  0.07 & 3.4625 &  0.18 $\pm$  0.08 & 4.0625 &  0.11 $\pm$  0.07 \\
 2.2875 &  0.04 $\pm$  0.04 & 2.8875 &  0.07 $\pm$  0.07 & 3.4875 &  0.17 $\pm$  0.10 & 4.0875 &  0.05 $\pm$  0.07 \\
 2.3125 &  0.00 $\pm$  0.00 & 2.9125 &  0.03 $\pm$  0.08 & 3.5125 &  0.05 $\pm$  0.05 & 4.1125 &  0.05 $\pm$  0.06 \\
 2.3375 &  0.00 $\pm$  0.00 & 2.9375 &  0.22 $\pm$  0.12 & 3.5375 &  0.13 $\pm$  0.11 & 4.1375 & -0.03 $\pm$  0.05~ \\
 2.3625 & -0.02 $\pm$  0.02~ & 2.9625 &  0.06 $\pm$  0.10 & 3.5625 &  0.01 $\pm$  0.08 & 4.1625 & -0.07 $\pm$  0.05~ \\
 2.3875 &  0.07 $\pm$  0.05 & 2.9875 &  0.34 $\pm$  0.13 & 3.5875 & -0.01 $\pm$  0.08~ & 4.1875 &  0.00 $\pm$  0.06 \\
 2.4125 & -0.02 $\pm$  0.05~ & 3.0125 &  0.10 $\pm$  0.09 & 3.6125 &  0.09 $\pm$  0.09 & 4.2125 &  0.00 $\pm$  0.05 \\
 2.4375 &  0.00 $\pm$  0.00 & 3.0375 &  0.16 $\pm$  0.10 & 3.6375 &  0.09 $\pm$  0.10 & 4.2375 &  0.10 $\pm$  0.08 \\
 2.4625 &  0.17 $\pm$  0.09 & 3.0625 &  0.25 $\pm$  0.13 & 3.6625 &  0.10 $\pm$  0.07 & 4.2625 &  0.06 $\pm$  0.04 \\
 2.4875 &  0.02 $\pm$  0.04 & 3.0875 &  1.63 $\pm$  0.26 & 3.6875 &  0.18 $\pm$  0.10 & 4.2875 &  0.12 $\pm$  0.08 \\
 2.5125 &  0.09 $\pm$  0.06 & 3.1125 &  0.95 $\pm$  0.22 & 3.7125 &  0.21 $\pm$  0.09 & 4.3125 &  0.04 $\pm$  0.06 \\
 2.5375 &  0.08 $\pm$  0.08 & 3.1375 &  0.24 $\pm$  0.10 & 3.7375 &  0.06 $\pm$  0.05 & 4.3375 &  0.15 $\pm$  0.07 \\
 2.5625 &  0.09 $\pm$  0.06 & 3.1625 &  0.20 $\pm$  0.11 & 3.7625 & -0.01 $\pm$  0.08~ & 4.3625 &  0.02 $\pm$  0.07 \\
 2.5875 &  0.16 $\pm$  0.08 & 3.1875 &  0.05 $\pm$  0.08 & 3.7875 &  0.00 $\pm$  0.07 & 4.3875 &  0.08 $\pm$  0.07 \\
 2.6125 &  0.05 $\pm$  0.05 & 3.2125 &  0.36 $\pm$  0.14 & 3.8125 &  0.08 $\pm$  0.08 & 4.4125 &  0.06 $\pm$  0.07 \\
 2.6375 &  0.26 $\pm$  0.10 & 3.2375 &  0.17 $\pm$  0.13 & 3.8375 & -0.04 $\pm$  0.09~ & 4.4375 & -0.02 $\pm$  0.08~ \\
 2.6625 &  0.14 $\pm$  0.07 & 3.2625 &  0.20 $\pm$  0.11 & 3.8625 &  0.15 $\pm$  0.07 & 4.4625 &  0.03 $\pm$  0.06 \\
 2.6875 & -0.04 $\pm$  0.06~ & 3.2875 &  0.04 $\pm$  0.09 & 3.8875 &  0.15 $\pm$  0.07 & 4.4875 &  0.04 $\pm$  0.11 \\

\end{tabular}
\end{ruledtabular}
\end{table*}

\begin{figure}[tbh]
\includegraphics[width=0.9\linewidth]{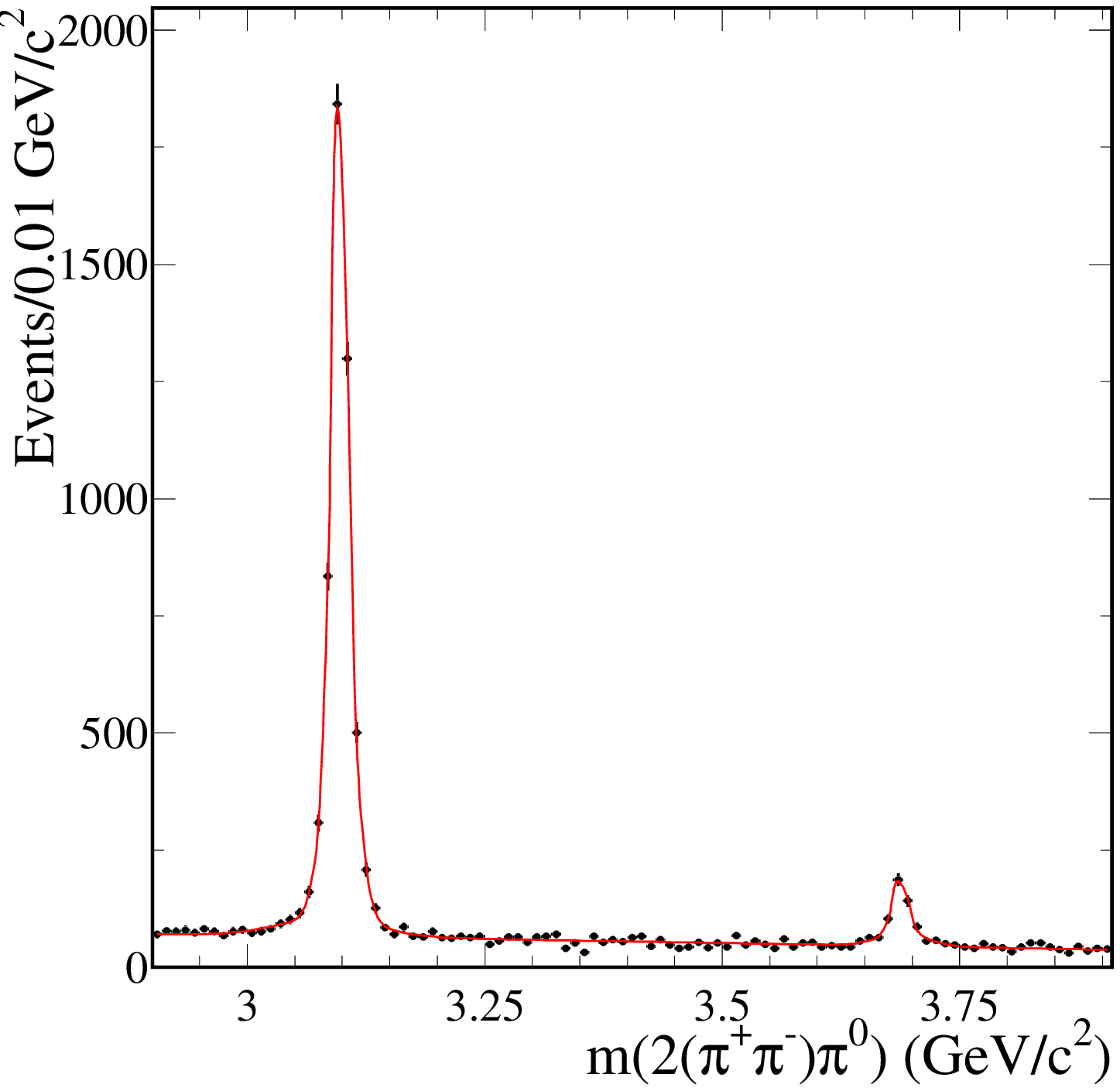}
\vspace{-0.4cm}
\caption{
  Raw invariant mass distribution for all selected 
  $\epem \!\!\to\! \fourpipn$ events in the charmonium region.
  The line represents the result of the fit described in the text.
  }
\label{jpsi}
\end{figure}
\section{\boldmath The Charmonium Region}
\label{sec:charmonium}
\begin{figure}[tbh]
\includegraphics[width=0.9\linewidth]{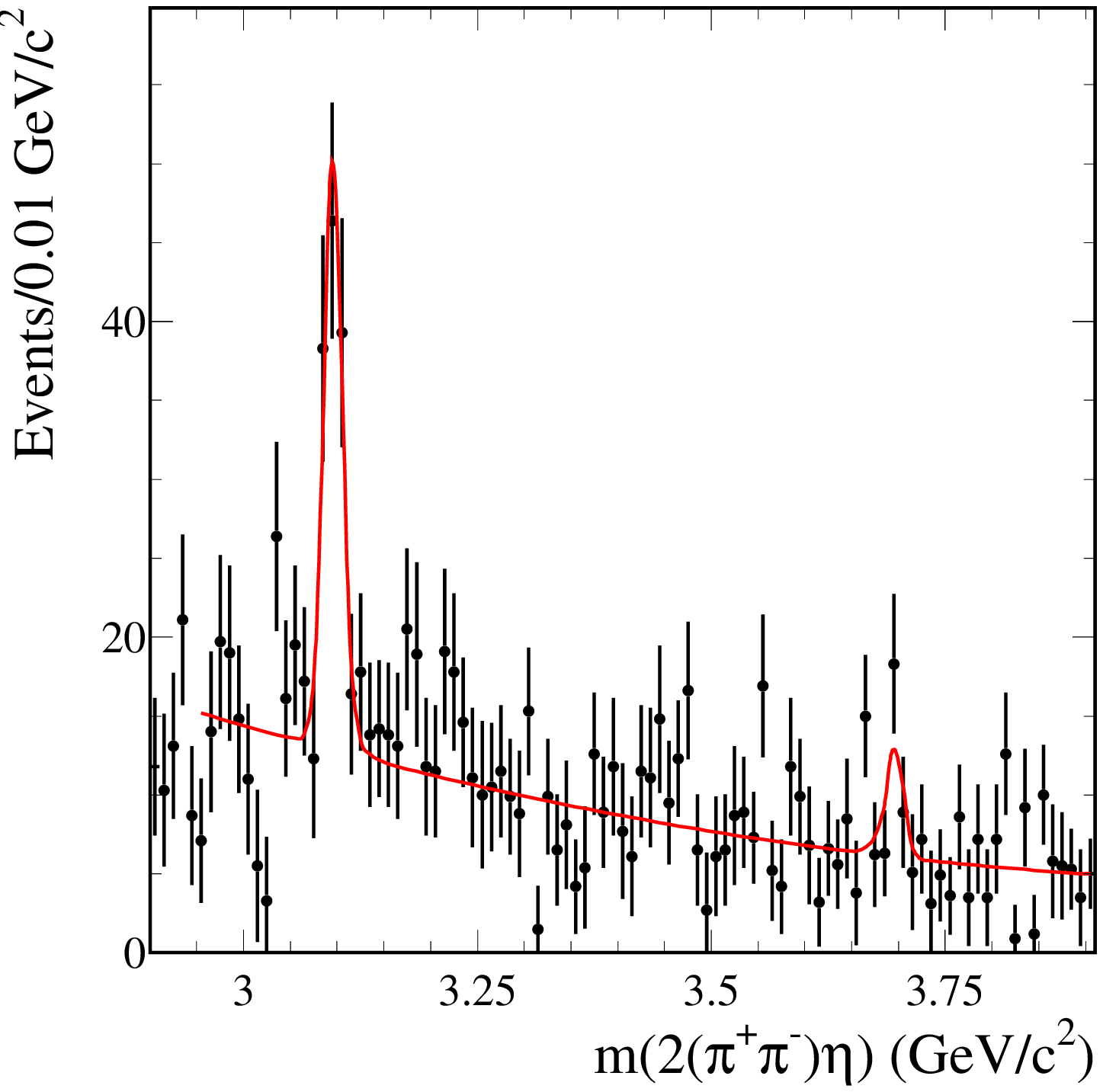}
\vspace{-0.4cm}
\caption{
  Raw invariant mass distribution for all selected 
  $\epem \!\!\to\! \fourpieta$ events in the charmonium region.
  The line represents the result of the fit described in the text.
  }
\label{jpsieta}
\end{figure}

\begin{figure}[tbh]
\includegraphics[width=0.9\linewidth]{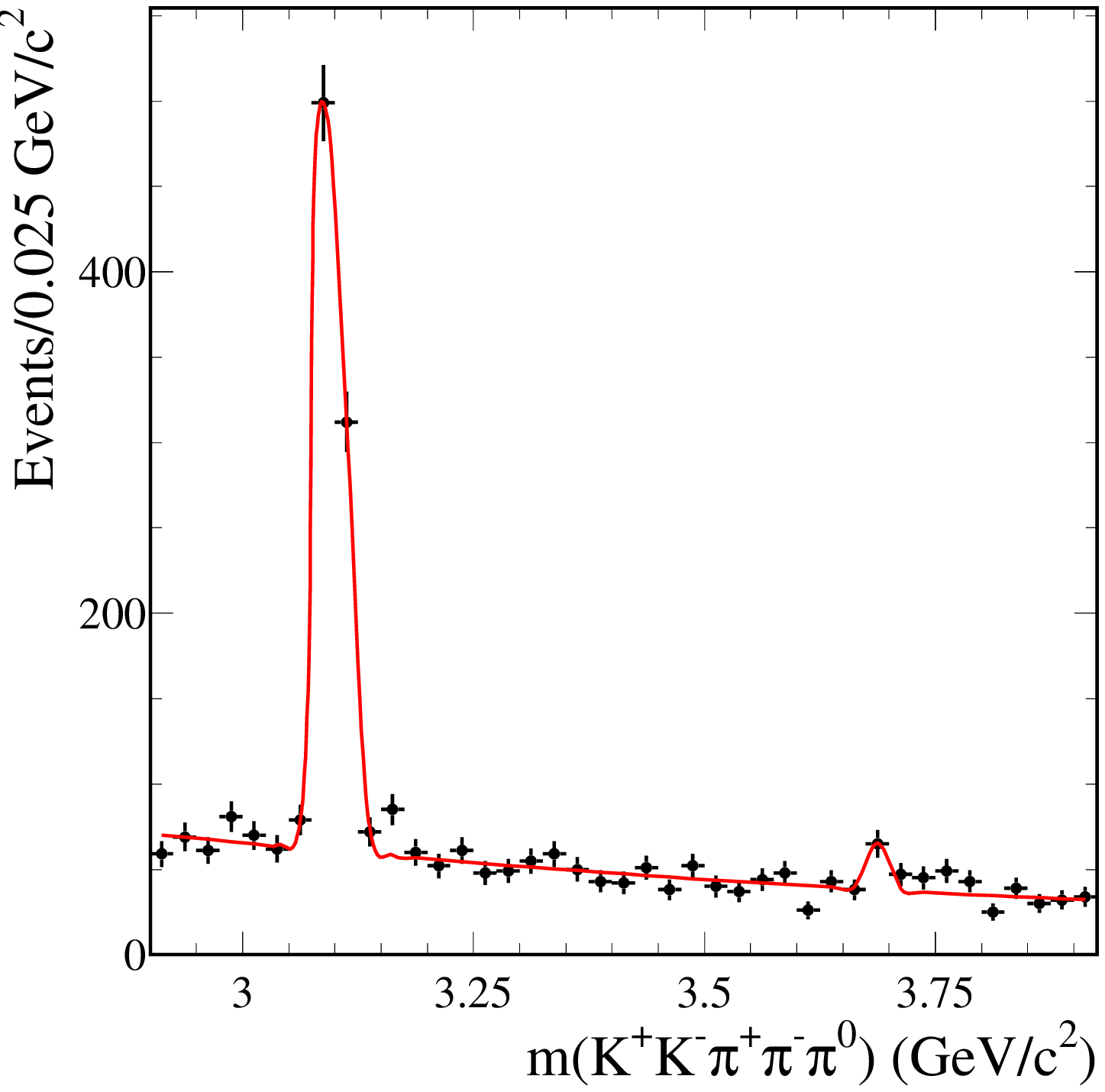}
\vspace{-0.4cm}
\caption{
  Raw invariant mass distribution for all selected 
  $\epem \!\!\to\! \KKpppn$ events in the charmonium region.
  The line represents the result of the fit described in the text.
  }
\label{jpsi2k3pi}
\end{figure}

\begin{figure}[tbh]
\includegraphics[width=0.9\linewidth]{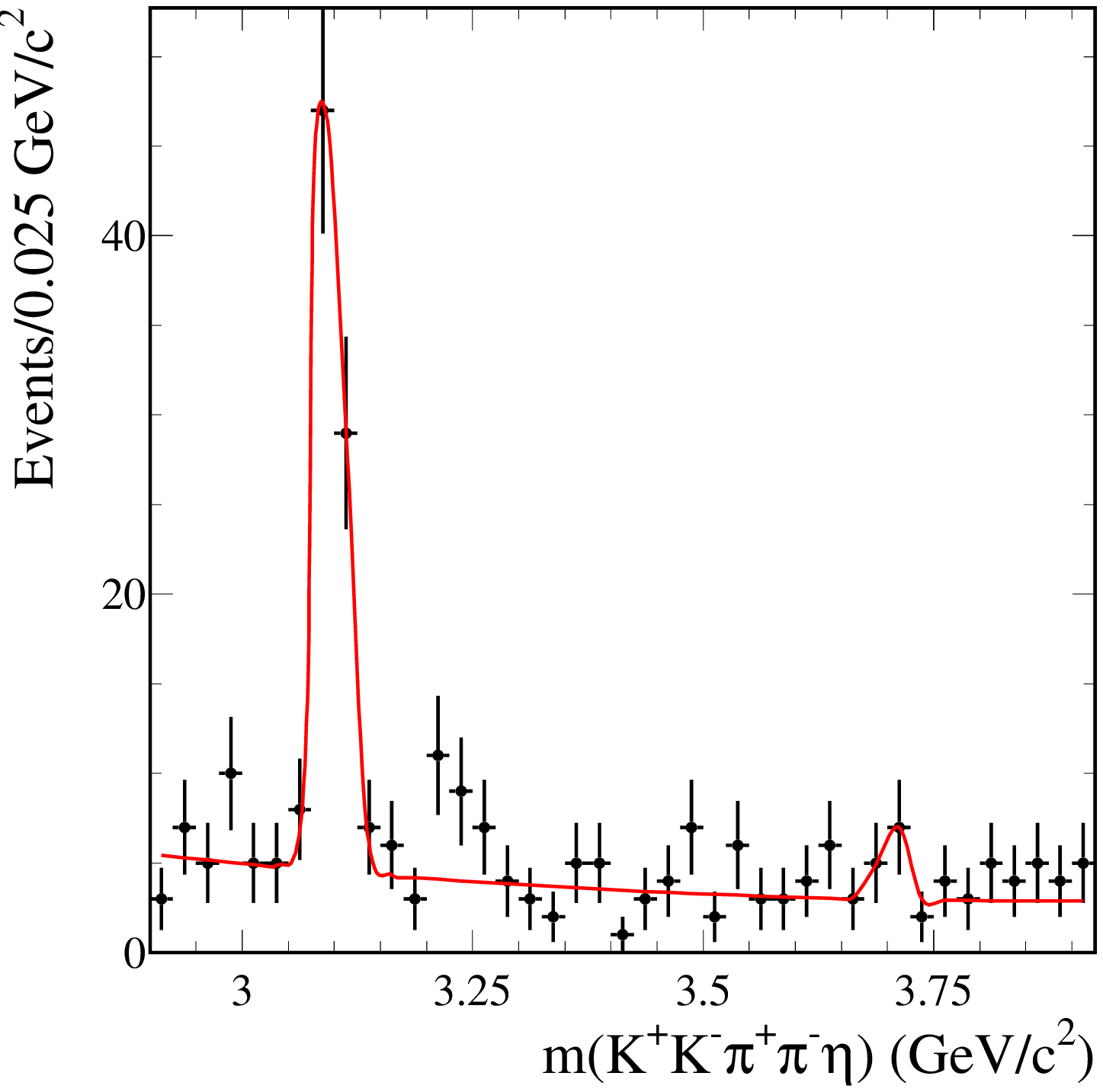}
\vspace{-0.4cm}
\caption{
  Raw invariant mass distribution for all selected 
  $\epem \!\!\to\! \KKppeta$ events in the charmonium region.
  The line represents the result of the fit described in the text.
  }
\label{jpsi2k2pieta}
\end{figure}

The data at masses above 3~\gevcc can be used to measure or set limits
on the branching fractions of narrow resonances, such as charmonia,
and the narrow $J/\psi$ and $\psi(2S)$ peaks allow measurements of our mass
scale and resolution.
Figures~\ref{jpsi}, \ref{jpsieta}, \ref{jpsi2k3pi} and~\ref{jpsi2k2pieta} 
show the invariant mass distributions for the selected \fourpipn, 
\fourpieta, \KKpppn and \KKppeta events, respectively, in this region, 
with finer binning than in the corresponding Figs.~\ref{4pipi0_babar}, 
~\ref{4pieta_babar}, ~\ref{2k3pi_babar} and~\ref{2k2pieta_ee_babar}.
We do not subtract any background,
since it is small and nearly uniformly distributed.
Signals from the $J/\psi$  and $\psi(2S)$ are visible in all four
distributions.


We fit these distributions using a sum of two Gaussian functions 
to describe each of the $J/\psi$ and $\psi (2S)$ signals plus a polynomial to 
describe the remainder of the distribution.
We fix the two Gaussian combination shape by fitting simulated events, 
but let the overall mean and width float in the fit, along with the amplitude
and the coefficients of the polynomial.
In all cases, the fitted mean values are within 1~\mevcc of the
PDG~\cite{PDG} $J/\psi$ and $\psi(2S)$ masses,
and the widths are consistent with the simulated resolutions
within 10\%.

In the $J/\psi$ peak the fits yield 
4990$\pm$79 \fourpipn events,
  85$\pm$14 \fourpieta events,
 768$\pm$31 \KKpppn events and
 72.9$\pm$9.4 \KKppeta events.
From the number of observed events in each final state $f$,
$N_{J/\psi \!\to\! f}$, 
we calculate the product of the $J/\psi$ branching fraction to $f$ and
the $J/\psi$ electronic width: 
\begin{equation}
     \BR_{J/\psi \!\to\! f} \cdot \Gamma^{J/\psi}_{ee}  =
 \frac{N_{J/\psi \!\to\! f} \cdot m_{J/\psi}^2}
      {6\pi^2 \cdot d{\cal L}/dE \cdot \epsilon_f(m_{J/\psi}) \cdot C}
 ~~~, \\
\label{jpsicalc}
\end{equation}
where
$d{\cal L}/dE\! =\! 65.6\pm 2.0~\invnb/\mev$ and $\epsilon_f(m_{J/\psi})$ 
are the ISR luminosity and corrected selection efficiency, respectively,
at the $J/\psi$ mass
and $C$ is a conversion constant.  
We estimate 
efficiencies of 0.105 for the \fourpipn and \fourpieta final states,
and 0.046 for the \KKpppn and \KKppeta modes,
with systematic uncertainties of about 5\% for modes with $\pi^0$ and
about 7\% for modes with $\eta$.
Adding the error on the ISR luminosity in quadrature,
we assign a 6\% (7.5\% for $\eta$) overall systematic uncertainty on each product.

Using $\Gamma^{J/\psi}_{ee}\! =\! 5.55\pm0.14~\kev$~\cite{PDG}, 
we obtain the branching fractions listed in Table~\ref{jpsitab},
along with the measured products and the current PDG values.
The systematic errors include the 2.5\% uncertainty on $\Gamma^{J/\psi}_{ee}$.
The $J/\psi\to\KKppeta$ channel has not been previously observed
and the \KKpppn and \fourpieta branching fractions are consistent and
competitive with the PDG values.
However, we find a \fourpipn branching fraction 4.8 standard
deviations higher than the PDG value.

In the $\psi(2S)$ peak the fits yield 
 410$\pm$30 \fourpipn events,
15.6$\pm$7.6 \fourpieta events,
31.8$\pm$11.9 \KKpppn events and
 7.0$\pm$4.0 \KKppeta events.
Using a calculation analogous 
to Eq.~\ref{jpsicalc}, with $d{\cal L}/dE\! =\! 84.0\pm 2.5~\invnb/\mev$ and
$\epsilon(\psi(2S))\! =\! 0.0965$ (0.0400) for the \fourpipn and
\fourpieta (\KKpppn and \KKppeta) modes, 
we obtain the product of the $\psi(2S)$ branching fractions to these
final states and its electronic width.
Dividing by the world average value of $\Gamma^{\psi(2S)}_{ee}$~\cite{PDG},
we obtain the branching fractions listed in Table~\ref{jpsitab}.
The \fourpieta and \KKppeta branching fractions are first measurements, 
and the \KKpppn branching fraction is consistent with the PDG value.

\begin{figure}[tbh]
\includegraphics[width=0.48\linewidth]{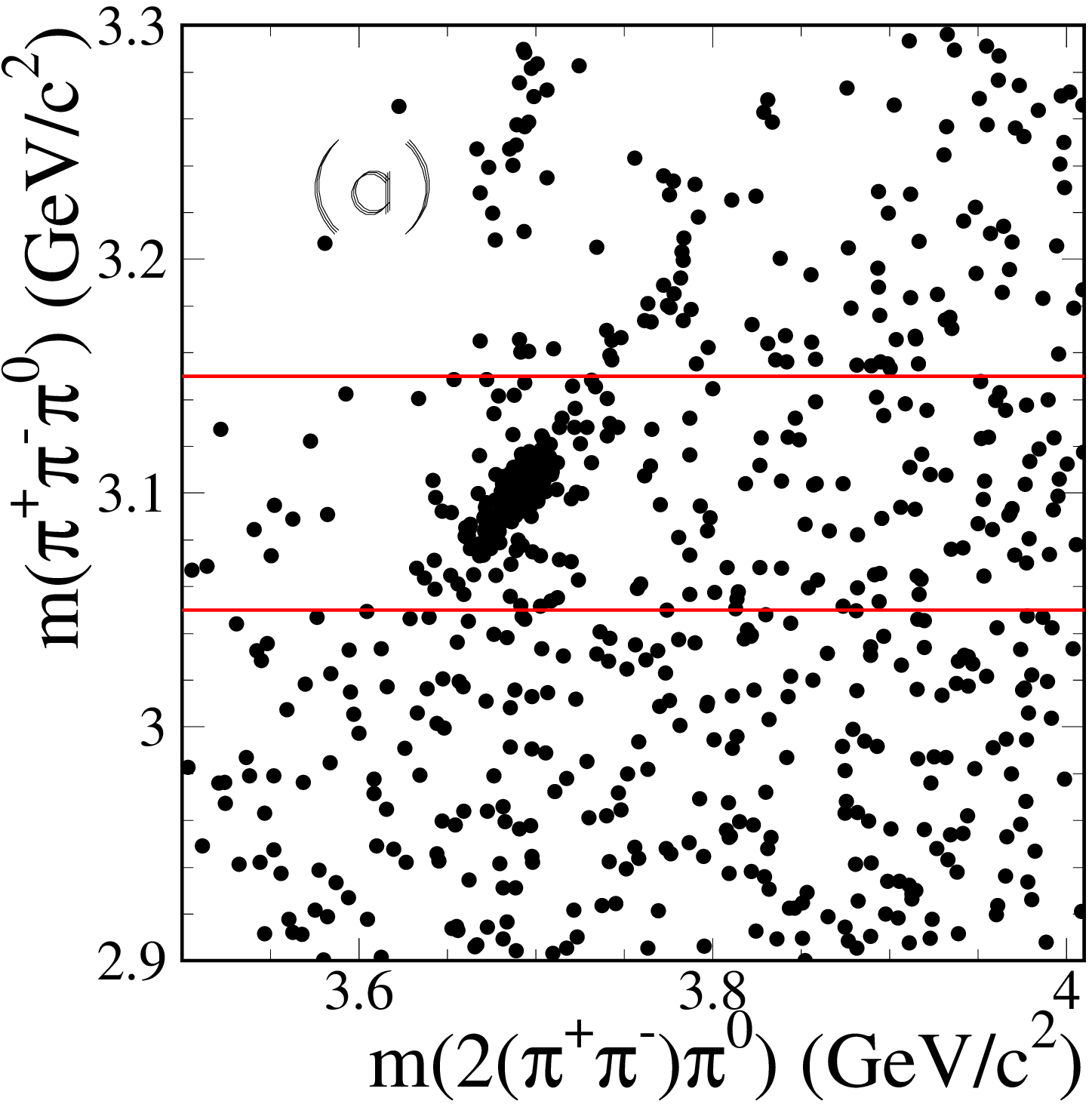}
\includegraphics[width=0.48\linewidth]{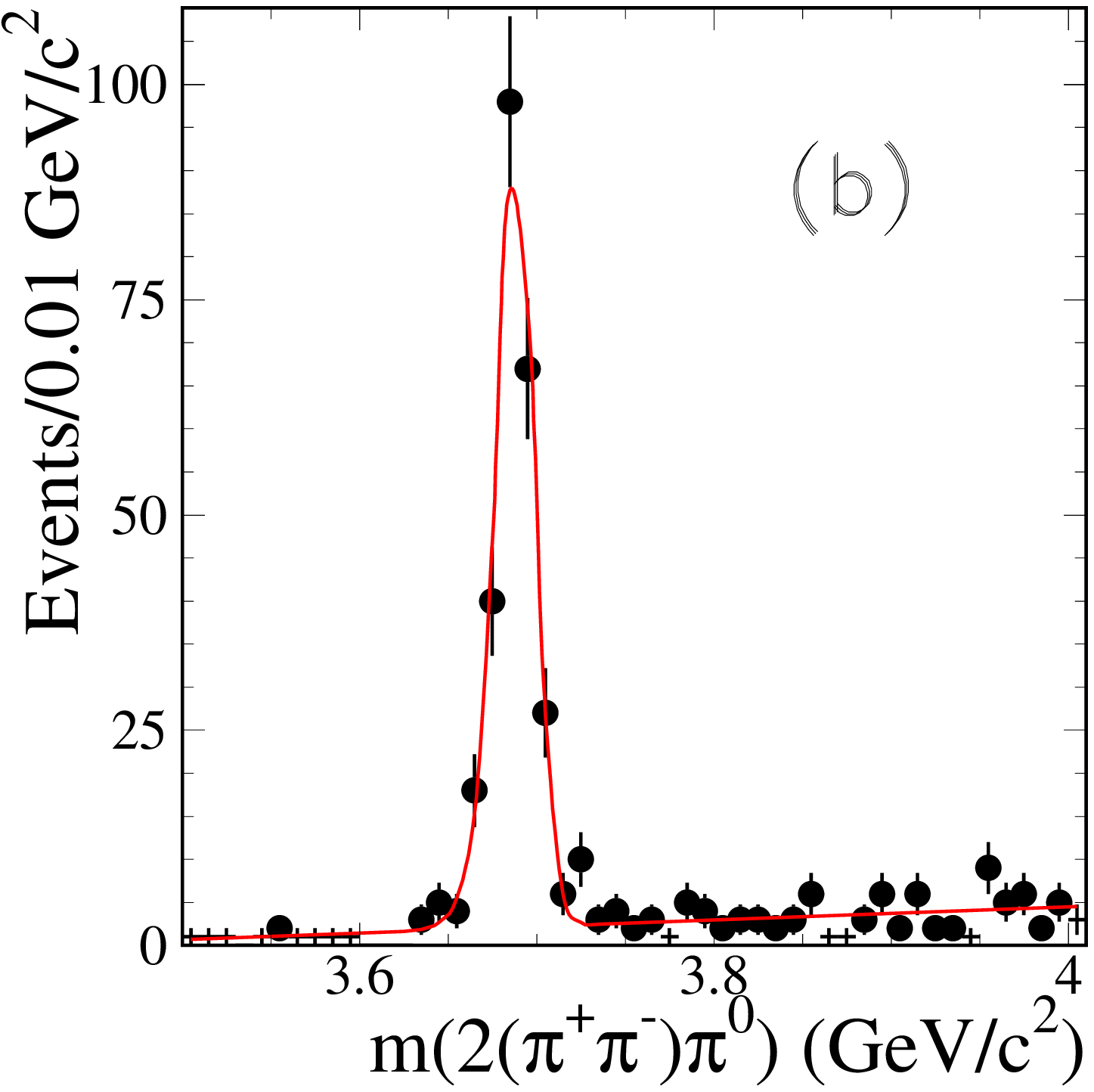}
\caption{
   (a) The $\pipi\piz$ mass closest to $J/\psi$ mass versus the
   five-pion mass for selected \fourpipn events.
   (b) The five-pion mass for events with a three-pion mass within
   50~\mev of the $J/\psi$ mass.
   }
\label{psi2s_chain}
\end{figure}

However, we find a \fourpipn branching fraction 7.6 standard
deviations higher than the PDG value.
We note that some of the observed $\psi(2S)$ could be due to
the decay chain $\psi(2S)\! \to\! J/\psi\pipi$, $J/\psi\! \to\! \pipi\piz$,
and we use this chain to check our result.
The scatter plot of the $\pipi\piz$ mass closest to the $J/\psi$ mass 
versus the \fourpipn mass in Fig.~\ref{psi2s_chain}(a) shows 
a cluster corresponding to this decay chain.
We select events with a three-pion mass within 50~\mevcc of the
$J/\psi$ mass (lines in Fig.~\ref{psi2s_chain}(a)) and plot their \fourpipn 
mass in Fig.~\ref{psi2s_chain}(b). 
A fit yields 256$\pm$17 $\psi(2S)$ events,
and using the well measured $\psi(2S)\! \to\! J/\psi\pipi$ branching
fraction of 0.318$\pm$0.06~\cite{PDG}, 
we calculate a $J/\psi\! \to\! \pipi\piz$ branching fraction 
$\BR_{J/\psi\to\pipi\pi^0} = (2.36\pm0.16\pm0.16)$\%
that is consistent with our previous measurement 
$\BR_{J/\psi\to\pipi\pi^0} = (2.19\pm0.19)$\%~\cite{isr3pi} as well
as with the current PDG value.
We obtain significantly higher values for both the $J/\psi$ and $\psi(2S)$ 
branching fractions to \fourpipn compared to previous
experiments~\cite{PDG}.
A similar difference was reported for the $J/\psi\to\pipi\pi^0$ decay in
recent experiments~\cite{bes,isr3pi}.

\begin{figure}[tbh]
\includegraphics[width=0.45\linewidth]{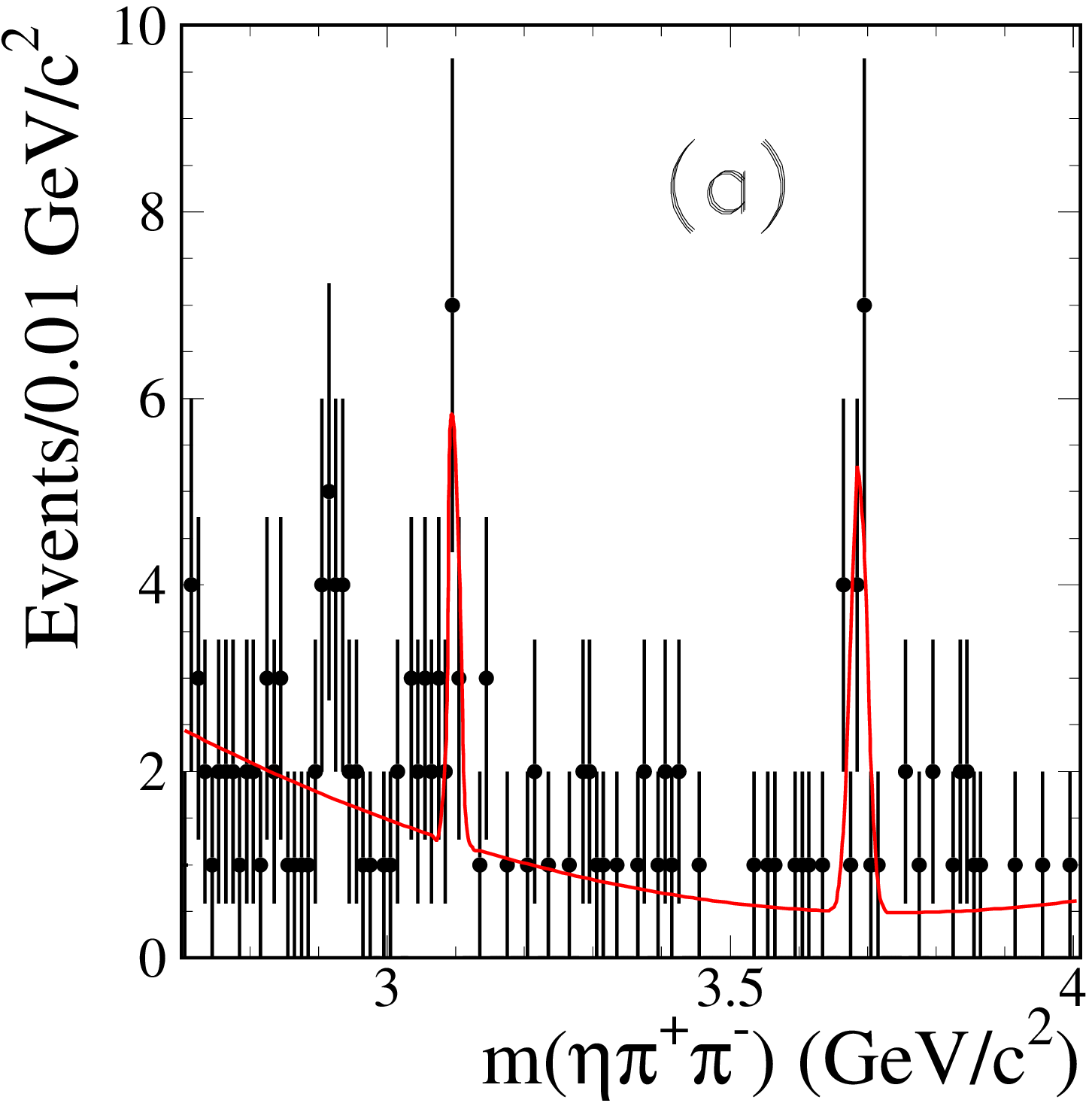}
\includegraphics[width=0.45\linewidth]{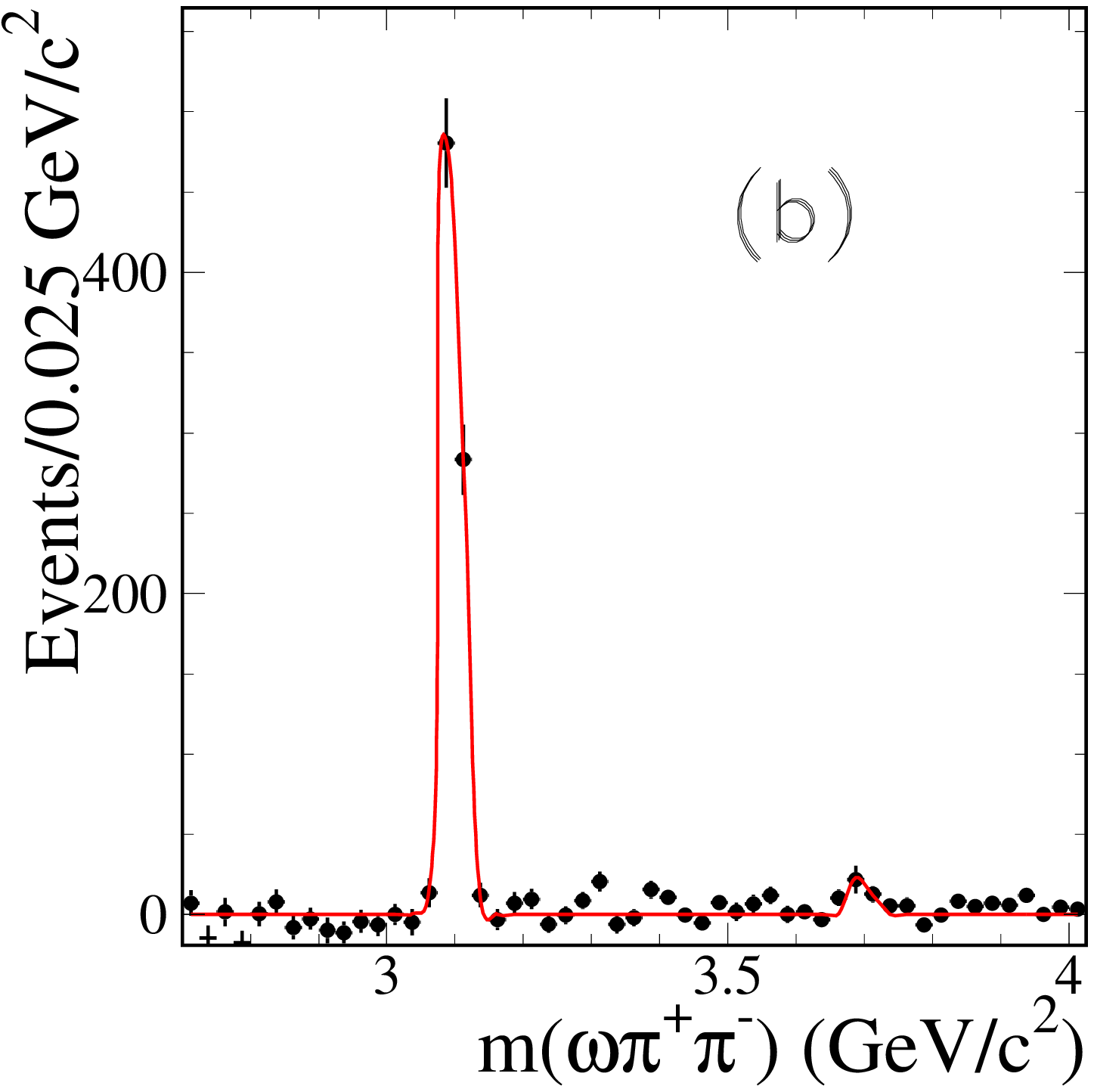}
\caption{
  Background-subtracted \fourpipn invariant mass distributions in 
  the charmonium region for events with a $\pipi\piz$ mass in the
  (a) $\eta$ and (b) $\omega(782)$ mass region.
  }
\label{etaomega2pi}
\end{figure}    

\begin{table*}[tbh]
\caption{
  Measurements of the $J/\psi$ and $\psi(2S)$ branching fractions.
  }
\label{jpsitab}
\begin{ruledtabular}
\begin{tabular}{r@{$\cdot$}l  r@{.}l@{$\pm$}l@{$\pm$}l 
                              r@{.}l@{$\pm$}l@{$\pm$}l
                              r@{.}l@{$\pm$}l } 
\multicolumn{2}{c}{Measured} & \multicolumn{4}{c}{Measured}    &  
\multicolumn{7}{c}{$J/\psi$ or $\psi(2S)$ Branching Fraction  (10$^{-3}$)}\\
\multicolumn{2}{c}{Quantity} & \multicolumn{4}{c}{Value (\ev)} &
\multicolumn{4}{c}{Calculated, this work}    & 
\multicolumn{3}{c}{PDG2006} \\
\hline
$\Gamma^{J/\psi}_{ee}$ &  $\BR_{J/\psi  \to 2(\pipi)\pi^0}$  &
\hspace*{0.6cm}  303&  & 5   & 18  \hspace*{0.7cm}  & 
\hspace*{0.6cm}   54&6 & 0.9 & 3.4  \hspace*{0.7cm}  & 
\hspace*{0.6cm}   33&7 & 2.6        \hspace*{0.7cm}  \\ 

$\Gamma^{J/\psi}_{ee}$  &  $\BR_{J/\psi  \to \omega\pipi}
                           \cdot \BR_{\omega    \to 3\pi}$  &
  47&8 &3.1& 3.2  &   9&7 & 0.6 & 0.6  &   7&2 & 1.0  \\ 

$\Gamma^{J/\psi}_{ee}$  &  $\BR_{J/\psi  \to \eta\pipi}
                            \cdot \BR_{\eta \to 3\pi}$  &
  0&51 &0.22& 0.03  &   0&40 & 0.17 & 0.03  &   0&193 & 0.023  \\ 

$\Gamma^{J/\psi}_{ee}$  &  $\BR_{J/\psi  \to   2(\pipi)\eta}
                             \cdot \BR_{\eta \to \gamma\gamma} $&
  5&16 &0.85& 0.39  &   2&35 & 0.39 & 0.20  & 2&26 & 0.28 \\   

$\Gamma^{J/\psi}_{ee}$  &  $\BR_{J/\psi  \to \Kp\Km\pipi\pi^0}  $  &
   107&0&4.3& 6.4 &   19&2 & 0.8 & 1.5  &   12&0 & 3.0\\      

$\Gamma^{J/\psi}_{ee}$  &  $\BR_{J/\psi  \to \phi\eta}  
                        \cdot \BR_{\phi\to\Kp\Km} \cdot \BR_{\eta\to 3\pi} $  &
   0&84&0.37& 0.05 &   1&4 & 0.6 & 0.1  &   0&74 & 0.08\\     

$\Gamma^{J/\psi}_{ee}$  &  $\BR_{J/\psi  \to \omega \Kp\Km}  
                        \cdot \BR_{\omega\to 3\pi} $  & 
   3&3& 1.3 & 0.2 &   1&36 & 0.50 & 0.10  &   1&9 & 0.4\\ 

$\Gamma^{J/\psi}_{ee}$  &  $\BR_{J/\psi  \to \Kp\Km\pipi\eta}
                             \cdot \BR_{\eta \to \gamma\gamma}   $  &
   10&2& 1.3& 0.8 &   4&7  & 0.6 & 0.4  &  \multicolumn{3}{c}{no entry} \\ 

$\Gamma^{\psi(2S)}_{ee}$  &  $\BR_{\psi(2S) \to 2(\pipi)\pi^0} $  &
   29&7& 2.2& 1.8 &   12&0  & 0.9  & 0.7   &   2&66 & 0.29 \\  

$\Gamma^{\psi(2S)}_{ee}$  &  $\BR_{\psi(2S) \to J/\psi\pipi} 
                             \cdot \BR_{J/\psi \to 3\pi}  $  &
   18&6& 1.2& 1.1 &   23&6  & 1.6  & 1.6   &   20&2 & 1.4 \\ 

$\Gamma^{\psi(2S)}_{ee}$  &  $\BR_{\psi(2S) \to \omega\pipi}
                        \cdot \BR_{\omega     \to 3\pi} $  &
   2&69& 0.73& 0.16 &   1&22 & 0.33 & 0.07  &   0&66& 0.17\\ 

$\Gamma^{\psi(2S)}_{ee}$  &  $\BR_{\psi(2S) \to J/\psi\eta} 
                             \cdot \BR_{\eta \to 3\pi} 
                             \cdot \BR_{J/\psi \to \mumu} $  &
   1&11& 0.33& 0.07 &   33&4  & 9.9  & 2.0   &   30&9 & 0.8 \\ 

$\Gamma^{\psi(2S)}_{ee}$  &  $\BR_{\psi(2S) \to 2(\pipi)\eta}
                             \cdot \BR_{\eta \to \gamma\gamma} $  &
   1&13& 0.55& 0.08 &   1&2 & 0.6 & 0.1  &   \multicolumn{3}{c}{no entry}\\  

$\Gamma^{\psi(2S)}_{ee}$  &  $\BR_{\psi(2S) \to \Kp\Km\pipi\pi^0} $ &
   4&4& 1.3  & 0.3 &   1&8 & 0.5 & 0.1  &   1&24& 0.10\\        

$\Gamma^{\psi(2S)}_{ee}$  &  $\BR_{\psi(2S) \to \Kp\Km\pipi\eta} 
                             \cdot \BR_{\eta \to \gamma\gamma}$ &
   1&2 & 0.7 & 0.1  &   1&3  & 0.7  & 0.1   &  \multicolumn{3}{c}{no entry} \\ 

\end{tabular}
\end{ruledtabular}
\end{table*}

We are also able to measure $J/\psi$ and $\psi(2S)$ branching fractions
for some of the submodes studied above. 
Figure~\ref{etaomega2pi} shows expanded views of the \fourpipn mass
distribution in the charmonium region for the $\eta\pipi$ and
$\omega\pipi$ intermediate states. 
Our fits yield 8.9$\pm$3.8 and 788$\pm$51 $J/\psi$ decays to
$\eta\pipi$ and $\omega\pipi$, respectively,
and 14.2$\pm$4.2 and 37$\pm$10 $\psi(2S)$ decays.
We list the corresponding products and branching fractions
in Table~\ref{jpsitab}.
The $\psi(2S)\! \to\! \eta\pipi$ branching fraction is very
small~\cite{PDG} and the observed events are from the decay chain 
$\psi(2S)\! \to\! J/\psi\eta$, $J/\psi\! \to\! \mumu$, 
$\eta\! \to\! \pipi\piz$.
The result in Table~\ref{jpsitab} assumes this decay chain. 
We also observe 6.0$\pm$2.7 and 24$\pm$9 events in the $J/\psi$ peaks
for the $\phi\eta$ and $\Kp\Km\omega$ modes, respectively. 
The corresponding products are also listed in Table~\ref{jpsitab}.
\section{Summary}
\label{sec:Summary}
The photon energy and charged particle momentum 
resolutions together with the particle
identification capabilities of the \babar\ detector permit the
reconstruction of
$\epem\!\! \to\! \fourpipn$, \fourpieta, \KKpppn and \KKppeta
events produced at low effective \epem c.m.\@ energy 
via ISR in data taken in the $\Upsilon(4S)$  mass region.
Luminosity and efficiency can be understood with 6--10\% accuracy, 
so that ISR production yields useful measurements of $R$, 
the ratio of the hadronic to di-muon cross section values, 
used for the \ammm calculations.

Our measurements of the $\epem\!\! \to\! \fourpipn$ cross section
represent a significant improvement upon existing data in both energy
range and precision.
In addition, these data provide new information on hadron spectroscopy.
The observed $\epem\!\! \to\! \omega\pipi$ and $\eta\pipi$ cross
sections show evidence of resonant structures around 1.4--1.7~\gevcc,
which were previously observed by DM2 and interpreted as 
$\omega(1420)$ and $\omega(1650)$ resonances. 
We obtain new measurements of the parameters of these resonances, 
which confirm the results of our previous study of ISR $\pipi\piz$ events,
that the $\omega(1650)$ is substantially narrower than currently
listed in PDG.

We also use this final state to make the first measurements of the 
$\epem\!\! \to\! \omega f_0(980)$ and
$\rho(770)3\pi$ cross sections.
In the latter events, there is an iso-vector resonant structure with 
$m\! =\! 1.243\pm0.012\pm0.020~\gevcc$ and 
$\Gamma\! =\! 0.410\pm0.031\pm0.030~\gev$ 
in the three-pion system recoiling against the $\rho(770)$.

We present the first measurements of the cross sections for
$\epem\!\! \to\! \fourpieta$, $\eta'(958)\pipi$ and 
$f_1 (1285)\pipi$.
We measure the mass and width of the $f_1 (1285)$,
and observe a candidate $\rho(2150)$ resonance 
with $m_{\rho(2150)}\! =\! 2.15\pm0.04\pm0.05~\gevcc$ and
$\Gamma_{\rho(2150)}\! =\! 0.35\pm0.04\pm0.05~\gev$. 

We present the first measurements of the cross sections for
$\epem\!\! \to\! \KKpppn$, $\Kp\Km\omega$ and \KKppeta.
Using the latter final state, we measure a contribution  from 
$\epem\!\! \to\! \phi\eta$ consistent with our 
measurement in the $\Kp\Km\gamma\gamma$ final state~\cite{isrphieta}.

The final states
analyzed in this paper, based on 232~\invfb of \babar\ data in the
1.0--4.5~\gevcc mass range, are 
already better in quality and precision than the direct measurements from
the DCI and ADONE machines, and do not suffer from the relative
normalization uncertainties which seem to exist for direct
measurements of these final states.  

The ISR events also allow a study of $J/\psi$ and $\psi(2S)$ production,
and the measurement of fifteen products of branching fractions into 
observed modes and the \epem width of the $J/\psi$ or $\psi(2S)$.
Three of these are first measurements, 
and two others are the most precise measurements to date, thanks to
our relatively small systematic error due to acceptance.
We observe substantial discrepancies with respect to the previous experiments in the 
$J/\psi\! \to\! \fourpipn$ and
$\psi(2S)\! \to\! \fourpipn$ decay modes.

\section{Acknowledgments}
\label{sec:Acknowledgments}
We are grateful for the 
extraordinary contributions of our \pep2\ colleagues in
achieving the excellent luminosity and machine conditions
that have made this work possible.
The success of this project also relies critically on the 
expertise and dedication of the computing organizations that 
support \babar.
The collaborating institutions wish to thank 
SLAC for its support and the kind hospitality extended to them. 
This work is supported by the
US Department of Energy
and National Science Foundation, the
Natural Sciences and Engineering Research Council (Canada),
the Commissariat \`a l'Energie Atomique and
Institut National de Physique Nucl\'eaire et de Physique des Particules
(France), the
Bundesministerium f\"ur Bildung und Forschung and
Deutsche Forschungsgemeinschaft
(Germany), the
Istituto Nazionale di Fisica Nucleare (Italy),
the Foundation for Fundamental Research on Matter (The Netherlands),
the Research Council of Norway, the
Ministry of Education and Science of the Russian Federation, 
Ministerio de Educaci\'on y Ciencia (Spain), and the
Science and Technology Facilities Council (United Kingdom).
Individuals have received support from 
the Marie-Curie IEF program (European Union) and
the A. P. Sloan Foundation.


\newpage
\begin{thebibliography}{99}

\bibitem{baier} V.~N.~Baier and V.~S.~Fadin, Phys.~Lett.~B~{\bf 27}, 223 
(1968).

\bibitem{arbus} A.~B.~Arbuzov {\em et al.}, J. High Energy Phys. {\bf 9812}, 
009 (1998).

\bibitem{kuehn} S.~Binner, J.~H.~K\"uhn and K.~Melnikov, 
Phys. Lett. B~{\bf 459}, 279 (1999).

\bibitem{ivanch} M.~Benayoun {\em et al.}, 
Mod. Phys. Lett. A~{\bf 14}, 2605 (1999).

\bibitem{PDG}Review of Particle Physics, W.-M. Yao {\em et al.},
 J. Phys.:Nucl. Part. Phys. G~{\bf 33}, 1 (2006).

\bibitem{g2new} 
M.~Davier,
Nucl.\ Phys.\ Proc.\ Suppl.\  {\bf 169} (2007) 288. 

\bibitem{Druzhinin1}
\babar\ Collaboration, B. Aubert {\em et al.},
Phys. Rev. D~{\bf 69}, 011103 (2004).

\bibitem{isr3pi}
\babar\ Collaboration, B.\ Aubert {\em et al.},
Phys. Rev. D~{\bf 70}, 072004 (2004). 

\bibitem{isr4pi}
\babar\ Collaboration, B.\ Aubert {\em et al.},
Phys. Rev. D~{\bf 71}, 052001 (2005).

\bibitem{isr6pi}
\babar\ Collaboration, B.\ Aubert {\em et al.},
Phys. Rev. D~{\bf 73}, 052003 (2006). 

\bibitem{isr2K2pi}
\babar\ Collaboration, B.\ Aubert {\em et al.},
Phys. Rev. D.~{\bf 76}, 012008 (2007).


\bibitem{omegadm1}DM1 Collaboration, A.\ Cordier {\em et al.}, 
Phys. Lett. B~{\bf 106}, 155  (1981).

\bibitem{etadm2}DM2 Collaboration, A.\ Antonelli {\em et al.}, 
Phys. Lett. B~{\bf 212}, 133  (1988).

\bibitem{omegadm2}DM2 Collaboration, A.\ Antonelli {\em et al.}, 
Z. Phys. C~{\bf 56}, 15  (1992).

\bibitem{5picmd2}CMD2 Collaboration, A.\ Akhmetshin {\em et al.}, 
Phys. Lett. B~{\bf 489}, 125  (2000).

\bibitem{etand}ND Collaboration, V.\ Druzhinin {\em et al.}, 
Phys. Lett. B~{\bf 174}, 115  (1986).

\bibitem{babar} \babar\ Collaboration, B.\ Aubert {\em et al.}, 
\nima{479}, 1 (2002).

\bibitem{kuehn2}
H.~Czy\.z and J.~H.~K\"uhn, Eur. Phys. J. C~{\bf 18}, 497 (2001).

\bibitem{kuraev} A.~B.~Arbuzov {\em et al.},
J. High Energy Phys. {\bf 9710}, 001 (1997).

\bibitem{strfun} M.~Caffo, H.~Czy\.z, E.~Remiddi, Nuovo Cim. A~{\bf 110},
 515  (1997); Phys. Lett. B~{\bf 327}, 369 (1994). 

\bibitem{PHOTOS} E.~Barberio, B.~van~Eijk and Z.~Was, Comput. Phys.
Commun. {\bf 66}, 115 (1991).

\bibitem{GEANT4} GEANT4 Collaboration, S.\ Agostinelli {\em et al.},
\nima{506}, 250 (2003).

\bibitem{jetset} T.~Sjostrand, Comput. Phys. Commun. {\bf 82}, 74 (1994).

\bibitem{koralb} S.~Jadach and Z.~Was, Comput. Phys. Commun. {\bf 85}, 453 (1995).

\bibitem{etaprimgamma}
\babar\ Collaboration, B.\ Aubert {\em et al.},
Phys. Rev. D~{\bf 74}, 012002 (2006). 

\bibitem{5pind}ND Collaboration, A.\ Dolinsky {\em et al.}, 
Phys. Rep. {\bf 202}, 99  (1991).

\bibitem{isrphieta}
\babar\ Collaboration, B.\ Aubert {\em et al.},
To be submitted to PRD.

\bibitem{bes}
BES Collaboration, J.~Z.~ Bai {\em et al.},
Phys. Rev. D~{\bf 70}, 012005 (2004).



\end{thebibliography}
\end{document}